\documentclass[preprint]{aastex}

\newcommand{\kms}{km s$^{-1}$}
\newcommand{\Bband}{B}
\newcommand{\Vband}{V}
\newcommand{\BminusV}{({\Bband}{\rm -}{\Vband})}
\newcommand{\bminusv}{[{\Bband}{\rm -}{\Vband}]}
\newcommand{\EBV}{E\bminusv}
\newcommand{\Ebv}{E\BminusV}
\newcommand{\ebv}{$E\BminusV$}
\newcommand{\halpha}{H$\alpha$}
\newcommand{\fetwo}{\protect\ion{Fe}{2}\ $\lambda\lambda4924,~5018,~5169$}
\newcommand{\hbeta}{H$\beta$}
\newcommand{\ubvri}{\protect\hbox{$U\!BV\!RI$}}		
\newcommand{\bvri}{\protect\hbox{$BV\!RI$}}		
\newcommand{\bvi}{\protect\hbox{$BV\!I$}}		
\newcommand{\vri}{\protect\hbox{$V\!RI$}}		
\newcommand{\ubv}{\protect\hbox{$U\!BV$}}		
\newcommand{\ssp}{\def\baselinestretch{1.0}\large\normalsize}
\newcommand{\gtrsi}{\mathrel{\hbox{\rlap{\hbox{\lower4pt\hbox{$\sim$}}}\hbox{$>$}}}}
\newcommand{\vi}{\mbox{$V\!-\!I$}}

\shorttitle{EPM Distance to SN 1999em} \shortauthors{Leonard et al.}
\slugcomment{Accepted for Publication in {\it The Publications of the Astronomical Society of the Pacific}}

\begin{document}

\title{The Distance to SN~1999em in NGC 1637 from the Expanding Photosphere Method}

\vspace{2cm}

\author{Douglas C. Leonard\altaffilmark{1,2}, 
Alexei V. Filippenko\altaffilmark{1}, 
Elinor L. Gates\altaffilmark{3}, 
Weidong Li\altaffilmark{1}, 
Ronald G. Eastman\altaffilmark{4,5}, 
Aaron J. Barth\altaffilmark{6},
Schelte J. Bus\altaffilmark{7}, 
Ryan Chornock\altaffilmark{1}, 
Alison L. Coil\altaffilmark{1}, 
Sabine Frink\altaffilmark{8}, 
Carol A. Grady\altaffilmark{9}, 
Alan W. Harris\altaffilmark{10}, 
Matthew A. Malkan\altaffilmark{11}, 
Thomas Matheson\altaffilmark{1,12}, 
Andreas Quirrenbach\altaffilmark{9}, and 
Richard R. Treffers\altaffilmark{1}
}

\altaffiltext{1}{Department of Astronomy, University of California, Berkeley,
California 94720-3411}

\altaffiltext{2}{Present address: Department of Astronomy, University of
Massachusetts, Amherst, MA 01003-9305; leonard@nova.astro.umass.edu}

\altaffiltext{3}{Lick Observatory, PO Box 82, Mount Hamilton, CA 95140}

\altaffiltext{4}{Lawrence Livermore National Laboratory, Livermore, CA 94551}

\altaffiltext{5}{Department of Astronomy and Astrophysics, University of
California, Santa Cruz, CA 95064}

\altaffiltext{6}{Harvard-Smithsonian Center for Astrophysics, 60 Garden St.,
MS-20, Cambridge, MA 02138}

\altaffiltext{7}{University of Hawaii, Institute for Astronomy, 640 N. A'ohoku Place
\#209, Hilo, HI 96720}

\altaffiltext{8}{Department of Physics, University of California, San Diego,
Center for Astrophysics and Space Sciences, 9500 Gilman Drive, La Jolla, CA
92093-0424}  

\altaffiltext{9}{NASA Goddard Space Flight Center, Code 685, Greenbelt, MD
20771}

\altaffiltext{10}{Jet Propulsion Laboratory, MS 183-501, 4800 Oak Grove
Dr., Pasadena, CA 91109-8099}

\altaffiltext{11}{Department of Physics and Astronomy, University of
California, Los Angeles, CA 90095-1562}

\altaffiltext{12}{Present address: Harvard-Smithsonian Center for Astrophysics,
60 Garden St., MS-20, Cambridge, MA 02138}

\vspace{1cm}

\begin{abstract}
We present 30 optical spectra and 49 photometric epochs sampling the first 517
days after discovery of supernova (SN) 1999em, and derive its distance through
the expanding photosphere method (EPM).  SN~1999em is shown to be a Type
II-plateau (II-P) event, with a photometric plateau lasting until about 100
days after explosion.  We identify the dominant ions responsible for most of
the absorption features seen in the optical portion of the spectrum during the
plateau phase.  Using the weakest unblended absorption features to estimate
photospheric velocity, we find the distance to SN~1999em to be $D =
8.2^{+0.6}_{-0.6}$ Mpc, with an explosion date of HJD $2,451,475.6\pm{1.4}$, or
$5.3\pm{1.4}$ days before discovery.  We show that this distance estimate is
about 10\% closer than the distance that results if the strong \fetwo\
absorption features, which have often been used in previous EPM studies, are
used to estimate photospheric velocity.

We examine potential sources of systematic error in EPM-derived distances, and
find the most significant to result from uncertainty in the theoretical
modeling of the flux distribution emitted by the SN photosphere (i.e., the
``flux dilution factor'').  We compare previously derived EPM distances to 8
SNe II in galaxies (or members of the same group) for which a recently revised
Cepheid distance exists from the $HST$ Key Project and find $D_{\rm
Cepheids}/D_{\rm EPM} = 0.87 \pm 0.06$ (statistical); eliminating the 3 SNe II
distances for which a Cepheid distance only exists to a group member, and not
the host galaxy itself, yields $D_{\rm Cepheids}/D_{\rm EPM} = 0.96 \pm 0.09$.
Additional direct comparisons, especially to spectroscopically and
photometrically normal SNe II-P, will certainly help to produce a more robust
comparison.

Finally, we investigate the possible use of SNe~II-P as standard candles and
find that for 8 photometrically confirmed SNe~II-P with previously derived EPM
distances and SN~1999em, the mean plateau absolute brightness is
$\overline{M}_V~{\rm (plateau)} = -16.4\pm{0.6}$ mag, implying that distances
good to $\sim 30\%$ ($1\sigma$) may be possible without the need for a complete
EPM analysis.  At $\overline{M}_V~{\rm (plateau)} = -15.9\pm{0.2}$ mag,
SN~1999em is somewhat fainter than the average SN~II-P.  The general
consistency of absolute SNe II-P brightness during the plateau suggests that
the standard candle assumption may allow SNe~II-P to be viable cosmological
beacons at $z > 2$.

\end{abstract}

\medskip
\keywords {cosmology: observations --- distance scale --- galaxies: individual
(NGC 1637) --- supernovae: individual (SN 1999em)}

\section{INTRODUCTION}
\label{sec:introduction}

``Normal'' core-collapse supernovae (SNe) are thought to result from isolated,
massive stars (initial mass $\gtrsi 8-10\ M_{\odot}$) with thick hydrogen
envelopes (generally several solar masses) intact at the time of the explosion.
Their light curves show a distinct plateau (hence the moniker ``SN II-P''),
resulting from an enduring period (sometimes as long as 150 days) of nearly
constant luminosity as the hydrogen recombination wave recedes through the
envelope and slowly releases the energy deposited by the shock and by
radioactive decay.  It is to these events that the expanding photosphere method
(EPM) of extragalactic distance determination is most accurately applied (e.g.,
Eastman, Schmidt, \& Kirshner 1996, hereafter E96).

SNe II have not traditionally been treated as standard candles.  Through
calibration with the EPM, however, they have provided direct measurements of
distances independent of all the uncertain rungs of the extragalactic distance
ladder.  While EPM distances have been derived to both SNe II-P and SNe
II-L,\footnote{``L'' for their linearly declining light curves, lacking a
plateau; these events are generally believed to result from progenitors that
have lost a substantial fraction of their hydrogen envelope prior to exploding
(e.g., Filippenko 1997).} it is thought to be most securely applied to SNe II-P
since the theoretical modeling of their flux distribution is much simpler and
better understood.  Thus far, the EPM has been applied to 18 SNe~II, including 10
SNe~II-P, spanning distances from 0.049 Mpc (SN 1987A) to 180 Mpc (SN~1992am;
Schmidt et al. 1994b).  Using SNe~II alone, a Cepheid-independent value of
$H_{\circ} = 73 \pm 7 {\rm\ km\ s^{-1}\ Mpc^{-1}} $ has been derived (Schmidt
et al. 1994a).  Despite EPM's apparent success, however, there is considerable
debate concerning possible sources of systematic error.  In this paper, we
derive an EPM distance to the most thoroughly observed SN II-P to date,
SN~1999em, and critically evaluate possible sources of systematic error in the
EPM technique.

EPM is fundamentally a geometric technique, a variant of the Baade (1926)
method used to determine the distance to variable stars: the linear radius of
the expanding photosphere, $R$, is compared with the photosphere's angular
size, $\theta$, to derive the distance to the SN, $D$.  Naturally, since all
extragalactic SNe~are unresolved during the ``photospheric'' phase (the plateau
in a SN~II-P), $R$ and $\theta$ must be derived rather than measured directly.

In the original formulation of EPM (Kirshner \& Kwan 1974), the theoretical
angular size of the SN photosphere is approximated as
\begin{equation}
\theta = \frac{R}{D},
\label{eqn:smallangle}
\end{equation}

\noindent where $R$ the photosphere's radius and $D$ the distance to the SN.
If spherical geometry is assumed and the ejecta are in free expansion (i.e.,
deceleration due to gravity or swept-up interstellar material is negligible,
and the radius of the progenitor is small compared with the size of the
photosphere) equation~(\ref{eqn:smallangle}) may be written as
\begin{equation}
\theta = \frac{v_{\rm phot}(t - t_\circ)}{D},
\label{eqn:theta1}
\end{equation}

\noindent where $v_{\rm phot}$ is the velocity of material instantaneously at the
photosphere at time $t$, and the explosion occurred at time $t_\circ$.
Rewriting this equation in the suggestive form
\begin{equation}
t = D\left(\frac{\theta}{v_{\rm phot}}\right) + t_\circ,
\label{eqn:epmdistance}
\end{equation}

\noindent we see that plotted measurements of at least two epochs of
$\theta/v_{\rm phot}$ against $t$ should result in a line with slope $D$ and
$y$-intercept $t_\circ$.  A particularly powerful feature of this method is
that if the date of explosion ($t_\circ$) is known (either from pre-discovery
images or derived from the EPM technique itself if more than one epoch of data
is available), then each measurement of $\theta/v_{\rm phot}$ provides an {\it
independent} estimate for $D$.  To determine the distance to a SN, then,
requires measurements of the theoretical angular size, $\theta$, and the
photospheric velocity, $v_{\rm phot}$.

Assuming a spherically symmetric SN photosphere that radiates isotropically as
a blackbody, conservation of flux demands (temporarily neglecting extinction) 
\begin{equation}
4\pi R^2 \pi B_\nu(T_c) = 4\pi D^2 f_\nu,
\label{eqn:fluxconserve}
\end{equation}

\noindent where $B_\nu(T_c)$ is the Planck function at color temperature $T_c$, 
and $f_\nu$ is the flux received at Earth.  Solving for $\theta$, this becomes
\begin{equation}
\theta = \sqrt{\frac{R^2}{D^2}} = \sqrt{\frac{f_\nu}{\pi B_\nu (T_c)}} .
\label{eqn:theta2}
\end{equation}

The observables thus required to derive an EPM distance are $v_{\rm phot}$,
$f_\nu$, and $T_c$.  Traditionally, $v_{\rm phot}$ is estimated by measuring
the blueshift of the P-Cygni absorption troughs of weak lines seen in the
optical spectrum, typically \fetwo\ and occasionally \ion{Sc}{2} $\lambda 5527$
and \ion{Sc}{2} $\lambda 5658$ as well.  The implicit assumption is that these
lines are optically thin above the photosphere and that the density of the
absorbing material is a steeply declining function of radius, with the densest
material (which causes the greatest amount of absorption) being instantaneously
at the location of the photosphere.  Since strong, optically thick lines may
form absorption minima at higher velocities (i.e., well above the photosphere),
velocities derived from the weakest features are the best indicators of
photospheric velocity.  In principle, $f_\nu$ and $T_c$ could be determined
with accurate spectrophotometry.  In practice, however, spectrophotometry is
generally not available, so broadband photometry (usually some subset of
$BVIJHK$) is used.  From several measurement epochs of $v_{\rm phot}$, $T_c$,
and $f_\nu$ during the photospheric phase, then, the distance and time of
explosion of the SN are derived.

The current debate over possible sources of systematic error in the EPM
technique generally focuses on three main assumptions:  

\begin{enumerate}

\item {\it The photosphere radiates as a blackbody.}  On applying the
Baade (1926) technique to measure the distance to variable stars, Walter Baade
himself wrote: ``...the only assumption we had to make [in formulating the
technique] was that Cepheids radiate as a blackbody.  Further experience has to
show how far we can hold this assumption.''\footnote{Passage translated from
the original German by M. Modjaz.} It was not until decades later that Wagoner
(1982) first demonstrated that a Type II SN photosphere in fact radiates as a
``dilute'' blackbody due to the dominance of electron scattering over
absorption processes.  That is, the continuum spectrum that is ultimately
released from the electron-scattering photosphere (defined as the surface of
last scattering, $\tau_e = 2/3$) is produced in the deeper layer at which the
radiation field thermalizes to the local gas temperature, known as the
``thermalization depth'' (see, e.g., E96).  This smaller radius is the location of
the last true absorption and reemission of the photons that make up the thermal
continuum.  The blackbody spectrum ultimately released by the
electron-scattering photosphere thus possesses a luminosity appropriate for a
smaller radiating surface.  The amount the flux is ``diluted'' is parameterized
by the relation
\begin{equation}
\zeta = \frac{R_{\rm therm}}{R_{\rm phot}},
\label{eqn:zeta}
\end{equation}

\noindent where $R_{\rm therm}$ is the radius at the thermalization depth,
$R_{\rm phot}$ is the photosphere's radius (i.e., the surface of last
scattering), and $\zeta$ is called the ``distance correction factor,'' since
its inclusion in equation~(\ref{eqn:fluxconserve}) ``corrects'' derived distances
such that 
\begin{equation}
D_{\rm actual} = \zeta D_{\rm measured}.
\label{eqn:distancezeta}
\end{equation}

\noindent That is, distances derived without taking into account flux dilution
will overestimate the distance by a factor $\zeta$.  In principle, $\zeta$
could depend on many things, including the chemical composition (i.e., the
metallicity) and density structure of the progenitor star, and the expansion
rate and luminosity of the SN explosion.  Studies of theoretical models of
realistic SN atmospheres with a wide range of properties, however, have
demonstrated that $\zeta$ is in fact a nearly one-dimensional function of color
temperature, $T_c$, with only a small density dependence at short wavelengths
(E96; Montes \& Wagoner 1995).  

From a study of 63 model atmospheres, E96 provide convenient analytic
approximations for $\zeta$ as a function of color temperature for a SN~II-P,
determined for 4 broadband filter combinations, $BV, BV\!I, VI,\ {\rm and\ }
JHK$ (the $R$ band is generally not used, since it is dominated by the
H$\alpha$ line, which can vary significantly among SNe~II-P).  In the models of
E96, $\zeta$ is shown to vary smoothly as a function of $T_c$, with a $1\sigma$
scatter among the models of $0.11\%, 0.05\%, 0.08\%, {\rm and\ } 0.07\%$ for
the $BV, BV\!I, VI,\ {\rm and\ } JHK$\ filter combinations, respectively.
Recently, by taking into account the effects that telluric features have on
observed photometric data, as well as a subtle correction to the $\bvri$ filter
functions of Bessell [1990] to make them appropriate for use with photon, and
not energy, distributions, Hamuy et al. (2001) has provided improved estimates
of $\zeta(T_c)$, which we adopt in this paper.\footnote{Hamuy et al. (2001)
also derive an EPM distance to SN~1999em.  We note that other than adopting the
improved dilution factors our analysis here is completely independent of this
work, which was posted on astro-ph after this paper was submitted; we do,
however, make a brief comparison of the two derived distances, and comment on
differences in the techniques used, in \S~\ref{sec:sysunc}.}  With \bvi\
photometry of SN~1999em, we shall be able to test the self-consistency of the
flux-dilution factors by comparing the distances derived using the three filter
combinations $BV, BV\!I, {\rm and\ } VI$ individually.

The dilution factors of E96 have been used to determine distances to 18 SNe II
(Schmidt, Kirshner, \& Eastman 1992; Schmidt et al. 1994a, b; see also
Clocchiatti et al. 1995).  A first-order check on the precision of the
predicted {\it shape} of the function relating $\zeta$ and $T_c$ is whether
multiple observations at different epochs, and therefore different $T_c$ and
$\zeta$, are consistent.  As discussed by Leonard et al. (2001, hereafter L01),
previous EPM studies of SNe that had a broad temporal range of observational
epochs have yielded consistent distances (i.e., a relatively straight line when
$\theta/v$ is plotted against $t$ [Equation~(\ref{eqn:epmdistance})]).
Empirical evidence therefore suggests that the shape of the theoretically
derived function relating $\zeta$ and $T_c$ must be nearly correct.  However,
differences of opinion remain concerning the overall {\it level} of the
dilution factor, with discrepancies in $\zeta$ as high as 60\% claimed by some
groups (e.g., Baron et al. 1995).  In all, the uncertainty in the dilution
factor remains the main point of contention when applying EPM to derive
cosmological parameters.

\item {\it The photosphere is spherically symmetric.}  Recent
polarization studies of young core-collapse SNe (see Wheeler [2000] for a
comprehensive list of polarimetric observations of SNe; in general, higher
polarization implies greater departure from a sphere), the possible association
of SNe with some gamma-ray bursts (e.g., Bloom et al. 1999), and the high
velocity of pulsars (Cordes \& Chernoff 1998), all implicate a fundamentally
asymmetric explosion for core-collapse SNe.  From the polarization studies an
interesting trend has also emerged: the degree of asphericity at early times
appears to be inversely correlated with the amount of envelope material intact
at the time of the explosion.  That is, SNe that have lost a significant amount
of their hydrogen envelope prior to exploding are found to be more highly
polarized at early times than those that have hydrogen envelopes substantially
intact (Wheeler 2000; see also Wang et al. 2001).  The addition of envelope
material evidently serves to dampen the asymmetry at early times, in agreement
with the theoretical expectation that asymmetric explosions should turn
spherical as they expand through the envelope (e.g., Chevalier \& Soker 1989;
H\"{o}flich, Wheeler, \& Wang 1999).

The most thorough investigation of the sphericity assumption for a SN II-P is
that by L01 for SN~1999em itself.  From multiple spectropolarimetric epochs
spanning the first 163 days after discovery, L01 find a very low, but perhaps
slowly increasing polarization for SN~1999em.  Spectropolarimetric observations
of other SNe II-P also show weak evidence for intrinsic polarization,
suggesting that departures from spherical symmetry may not be substantial at
early times for this class of objects (Leonard \& Filippenko 2001).  Although
the translation from observed polarization to degree of asphericity is
obfuscated by uncertainties in the interstellar polarization and SN viewing
orientation, when the polarization of SN~1999em is interpreted in terms of the
oblate, electron-scattering models of H\"{o}flich (1991), a lower bound on the
asphericity of $7\%$ during the middle of the plateau results.  If the
asphericity implied by the spectropolarimetry results in
directionally-dependent flux, it would clearly lead to distance errors in the
EPM analysis for SN 1999em (see, e.g., Wagoner, 1991).  We caution, however,
that although global asphericity is the preferred interpretation of the
observed polarization, there are other mechanisms that can produce SN
polarization, including scattering by dust (e.g., Wang \& Wheeler 1996) and
asymmetrically distributed radioactive material within the SN envelope (e.g.,
H\"{o}flich 1995; H\"{o}flich, Khokhlov, \& Wang 2001).  Thus, while the
question is far from settled, available polarimetric evidence is consistent
with small asphericity during the period of interest for EPM.

Further support for early-time SN sphericity comes from the apparent success of
the EPM technique itself.  By quantifying the amount of systematic scatter in
the EPM Hubble diagram, L01 find that the agreement between previous EPM
distance measurements of SNe II and the distances to the host galaxies
predicted by a linear Hubble law restrict mean SN II asphericity to values less
than $30\%$ ($3\sigma$) during the photospheric epoch.  From the available
evidence, then, there is growing support for the claim that asphericity does
not significantly hamper the cosmological use of SNe II whose progenitors have
massive hydrogen envelopes intact at the time of explosion.

\item {\it The photospheric velocity is accurately determined by measuring the
blueshift of weak P-Cygni absorption lines.}\label{sec:blueshift} In all
previous EPM studies, photospheric velocity has been estimated by measuring the
blueshift of the flux minima in the P-Cygni absorption profiles of lines in the
optical spectrum (e.g., Kirshner \& Kwan 1974).  The main assumptions here are
that the lines in the SN envelope are optically thin above the photosphere, and
that density is a steeply declining function of radius.  We also note that due
to the multiple scatters that occur to photons as they travel from the location
of the thermalization depth to the surface of last scattering (i.e., $\tau_e =
2/3$, the optical photosphere), line features are not expected to be formed
significantly below the optical photosphere (i.e., Hatano et al. 1999; cf.,
Hamuy et al. [2001]).  With these ideas in mind, then, it is clear that the
greatest amount of absorption will occur just above the photosphere, and that
velocities measured with weak lines should be good photospheric velocity
indicators.

The very early-time (less than 10 days after core collapse) spectra of SN~1987A
were modeled in detail by Schmutz et al. (1990) and Eastman \& Kirshner (1989).
The hydrogen Balmer lines are shown by Eastman \& Kirshner (1989) to yield
velocities that become a progressively poorer representation of the
photospheric velocity with time: they are within 10\% of the photospheric
velocity two days after collapse, but by ten days \halpha, \hbeta, and
H$\gamma$ yield velocities 1.9, 1.5, and 1.4 times the photospheric velocity,
respectively.  The explanation for this behavior comes from examining the
density structure of the material overlying the photosphere.  At early times,
the atmosphere is highly ionized and the photosphere is situated in a region
with a density that is a very steeply declining function of radius (e.g.,
Matzner \& McKee 1999).  As the atmosphere cools and starts to recombine the
lines form in a much more extended region characterized by a flatter density
gradient (Eastman \& Kirshner 1989).  The flatter the density gradient the more
extended the line-forming region becomes, which produces the larger blueshift
of the optically thick lines.  At {\it very} early times, then, we expect all
lines to converge on the photospheric velocity, but quite soon thereafter
weaker features must be used.  Both Schmutz et al. (1990) and Eastman \&
Kirshner (1989) conclude that the weak \fetwo\ lines provide reliable
photospheric velocities during the early period studied for SN~1987A.
Encouraged by these results, these features have been used as photospheric
velocity indicators in most subsequent EPM distance determinations (e.g.,
Schmidt et al. 1992; Schmidt et al. 1994a, b; Clocchiatti et al. 1995), with
the occasional use of \ion{Sc}{2} $\lambda 5527$ and \ion{Sc}{2} $\lambda 5658$
as well.  Most of these EPM studies, however, used data from significantly
later evolutionary epochs than those used for SN~1987A.

It is possible that the \fetwo\ features may themselves become optically thick
above the photosphere as the temperature cools to the recombination temperature
of hydrogen.  Hatano et al. (1999) derive the expected local thermodynamic
equilibrium (LTE) optical depths at the photosphere as a function of
temperature for prominent absorption lines from each of the major ions thought
to be responsible for features in a SN atmosphere with solar metallicity.  The
predicted depths nicely reflect the order of appearance of lines generally seen
in SNe~II-P as they cool: first hydrogen Balmer (which is shown to become
optically thick below 12,000 K) and \ion{He}{1} $\lambda 5876$ followed by
\ion{Fe}{2}, \ion{Si}{2}, \ion{Sc}{2}, and then other weaker metal lines.
Using results derived in \S~\ref{sec:obsandred} and
\S~\ref{sec:sn1999emepmdistance}, Figure~\ref{fig:5.17} illustrates the
color-temperature (in this case, the blackbody temperatures determined by the
\bvi\ photometry, $T_{\rm BVI}$) evolution of line strengths for SN~1999em.
The line strengths and order of appearance are in good qualitative agreement
with the predictions of Hatano et al. (1999); of particular interest, Hatano et
al. (1999) predict \ion{Fe}{2} $\lambda 5018$ to be optically thick above the
photosphere at temperatures below $T_{\rm eff} \approx 7500$~K, suggesting that
velocities derived using this line during the recombination phase may slightly
overestimate photospheric velocity. By comparing the velocity derived from the
\fetwo\ features with that inferred from weaker lines, we shall empirically
test this for SN~1999em.


\begin{figure}
\ssp
\begin{center}
\rotatebox{90}{
\scalebox{0.8}{
\plotone{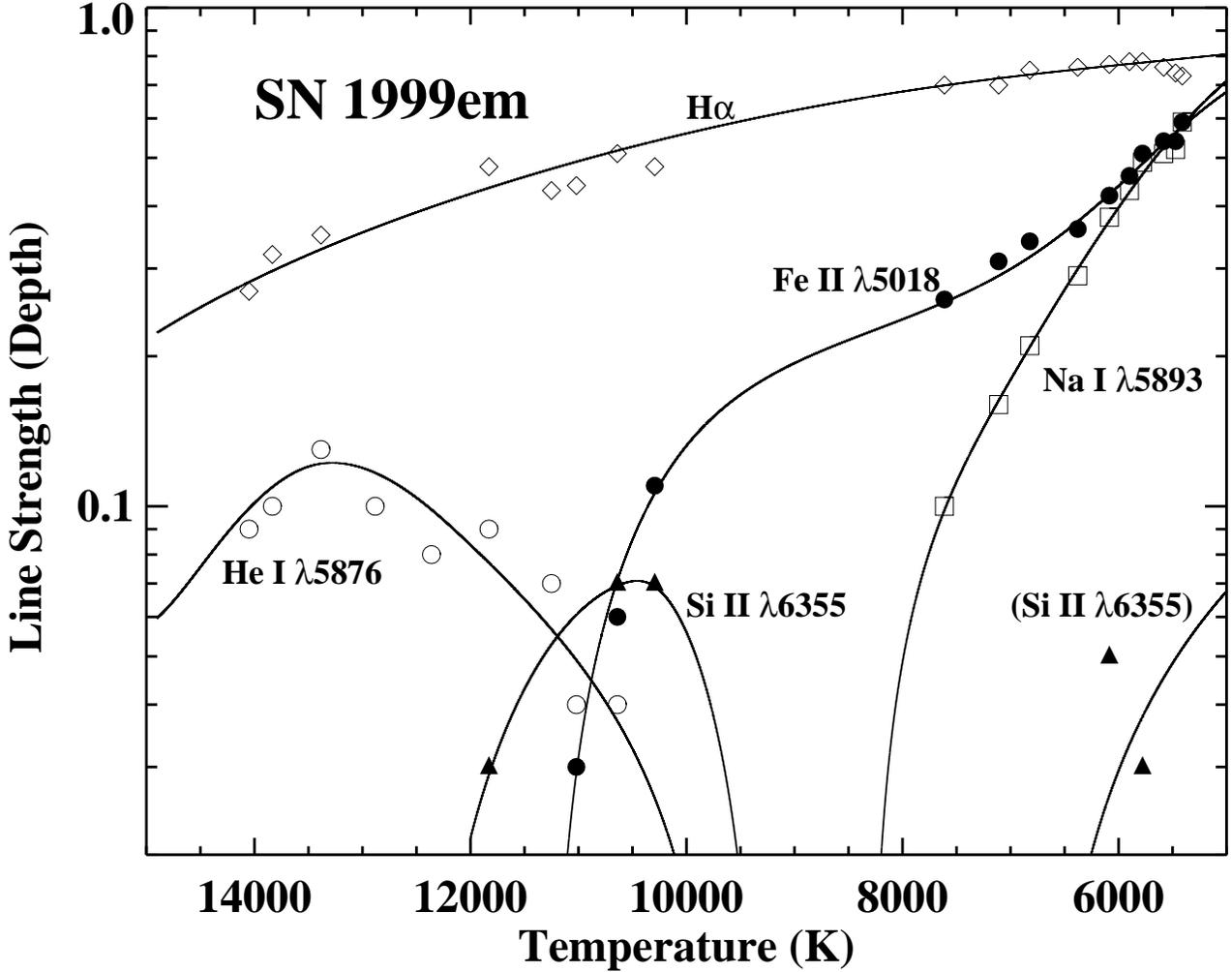}
}
}
\end{center}
\caption{The strength of several optical absorption lines plotted against \bvi\
color temperature for SN~1999em.  Approximate spline fits to the observations
are shown for clarity.  Line strength is set equal to the measured line depth,
defined as ($f_{\rm c} - f_{\rm min} ) / f_c$, where $f_{\rm min}$ is the flux
at the line's minimum and $f_c$ is the value of the flux in the surrounding
regions interpolated over the absorption line (\S~\ref{sec:inferringvel}).  The
order of appearance and relative strengths of the lines are in good qualitative
agreement with the theoretical predictions of Hatano et al. (1999).  Note the
presence of \ion{Si}{2} $\lambda 6355$ absorption at high photospheric
temperature ($T_{\rm BVI} > 10,000 $K) and then again when the photosphere has
cooled to below $T_{\rm BVI} \approx 6,000 $K.}
\label{fig:5.17}
\end{figure}

\end{enumerate}

With the preceding discussion in mind, we now explicitly include the
effects of extinction ($A_\nu$) and flux dilution ($\zeta$) in
equation~(\ref{eqn:fluxconserve}) and rewrite equation~(\ref{eqn:theta2}) as
\begin{equation}
f_\nu = \zeta_\nu^2\theta^2\pi B_\nu(T_c)10^{-0.4A_\nu}.
\label{theta2revised}
\end{equation}

\noindent If accurate spectrophotometry were available, we could solve for
$\theta$ and $T_c$ by minimizing the quantity ($f_\nu - \zeta_\nu^2\theta^2\pi
B_\nu[T_c]10^{-0.4A_\nu}$) over the observed frequency range.  Since such data
are seldom obtained we recast equation (\ref{theta2revised}) in terms of
broadband photometry.  Following E96, we replace $f_\nu$ with the observed
broadband magnitude (for brevity writing only \bvi\ here) to arrive at
\begin{equation}
m_{BV\!I} = -5\log \theta - 5 \log [\zeta(T_c)] + b_{BV\!I} + A_{BV\!I},
\label{eqn:myequation}
\end{equation}

\noindent where $b_{BVI}$ is the broadband magnitude of the Planck function integrated
over the transmission function for the $B,V, {\rm or\ } I$ bandpass (plus an
arbitrary constant).  Armed with an estimate of the extinction and a set of
photometric magnitudes, we solve for $\theta$ and $T_c$ by minimizing the
quantity
\begin{equation}
\varepsilon = \sum_{BVI}^{} [m_{BVI} - A_{BVI} + 5\log \theta + 5\log [\zeta(T_c)] - b_{BVI}(T_c)]^2.
\label{eqn:minimize}
\end{equation}

\noindent With $v_{\rm phot}$ measured from spectral lines and $\theta$
derived from photometry, the distance and explosion date of the SN are then
determined through equation (\ref{eqn:epmdistance}) by finding the best-fitting
line to describe the data, either through the method of least squares or a more
robust (i.e., less sensitive to outliers) estimator, such as the
criterion of least absolute deviations (e.g., Press et al. 1992).

SN 1999em was discovered on 1999 October 29.44 UT (UT dates are used throughout
this paper) by Li (1999) at an unfiltered magnitude of $m \approx 13.5$ mag in
the nearly face-on ($i \approx 32^{\circ}$, from
LEDA\footnote{http://www-obs.univ-lyon1.fr/leda/home\_leda.html.}) SBc galaxy
NGC~1637 (Fig.~\ref{fig:5.19a}).  Since an image of the same field on
Oct. 20.45 showed nothing at the position of SN 1999em (limiting mag about
19.0; Li 1999), it was likely discovered shortly after explosion.  It was
quickly identified as a Type II event, with prominent P-Cygni features of
hydrogen Balmer and \ion{He}{1} $\lambda$5876 and a blue continuum (Jha et
al. 1999).  Immediately after discovery a spectral, spectropolarimetric, and
photometric campaign was initiated at Lick Observatory.  Nearly daily optical
photometry and spectroscopy of the young SN were obtained with the 0.8 m
Katzman Automatic Imaging Telescope (KAIT; Treffers et al. 1997; Li et
al. 2000; A. V. Filippenko et al., in preparation) and the Lick 1-m Nickel
reflector, respectively, providing unprecedented temporal coverage of this
Type~II event during the photospheric phase.  The spectropolarimetric
observations are discussed in detail by L01; here we focus on the spectral and
photometric data.

In this paper we derive an EPM distance to SN~1999em, and present 30 optical
spectra and 49 photometric epochs spanning the first 517 days of its
development.  We present the photometric data in \S~\ref{sec:photometry}, and
the total flux spectra in \S~\ref{sec:spectra}, where we also identify and
discuss the dominant ions responsible for the absorption features seen in the
optical spectrum during the plateau phase.  We consider the reddening of
SN~1999em in \S~\ref{sec:reddening}.  We derive and compare the photospheric
velocities inferred from the P Cygni absorption minima of several line features
in \S~\ref{sec:photovel}, and apply the EPM to SN~1999em in
\S~\ref{sec:sn1999emepmdistance}.  We discuss the results in
\S~\ref{sec:discussion}, and summarize our main conclusions in
\S~\ref{sec:conclusions}.

\section{Reductions and Analysis}
\label{sec:obsandred}

\subsection{Photometry}
\label{sec:photometry}

All photometric data were collected using the KAIT, which is equipped with a
$512 \times 512$ pixel Apogee CCD camera (AP7) located at the $f/8.17$
Cassegrain focus, providing a field of view of $6\farcm7 \times 6\farcm7$ with
$0\farcs8$ per pixel.  The ``seeing'', estimated by the full width at half maximum
(FWHM) of stars on the CCD frame, generally ranged from $3\arcsec$ to
$4\farcs5$, and exposure times of 3 to 4 minutes in \vri\ and 5 to 7 minutes in
$UB$ were typical during the first 95 days after discovery while the SN was
bright and on the plateau portion of its light curve.  After SN~1999em dropped
off the plateau, exposure times were increased to about 10 minutes.

CCD frames were flatfielded using twilight-sky flats in the usual manner, and
cosmic rays were removed interactively by visual inspection of each frame. A
noticeable low-level pattern was removed from each image by subtracting a dark
frame scaled to the same exposure time; however, on some images, particularly
the longer $U$ and $B$ exposures, a pattern still remained, most likely caused
by temperature fluctuations of the thermoelectrically cooled CCD camera (Modjaz
et al. 2001).  This was removed by interactively adding or subtracting an
additional fraction (generally $< 10\%$) of the dark frame, until the pattern
disappeared.  Photometric tests performed on frames before and after the
additional dark subtraction indicate that the low-level pattern has a negligible
effect on the photometry.  Considerable fringing remained in $I$-band images
that did not properly flatten, likely produced by the varying intensity of
night-sky emission lines.  This also has minimal impact on the photometry since
both the SN and the comparison stars were much brighter than the background.


\begin{figure}
\ssp
\begin{center}
\rotatebox{0}{
\scalebox{1.0}{
\plotone{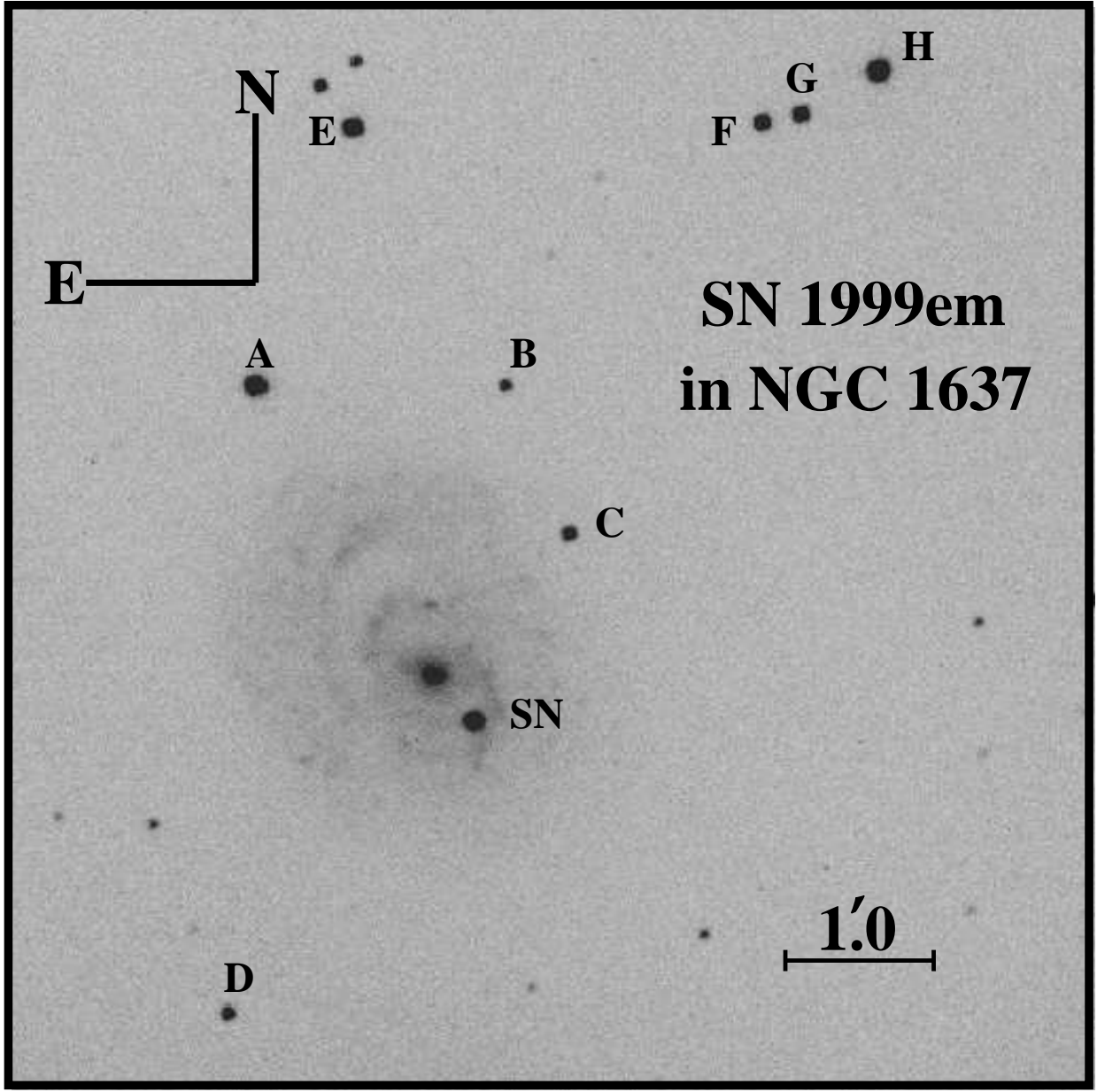}
}
}
\end{center}
\caption{$B$-band image of NGC 1637 taken on 1999 November 1 with the Katzman
Automatic Imaging Telescope.  SN~1999em ({\it SN}) is measured to be 15\farcs1
west and 17\farcs2 south of the galaxy nucleus (cf. Jha et al. 1999).  The
local standards listed in Table 1 are marked.}
\label{fig:5.19a}
\end{figure}

Figure~\ref{fig:5.19a} shows a KAIT $B$-band image of NGC 1637 taken on 1999
November 1.  Of the 8 ``local
standards'' identified in the field of SN~1999em, the best were judged to be
the three bright, isolated stars labeled {\it A, B,} and {\it C}, and they were
used to measure the relative SN brightness on non-photometric nights.  The
absolute calibration of the field was accomplished on the two photometric
nights of 2000 April 3 and September 25 by observing several fields of Landolt
(1992) standards over a range of airmasses.  The mean \ubvri\ magnitudes of the
stars on the two nights are given in Table 1.  The transformation coefficients
to the standard Johnson-Cousins (Johnson et al. 1966 for \ubv; Cousins 1981 for
$RI$) systems were derived using the solutions from these nights and two
additional photometric nights, 2000 September 5 and 27.  We determined the
instrumental magnitudes for the standards using aperture photometry with the
IRAF\footnote{IRAF is distributed by the National Optical Astronomy
Observatories, which are operated by the Association of Universities for
Research in Astronomy, Inc., under cooperative agreement with the National
Science Foundation.} DAOPHOT package (Stetson 1987, 1991), which yielded color
terms for the KAIT observations of the form
\begin{eqnarray}
U & = & u + 0.074(U - B) + C_U,\nonumber\\
B & = & b + 0.06(B - V) + C_B,\nonumber\\
V & = & v - 0.04(B - V) + C_V,\\
R & = & r - 0.08(V - R) + C_R,\nonumber\\
I & = & i + 0.01(V - I) + C_I,\nonumber
\label{eqn:photometric_solutions}\nonumber
\end{eqnarray}
\noindent where $ubvri$ are the instrumental and \ubvri\  the standard
Johnson-Cousins magnitudes. The terms $C_U, C_B, C_V, C_R,$ and $C_I$ are the
differences between the zero-points of the instrumental and standard
magnitudes, determined for each observation by measuring the offset between the
instrumental and standard magnitudes and colors of the local standard stars.

The location of SN~1999em at about $23\arcsec$ from the center of NGC~1637
(Figure~\ref{fig:5.19a}) on the inner edge of a spiral arm gives it a rather
complex background.  In such situations, it is desirable to wait until the SN
fades beyond detection to obtain a ``template'' observation of the host galaxy,
which can then be subtracted from the earlier observations (Filippenko et
al. 1986; Hamuy et al. 1994; \\

\begin{deluxetable}{lccccc}
\renewcommand{\arraystretch}{1.0}
\tablenum{1}
\tablewidth{235 pt}
\tablecaption{Magnitudes of Local Standards}
\tablehead{\colhead{Star}  &
\colhead{$U$}  &
\colhead{$B$} &
\colhead{$V$} &
\colhead{$R$} &
\colhead{$I$} } 

\startdata

A & 14.97 & 13.63 & 12.42 & 11.75 & 11.17 \\
B &\nodata& 16.75 & 15.77 & 15.18 & 14.69 \\
C & 15.66 & 15.74 & 15.12 & 14.76 & 14.38 \\
D & 15.96 & 16.21 & 15.61 & 15.25 & 14.89 \\
E & 15.44 & 14.16 & 12.99 & 12.37 & 11.84 \\
F & 15.21 & 14.99 & 14.30 & 13.94 & 13.59 \\
G &\nodata& 14.22 & 13.50 & 13.09 & 12.72 \\
H & 13.85 & 13.51 & 12.79 & 12.39 & 12.04 \\

\enddata



\end{deluxetable}
\clearpage

\noindent Richmond et al.  1995), leaving a ``clean'' image
of the SN with no galaxy contamination.  However, nearly a year after discovery
SN1999em still remained quite detectable in 10-min KAIT exposures, especially
in the \vri\ bandpasses.  Fortunately, during the period of main interest to
this study (the first 100 days), SN~1999em was over 6 mag brighter than the
galaxy background, reducing uncertainty due to improper background removal to
at most the few percent level (e.g., Suntzeff et al. 1999).  To obtain the most
accurate photometric measurements without the benefits of galaxy subtraction,
we used point-spread function (PSF) fitting by determining a PSF for each image
with DAOPHOT using bright (but unsaturated) isolated stars in the field.  We
used only the inner core of SN~1999em and the local standards to fit the PSF in
order to reduce errors that can be introduced when there is a strong gradient
in the background (e.g., Schmidt et al. 1993).  In practice, this core was
generally set to be about the FWHM of a given image.

One advantage of using the PSF fitting technique over simple aperture
photometry is that it produces an image with the SN and local standard stars
subtracted away, which allows a visual check on whether the brightness of the
SN has been accurately measured.  While the fitting radius of the SN and
comparison stars was varied from night to night to match the seeing, the sky
background of the SN and local standards was always set to an annulus with a
radius of $20 - 26$ pixels ($16\farcs0 - 20\farcs8$) to maintain consistency
throughout the observations.  We then subtracted the mode of the sky
background\footnote{See Da Costa (1992) for a discussion of the advantages of
using the mode of the background region rather than the mean or median.} to
derive the instrumental magnitudes for the SN and local standards, and
transformed them to the standard Johnson-Cousins system by taking the weighted
mean of the values obtained using the three calibrator stars.  Due to the
faintness of the field stars in the $U$ band, only star {\it A} was used to
transform the $U$-band magnitudes.  This limitation for $U$, coupled with the
general difficulty of calibrating the $U$ band using traditional CCDs with poor
blue response (see, e.g., Suntzeff 2000), makes us regard the $U$ photometry as
systematically suspect at the $0.1 - 0.2$ mag level.

\begin{deluxetable}{lcccccc}
\ssp
\tablenum{2}
\tablewidth{450pt}
\tablecaption{Photometric Observations of SN 1999em}
\tablehead{\colhead{UT Date\tablenotemark{a}}  &
\colhead{Day\tablenotemark{b}}  &
\colhead{$U$ ($\sigma_U$)} &
\colhead{$B$ ($\sigma_B$)} &
\colhead{$V$ ($\sigma_V$)} &
\colhead{$R$ ($\sigma_R$)} &
\colhead{$I$ ($\sigma_I$)} }

\startdata
1999-10-30 & 1.0 & 13.03 (0.27) & 13.87 (0.02) & 13.87 (0.01) & 13.67 (0.02) & 13.65 (0.08) \\ 
1999-10-31 & 2.0 & 13.00 (0.11) & 13.80 (0.03) & 13.80 (0.02) & 13.62 (0.03) & 13.58 (0.03) \\ 
1999-11-01 & 3.0 & 13.03 (0.12) & 13.80 (0.02) & 13.79 (0.02) & 13.60 (0.02) & 13.56 (0.03) \\ 
1999-11-02 & 4.0 & 13.11 (0.09) & 13.82 (0.03) & 13.79 (0.02) & 13.59 (0.02) & 13.56 (0.03) \\ 
1999-11-03 & 5.0 & 13.16 (0.14) & 13.85 (0.02) & 13.79 (0.01) & 13.57 (0.02) & 13.54 (0.02) \\ 
1999-11-05 & 7.0 & 13.29 (0.09) & 13.92 (0.03) & 13.84 (0.01) & 13.59 (0.02) & 13.53 (0.03) \\ 
1999-11-06 & 8.0 & 13.34 (0.09) & 13.95 (0.02) & 13.84 (0.01) & 13.58 (0.02) & 13.50 (0.03) \\ 
1999-11-07 & 9.0 & 13.41 (0.09) & 13.99 (0.01) & 13.86 (0.02) & 13.58 (0.02) & 13.51 (0.02) \\ 
1999-11-09 & 11.0 & \nodata & 14.02 (0.04) & 13.84 (0.04) & 13.54 (0.05) & 13.48 (0.04) \\ 
1999-11-10\tablenotemark{c} & 12.0 & \nodata & 14.07 (0.15) & 13.83 (0.10) & \nodata & \nodata \\ 
1999-11-11 & 13.0 & 13.90 (0.11) & 14.09 (0.01) & 13.80 (0.01) & 13.53 (0.02) & 13.44 (0.02) \\ 
1999-11-12 & 14.0 & 14.02 (0.14) & 14.15 (0.01) & 13.81 (0.01) & 13.52 (0.02) & 13.43 (0.02) \\ 
1999-11-13 & 15.0 & 14.18 (0.12) & 14.20 (0.01) & 13.81 (0.02) & 13.53 (0.01) & 13.42 (0.01) \\ 
1999-11-14 & 16.0 & 14.37 (0.12) & 14.25 (0.02) & 13.81 (0.04) & 13.56 (0.06) & 13.44 (0.03) \\ 
1999-11-16 & 18.0 & 14.56 (0.10) & 14.34 (0.02) & 13.85 (0.03) & 13.56 (0.07) & 13.44 (0.06) \\ 
1999-11-19 & 21.0 & 14.97 (0.11) & 14.47 (0.02) & 13.86 (0.03) & 13.56 (0.03) & 13.40 (0.03) \\ 
1999-11-26 & 28.0 & 15.46 (0.13) & 14.73 (0.02) & 13.91 (0.02) & 13.57 (0.02) & 13.35 (0.02) \\ 
1999-11-28 & 30.0 & 15.56 (0.16) & 14.79 (0.04) & 13.94 (0.03) & 13.59 (0.02) & 13.35 (0.04) \\ 
1999-12-02 & 34.0 & 15.77 (0.10) & 14.87 (0.03) & 13.93 (0.02) & 13.56 (0.02) & 13.31 (0.02) \\ 
1999-12-04 & 36.0 & 15.81 (0.12) & \nodata & 13.90 (0.03) & 13.54 (0.02) & 13.29 (0.02) \\ 
1999-12-06 & 38.0 & 15.96 (0.13) & 14.94 (0.02) & 13.93 (0.02) & 13.55 (0.03) & 13.29 (0.02) \\ 
1999-12-08 & 39.9 & 16.02 (0.09) & 14.97 (0.02) & 13.93 (0.02) & 13.54 (0.02) & 13.28 (0.03) \\ 
1999-12-11\tablenotemark{d} & 42.9 & 16.21 (0.11) & 14.98 (0.04) & 13.91 (0.03) & 13.51 (0.03) & 13.22 (0.05) \\ 
1999-12-14 & 45.9 & 16.31 (0.15) & 15.04 (0.04) & 13.93 (0.03) & 13.50 (0.03) & 13.23 (0.04) \\ 
1999-12-15 & 46.9 & 16.34 (0.10) & 15.05 (0.03) & 13.93 (0.04) & 13.52 (0.02) & 13.22 (0.04) \\ 
1999-12-18 & 49.9 & 16.45 (0.11) & 15.09 (0.03) & 13.94 (0.02) & 13.50 (0.03) & 
13.21 (0.03) \\ 
1999-12-29 & 60.9 & 16.96 (0.13) & 15.27 (0.02) & 13.99 (0.02) & 13.50 (0.03) & 13.19 (0.04) \\ 
2000-01-03\tablenotemark{e} & 65.9 & 16.96 (0.25) & 15.32 (0.03) & 14.02 (0.04) & 13.51 (0.02) & 13.17 (0.07) \\ 
2000-01-08 & 70.8 & 17.01 (0.10) & 15.38 (0.02) & 14.03 (0.02) & 13.50 (0.02) & 13.22 (0.03) \\ 
2000-01-13 & 75.8 & 17.26 (0.16) & 15.40 (0.04) & 14.04 (0.03) & 13.51 (0.02) & 13.22 (0.03) \\ 
2000-01-14 & 76.8 & 17.33 (0.15) & \nodata & \nodata & \nodata & \nodata \\ 
2000-01-15\tablenotemark{f} & 77.8 & \nodata & 15.52 (0.09) & 14.07 (0.05) & \nodata & 13.23 (0.09) \\ 
2000-01-27 & 89.8 & \nodata & 15.63 (0.05) & 14.17 (0.01) & 13.60 (0.04) & 13.31 (0.03) \\ 
2000-02-02 & 95.8 & \nodata & 15.74 (0.02) & 14.27 (0.02) & 13.69 (0.04) & 13.38 (0.03) \\ 

\enddata


\end{deluxetable}

\begin{deluxetable}{lcccccc}
\ssp
\tablenum{2}
\tablewidth{450pt}
\tablecaption{Photometric Observations of SN 1999em -- {\it Continued}}
\tablehead{\colhead{UT Date\tablenotemark{a}}  &
\colhead{Day\tablenotemark{b}}  &
\colhead{$U$ ($\sigma_U$)} &
\colhead{$B$ ($\sigma_B$)} &
\colhead{$V$ ($\sigma_V$)} &
\colhead{$R$ ($\sigma_R$)} &
\colhead{$I$ ($\sigma_I$)} }
\startdata
2000-02-07 & 100.8 & \nodata & 15.87 (0.03) & 14.38 (0.02) & 13.77 (0.02) & 13.46 (0.04) \\ 
2000-03-01\tablenotemark{d} & 123.8 & \nodata & \nodata & \nodata & 15.04 (0.12) & \nodata \\ 
2000-03-04 & 126.8 & \nodata & 17.77 (0.05) & 16.25 (0.03) & 15.27 (0.03) & \nodata \\ 
2000-03-11 & 133.7 & \nodata & 17.99 (0.05) & 16.37 (0.03) & 15.40 (0.04) & 14.93 (0.07) \\ 
2000-03-16 & 138.7 & \nodata & 17.97 (0.09) & 16.42 (0.03) & 15.42 (0.03) & 14.93 (0.03) \\ 
2000-03-21\tablenotemark{f} & 143.7 & \nodata & 18.06 (0.12) & 16.36 (0.05) & 15.41 (0.03) & 14.93 (0.04) \\ 
2000-03-26 & 148.8 & \nodata & 18.10 (0.10) & 16.49 (0.04) & 15.47 (0.03) & 15.01 (0.03) \\ 
2000-04-03 & 156.8 & \nodata & 17.98 (0.09) & 16.55 (0.04) & 15.61 (0.03) & 15.11 (0.03) \\ 
2000-08-25 & 301.1 & \nodata & \nodata & \nodata & 16.79 (0.06) & \nodata \\ 
2000-09-09 & 316.1 & \nodata & \nodata & \nodata & 16.82 (0.06) & 16.31 (0.07) \\ 
2000-09-16 & 323.1 & \nodata & \nodata & \nodata & 16.81 (0.06) & 16.37 (0.08) \\ 
2000-09-23 & 330.1 & \nodata & \nodata & \nodata & 16.85 (0.06) & 16.34 (0.07) \\ 
2000-09-25 & 332.1 & \nodata & 18.79 (0.08) & 17.84 (0.07) & 16.97 (0.06) & 16.39 (0.06) \\ 
2000-10-01 & 338.1 & \nodata & \nodata & \nodata & 17.05 (0.08) & 16.52 (0.09) \\ 
2000-10-13 & 350.0 & \nodata & \nodata & \nodata & 17.15 (0.07) & 16.63 (0.08) \\ 

\enddata

\tablenotetext{a}{yyyy-mm-dd.}
\tablenotetext{b}{Day since discovery, 1999-10-29 UT (HJD 2,451,480.94).}
\tablenotetext{c}{Cloudy.}
\tablenotetext{d}{Poor seeing ($\gtrsi 6^{\prime\prime}$).}
\tablenotetext{e}{Poor guiding -- images trailed.}
\tablenotetext{f}{Bright moon.}


\end{deluxetable}

\begin{figure}
\ssp
\begin{center}
\rotatebox{0}{
\scalebox{0.9}{
\plotone{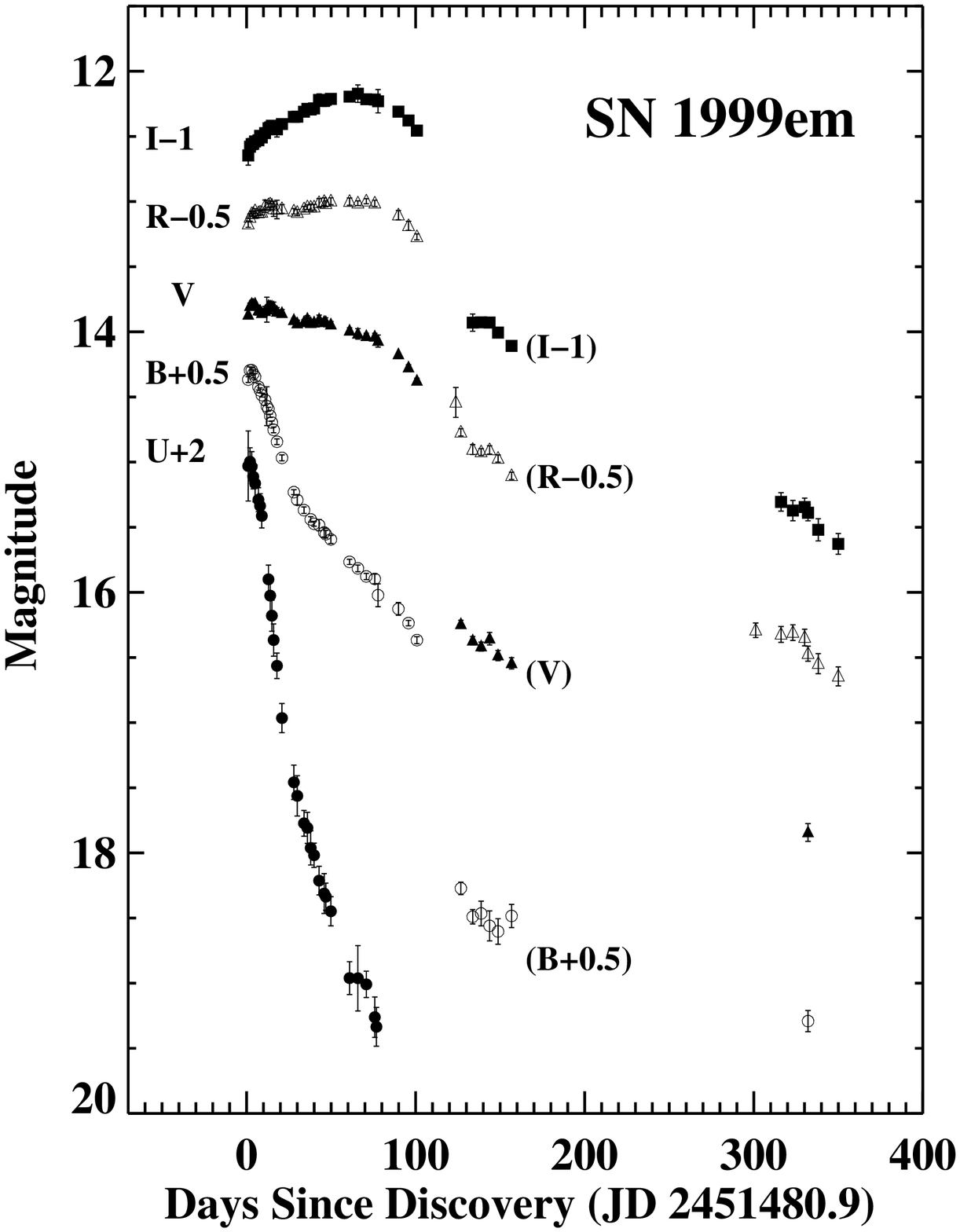}
}
}
\end{center}
\caption{\ubvri\ light curves for SN
1999em from Table 2.  For clarity, the magnitude scales for {\it UBRI} have
been shifted by the amounts indicated.}
\label{fig:5.15}
\end{figure}

\begin{figure}
\ssp
\begin{center}
\rotatebox{0}{
\scalebox{0.9}{
\plotone{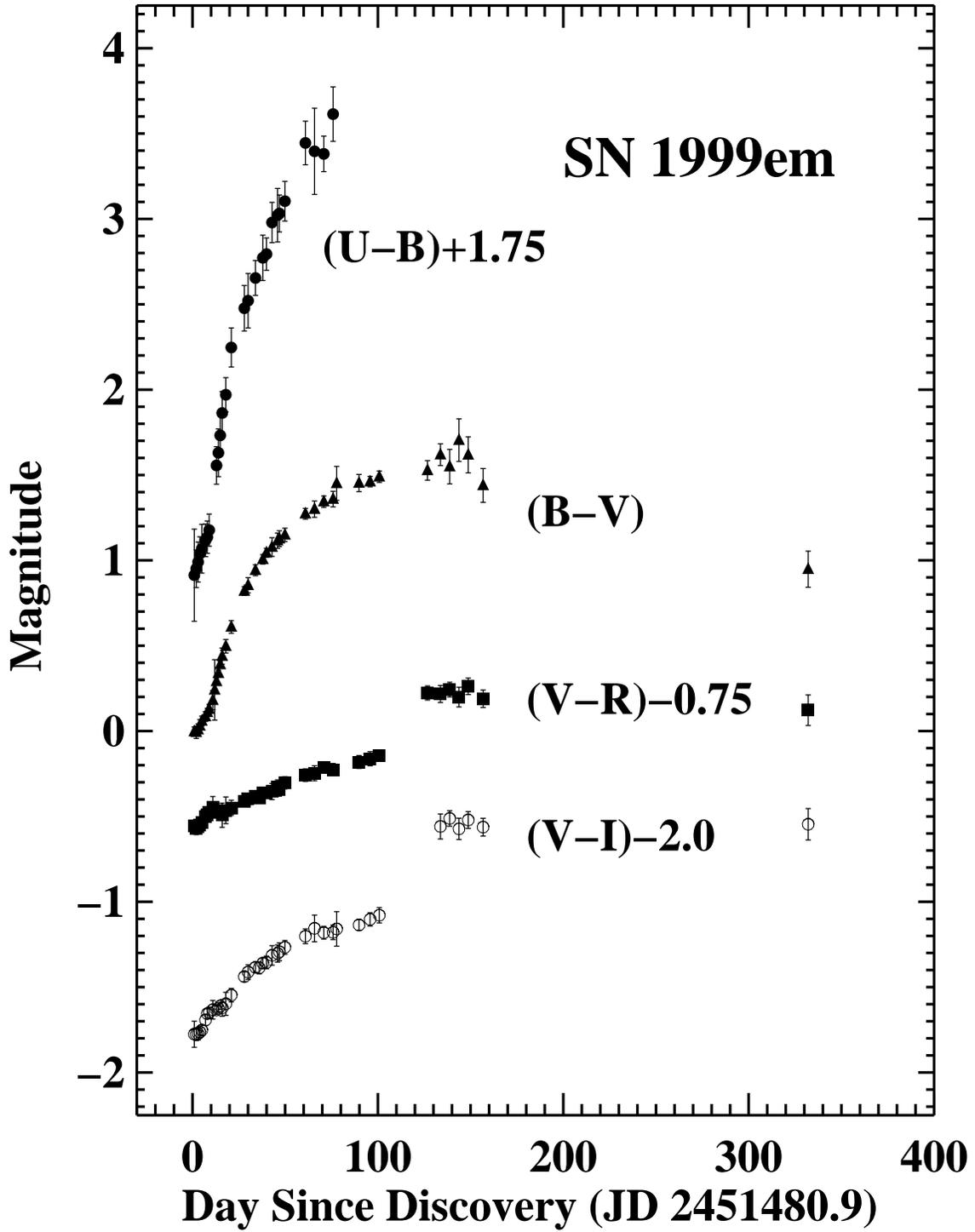}
}
}
\end{center}
\caption{\ub, \bv, \vr, and
\vi\ color curves of SN~1999em.  The magnitude scales for \ub, \vr, \vi\ have
been shifted by the amounts indicated.}
\label{fig:5.16}
\end{figure}

The results of our photometric observations are given in Table 2 and shown in
Figures \ref{fig:5.15} and \ref{fig:5.16}.  The reported uncertainties come
from two sources.  First, there is the photometric uncertainty reported by the
error analysis package in DAOPHOT from the statistics of the SN and background
region.  Second, there is uncertainty in the transformation to the standard
system.  We estimate the transformation error by taking the standard deviation
of the spread of the standard magnitudes obtained using each of the local
standard stars.  Since three values do not determine a robust
standard deviation, we opted to take a conservative estimate of the
transformation error by measuring the standard deviation of the values obtained
from all eight of the stars labeled in Figure~\ref{fig:5.19a}.  Due to the
faintness of the local standards in the $U$ band, the transformation error in
that bandpass was generally obtained using fewer stars.  The photometric and
transformation errors were then added in quadrature to obtain the uncertainty
reported in Table 2; in most cases, the total error was dominated by the
uncertainty in the transformation.


The \vri\ light curves of SN~1999em shown in Figure~\ref{fig:5.15} indicate a
clear plateau lasting until about 95 days from discovery.  The plateau phase is
characterized by a steeply declining $U$-band brightness, a declining $B$-band
brightness, a slightly declining $V$-band brightness, a nearly constant
$R$-band brightness, and an $I$-band brightness that increases through the
first 70 days and then slowly declines for the remainder of the plateau.  The
$U$-band and $B$-band light curves decline by an average of 0.06 and 0.02 mag
day$^{-1}$, respectively, during the plateau epoch after an initial quick rise
to maximum on about day 2 after discovery.  We measure the decline in the $B$
band over the first 100 days after maximum to be $\beta^B_{100} = 2.1$ mag,
firmly establishing SN~1999em as a Type II-P event according to the definition
of Patat et al. (1994), used to discriminate between SNe~II-P ($\beta^B_{100} <
3.5$ mag) and SNe~II-L ($\beta^B_{100} > 3.5$ mag).

\begin{figure}
\ssp
\begin{center}
\rotatebox{90}{
\scalebox{0.7}{
\plotone{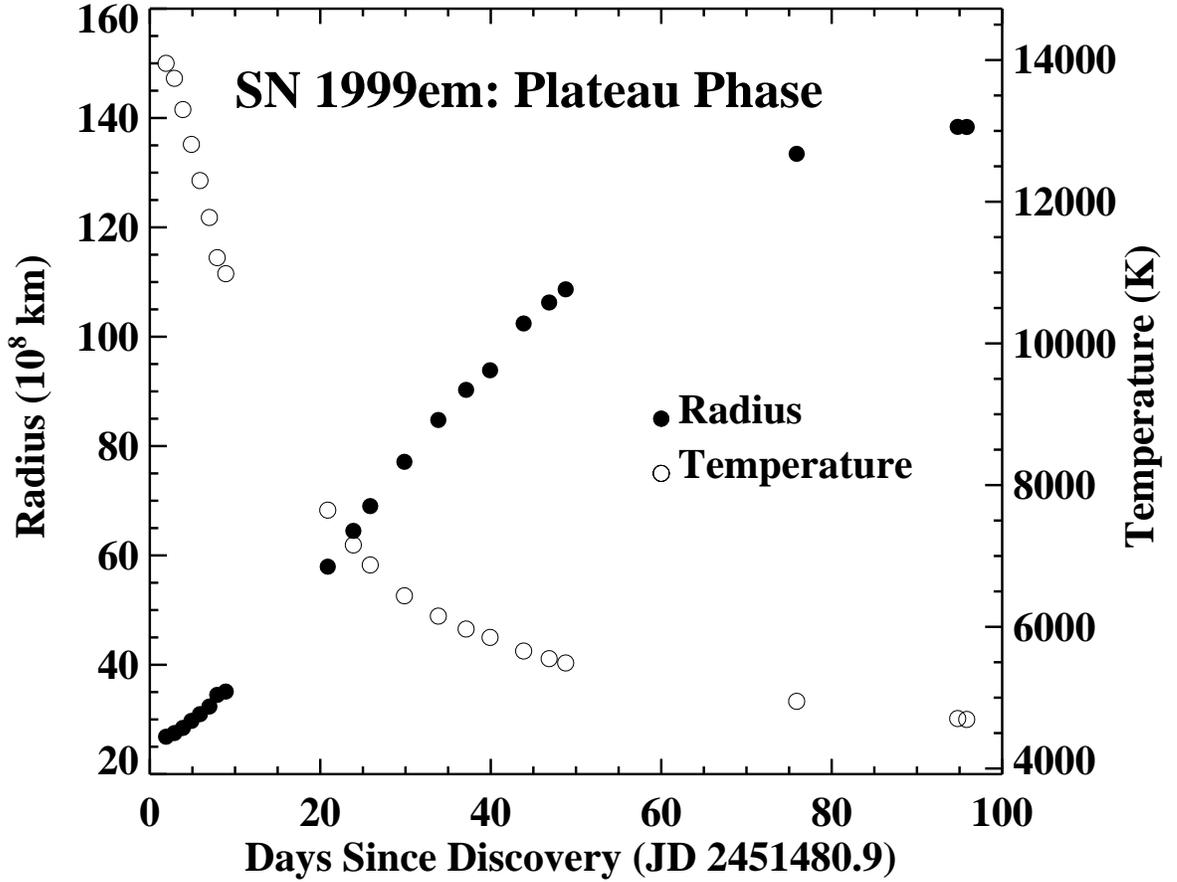}
}
}
\end{center}
\caption{ The temporal evolution of the radius of the radiating surface at the
thermalization depth and $BVI$ color temperature for SN~1999em during the
photospheric phase (i.e., $R = R_{\rm therm} \equiv \theta \zeta D$ with the
theoretical angular size [$\theta_{\rm BVI}$], dilution factor [$\zeta_{\rm
BVI}$], and color temperature [$T_{\rm BVI}$] taken from Table 7, and a
distance of $D = D_{\rm BVI} = 8.27$ Mpc [Table 8] assumed).}
\label{fig:5.24a}
\end{figure}

The \ubvri\ photometric behavior of SN~1999em is broadly understood by
considering the temporal evolution of the radius at the thermalization depth
and color temperature shown in Figure~\ref{fig:5.24a} (using quantities
subsequently derived by the EPM analysis in \S~\ref{sec:sn1999emepmdistance}).
The surface area at the thermalization depth is seen to increase throughout
most of the plateau as the overall expansion of the ejecta competes with the
inwardly moving recombination front.  If this were the only effect operating,
we would expect to observe increasing brightness throughout the majority of the
plateau phase. Countering this, however, is the decreasing temperature of the
photosphere as the SN~cools, which lowers the flux at all wavelengths.  The
effect of the decreasing photospheric temperature is least pronounced at longer
wavelengths (i.e., the $I$ band) since bandpasses significantly redward of the
blackbody peak are more nearly in the Rayleigh-Jeans regime of the Planck
function and thus only linearly dependent on temperature.  As one moves to
progressively bluer wavelengths the effect of the decreasing temperature
becomes more pronounced, and the plateau brightness changes from generally
increasing ($I$ band) to nearly constant ($R$ band), to slightly declining ($V$
band), to moderately declining ($B$ band), to steeply declining ($U$ band);
significant line blanketing by metals further reduces the flux in the $U$ and
$B$ bandpasses.  The precipitous drop observed in all bandpasses after about
day 95 signals the point at which the hydrogen recombination wave has receded
completely through the massive hydrogen envelope and the transition to the
nebular phase has begun.  The full nebular phase, usually defined as the point
at which the total electron-scattering optical depth to the center drops below
unity, is not expected until after about day 250 (Eastman et al. 1994).


One feature of the $V$-band light curve (and possibly $R$ as well) that
warrants special comment is the distinct ``double peak'' evident in the early
stages of development.  The first peak occurs near day 4 after discovery
(perhaps a bit later than the single peaks seen in $U$ and $B$) and the second
peak of nearly equal brightness occurs on day 13, with the dip between them
reaching a local minimum on day 9 (0.07 mag fainter than the peak brightness).
Similar photometric behavior can be discerned in the light curve presented by
Schmidt et al. (1993) for the Type II-P SN~1990E, although in that case the
second peak is brighter than the first, and occurs about two weeks after the
first peak.  In both SN~1999em and SN~1990E the rise to the second maximum
coincides with the appearance of \ion{Fe}{2} absorption lines in the spectrum
(\S~\ref{sec:spectra}) and the point at which the photospheric color
temperature has cooled down to near 10,000~K.  This effectively marks the onset
of the recombination phase.  It is possible that the additional liberation of
recombination energy slows the fall of the effective temperature enough to
allow the effects of the increasing photospheric surface area to temporarily
dominate the visual brightness of the SN.  Additional examples of well-sampled
early-time SNe~II-P light curves are needed to see if this effect is generic.

\subsection{Spectra}
\label{sec:spectra}
\subsubsection{Observations, Reductions, and Identification of Line Features}
Table 3 lists the spectral observations of SN~1999em.  With the exception of
the single echelle spectrum, whose reduction details will be given elsewhere
(D. C. Leonard et al., in preparation), all one-dimensional sky-subtracted
spectra were extracted optimally (Horne 1986) in the usual manner, generally
with a width of $\sim 10\arcsec$ along the slit, although this was modified
based on seeing conditions (usually a width of $\sim 4$ times the seeing was
employed).  Each spectrum was then wavelength and flux calibrated, as well as
corrected for continuum atmospheric extinction and telluric absorption bands
(Wade \& Horne 1988; Bessell 1999; Matheson et al. 2000).

In general, the position angle of the slit was not aligned along the
parallactic angle (the slit of the Lick 1-m spectrograph is fixed at $0^\circ$)
so the spectral shape may suffer from differential light loss (Filippenko
1982).  In addition, the shape of all spectra obtained with the Lick 1-m
reflector is suspect, especially at wavelengths greater than 5800~\AA, since
successive observations from the same night occasionally showed significant
variations.\footnote{Normalizing the spectrum at the blue end, the flux level
of the red end varied by as much as $\pm 30\%$ on the worst nights.}  Multiple
extractions of the data using apertures of varying width and different
background regions failed to remedy the problem, and its cause remains unknown.
Fortunately, this calibration uncertainty has minimal scientific impact here
since it is the {\it wavelengths} of absorption features, not absolute fluxes,
that are of primary interest.  In any case, the effect was evident on only a
few nights, and quantities derived using spectra obtained on these nights are
in good agreement with those derived from spectra taken at other telescopes.

\begin{figure}
\ssp
\vskip -0.5in
\hskip -0.3in
\rotatebox{0}{
\scalebox{1.0}{
\plotone{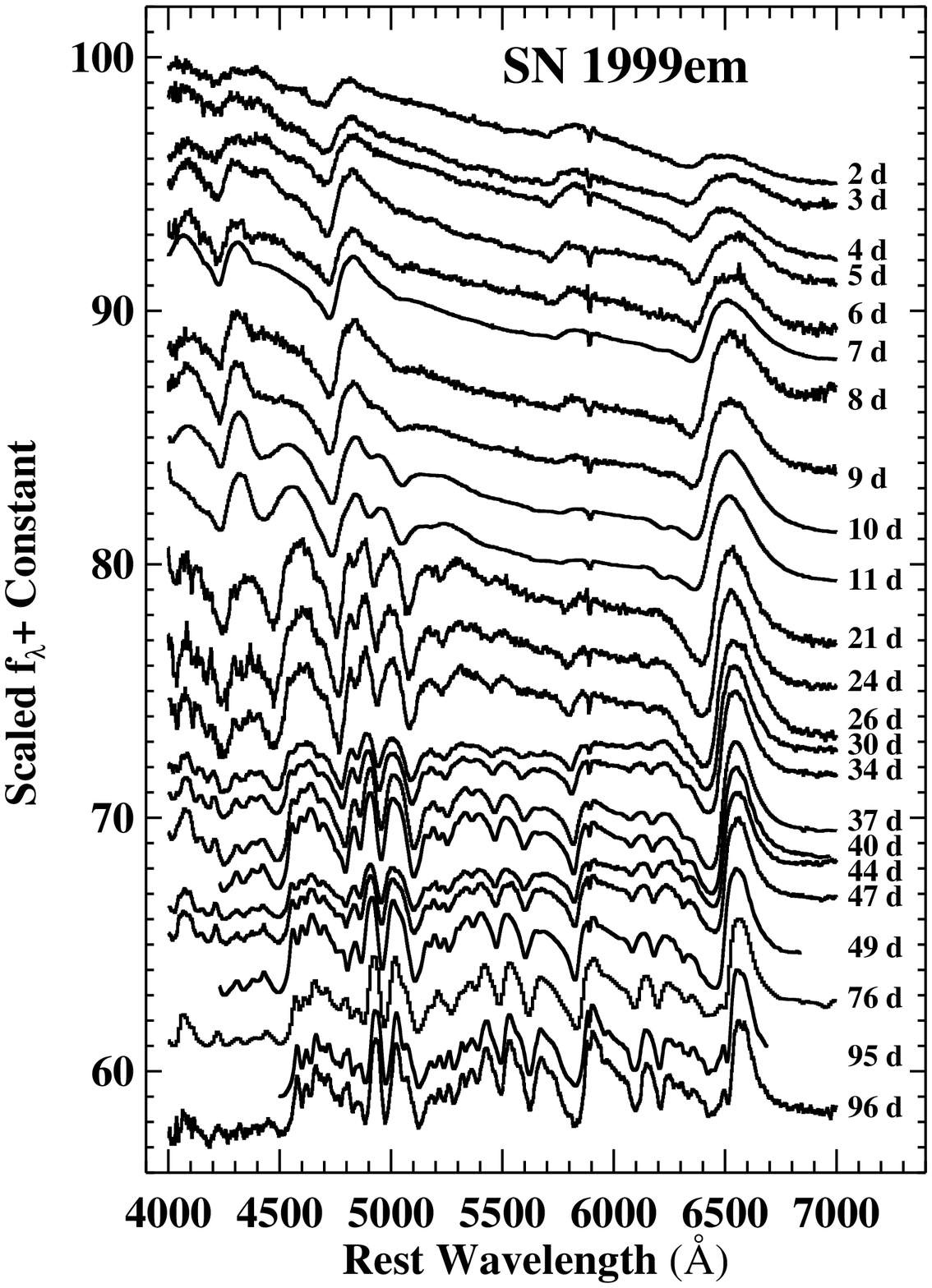}
}
}
\caption{ Optical flux spectra of SN~1999em spanning the first 96 days after
discovery.  In this and all figures a recession velocity of 800 \kms\ has been
removed from the observed spectrum (\S~\ref{sec:inferringvel}).  Note that the
continuum shape of the spectra on days 5 and 6 has been manually adjusted
redward of 5800 \AA\ due to irregularities introduced by the spectrograph.  }
\label{fig:5.8}
\end{figure}

\begin{figure}
\ssp
\vskip -0.5in
\hskip -0.5in
\rotatebox{0}{
\scalebox{1.0}{
\plotone{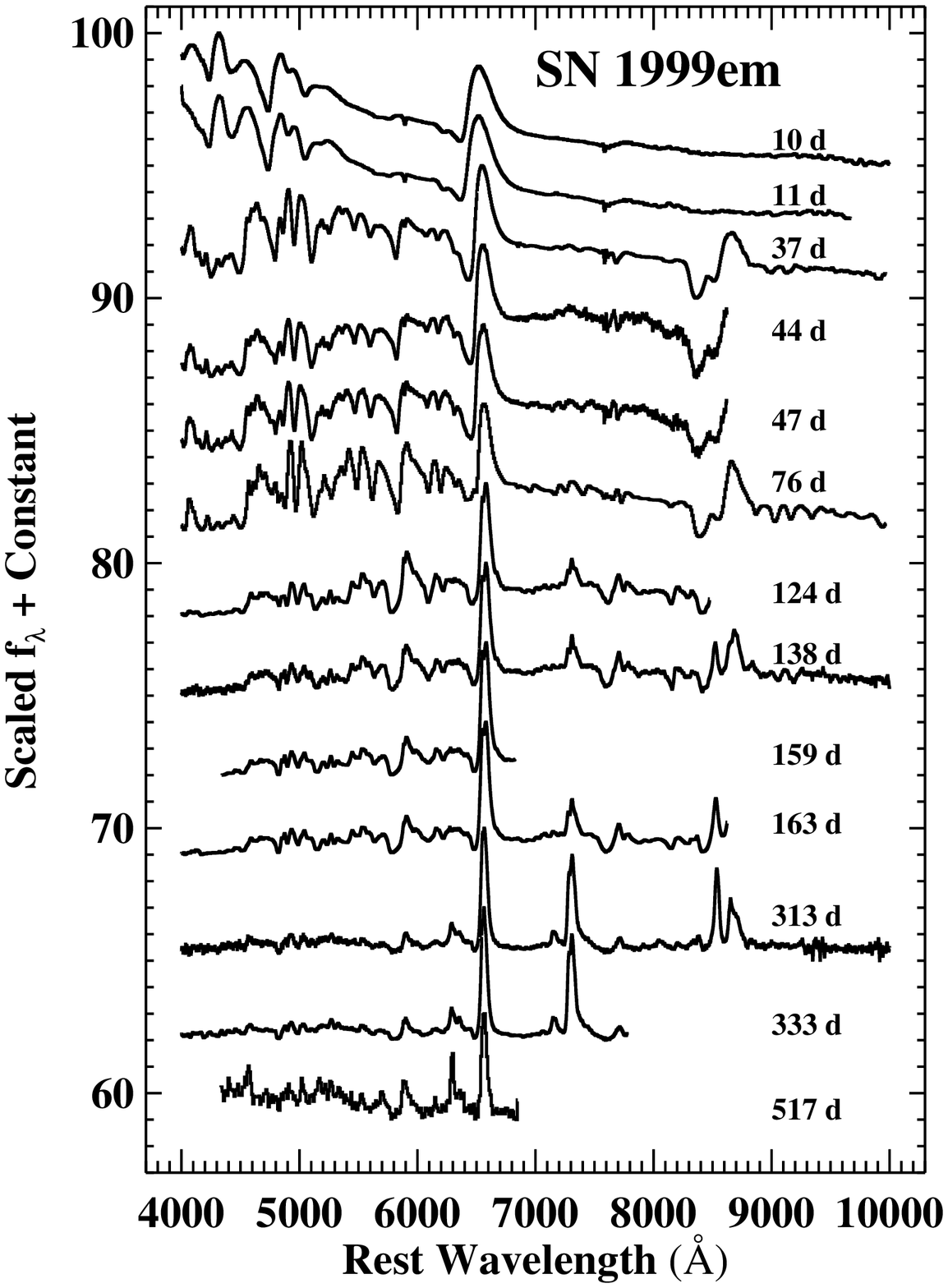}
}
}
\caption{ The optical and near-infrared spectral
development of SN 1999em during its first 517 days after discovery (1999
October 29).  Six plateau-phase spectra shown in Figure~\ref{fig:5.8} are
reproduced here as well, but with their full spectral range displayed. }
\label{fig:5.7}
\end{figure}

The spectral evolution of SN~1999em during the first 517 days of its
development is shown in Figures~\ref{fig:5.8}~and~\ref{fig:5.7}.  The early
spectra are characterized by a nearly featureless continuum with broad hydrogen
Balmer and \ion{He}{1} $\lambda 5876$ P-Cygni lines.  As early as day 7, a hint
of \ion{Fe}{2} $\lambda 5169$ absorption is visible, and it becomes quite
strong by day 11 along with \ion{Fe}{2} $\lambda 5018$.  During the next 90
days the strength of numerous metal lines continues to increase (see
Figure~\ref{fig:5.17}).  After the plateau ends, the spectrum becomes more
emission dominated and marked by the presence of such forbidden lines as
[\ion{Ca}{2}] $\lambda\lambda 7291, 7324$ and [\ion{O}{1}] $\lambda\lambda
6300, 6364$.

\begin{deluxetable}{lcccccccccccl}
\ssp
\tablenum{3}
\ptlandscape
\rotate
\tabletypesize{\scriptsize}
\tablewidth{0pt}
\tablecaption{Journal of Spectroscopic Observations of SN 1999em}
\tablehead{\colhead{Day\tablenotemark{a}} & 
\colhead{UT Date} &
\colhead{Tel.\tablenotemark{b}} &
\colhead{Range\tablenotemark{c}}  &
\colhead{Res.\tablenotemark{d}} &
\colhead{P.A.\tablenotemark{e}} &
\colhead{Opt. P.A.\tablenotemark{f}} &
\colhead{Air.\tablenotemark{g}} & 
\colhead{Flux Std.\tablenotemark{h}} &
\colhead{See.\tablenotemark{i}} &
\colhead{Slit} &
\colhead{Exp.} &
\colhead{Observer(s)\tablenotemark{j}} \\
\colhead{} &
\colhead{(yyyy-mm-dd)} &
\colhead{} &
\colhead{(\AA)} &
\colhead{(\AA)} &
\colhead{(deg)} &
\colhead{(deg)} &
\colhead{} &
\colhead{} &
\colhead{(arcsec)} &
\colhead{(arcsec)} &
\colhead{(s)} & 
\colhead{} }
\startdata

1.9       & 1999-10-31  & L1  & 3948-7449 & 6    &  0     & 150     & 1.5   & 
HD19 & 2.0         & 2.9 & 2x900            & EG	\\

2.9       & 1999-11-01  & L1  & 3948-7449 & 7    &  0     & 147     & 1.5   & 
HD19 & 2.2         & 2.9 & 3x900            & EG\\

3.9       & 1999-11-02  & L1  & 3948-7449 & 7    &  0     & 154     & 1.6   & 
HD19 & 1.4         & 2.9 & 2x900            & EG\\

4.0  	& 1999-11-02  & L3h & 3630-10100 & 0.1   & \nodata & \nodata & 1.3 & \nodata
& 2.2 & 2.0 & 5400 & AQ, SF, CG\\

4.9       & 1999-11-03  & L1  & 3948-7449 & 6    &  0     & 155     & 1.5   & 
HD19 & 2.4         & 2.9 & 3x900            & EG\\

5.9       & 1999-11-04  & L1  & 3948-7449 & 6    &  0     & 159     & 1.4   & 
HD19 & 1.8         & 2.9 & 3x900            & EG\\

7.0       & 1999-11-05  & L3p & 3400-5500 & 5    &  170   & 174     & 1.4   & 
F34& 2.3         & 2.0 & 4x1200           & AF, DL, RC\\

        &             &     & 5000-7700 & 6    &  170   & 174     & 1.4   & 
HD84 & 2.3         & 2.0 & 4x1200           & AF, DL, RC\\

        &             &     & 3400-5500 & 5    &  20    & 24      & 1.5   & 
F34& 2.3         & 2.0 & 4x1500           & AF, DL, RC\\

        &             &     & 5000-7700 & 6    &  20    & 24      & 1.5   & 
HD84 & 2.3         & 2.0 & 4x1500           & AF, DL, RC\\

7.9       & 1999-11-06  & L1  & 3948-7449 & 6    &  0     & 157     & 1.4   & 
HD19 & 2.1         & 2.9 & 3x900            & EG\\

8.9       & 1999-11-07  & L1  & 3948-7449 & 8    &  0     & 158     & 1.4   & 
HD19 & 4.2         & 2.9 & 3x900            & EG\\

10.2      & 1999-11-08  & KII  & 4000-10000& 14   &  63    & 61      & 1.6   & 
BD17 & 1         & 0.7 & 30+70+70         & AF, AC, AR\\

11.2      & 1999-11-09  & KII  & 4000-10000& 14   &  58    & 61      & 1.5   & 
BD17 & 1         & 0.7 & 60+70            & AF, AC, AR\\

20.9      & 1999-11-19  & L1  & 3948-7449 & 7    &  0     & 159     & 1.4   & 
HD19 & 3           & 2.9 & 3x900            & EG\\

23.9      & 1999-11-22  & L1  & 3948-7449 & 7    &  0     & 162     & 1.4   & 
HD19 & 2.2         & 2.9 & 4x900            & EG\\

25.9      & 1999-11-24  & L1  & 3948-7449 & 7    &  0     & 166     & 1.4   & 
HD19 & 2.7         & 2.9 & 4x900            & EG\\

29.9     & 1999-11-28  & L1  & 3948-7449 & 7    &  0     & 164     & 1.4   & 
HD19 & 3.0         & 2.9 & 4x900            & EG\\

33.9      & 1999-12-02  & L1  & 3948-7449 & 7    &  0     & 160     & 1.4   & 
HD19 & 3.5         & 2.9 & 4x900            & EG\\

37.1      & 1999-12-05  & KII  & 4000-10000& 14   &  61    & 61      & 1.0   & 
HD84 & 2.2         & 0.7 & 330              & AF, PG\\

39.9      & 1999-12-08  & L3p & 4230-7010 & 7    &  16    & 28      & 1.5   & 
HD84    & 3           & 3.0 & 4x1200           & DL, DA\\

        &             &     & 4230-7010 & 7    &  43    & 45      & 2.1   & 
HD84    & 4           & 3.0 & 4x1200           & DL, DA\\

43.9      & 1999-12-12  & L1  & 3948-7449 & 7   &  0      & 177     & 1.4   & 
HD19    & 2.6         & 2.9 & 3x900            & EG    \\

        &             &     & 5550-8649 & 7   &  0      & 177     & 1.4   & 
HD19    & 2.6         & 2.9 & 3x900            & EG    \\

46.8      & 1999-12-15  & L1  & 3948-7449 & 7   &  0      & 177     & 1.4   & 
HD19    & 2.7         & 2.9 & 3x900            & EG    \\

        &             &     & 5550-8649 & 7   &  0      & 177     & 1.4   & 
HD19    & 2.7         & 2.9 & 3x900            & EG    \\

48.8      & 1999-12-17  & L3p & 4220-6882 & 6    &  161   & 170     & 1.4   & 
HD19    & 2.3         & 2.0 & 4x1800           & DL, RC, AC\\

        &             &     & 4220-6882 & 6    &  24    & 24      & 1.5   & 
HD19    & 2.5         & 2.0 & 4x1800           & DL, RC, AC\\

75.9      & 2000-01-13  & P   & 3500-5500 & 11   &  90    & 43      & 1.7   & 
HD84    & 2.1         & 1.0 & 3x120            & AH, SB\\

        &             &     & 5400-10000& 14   &  90    & 43      & 1.7   & 
HD84    & 2.1         & 1.0 & 3x120            & AH, SB\\

        &             &     & 3500-5500 & 27   &  90    & 43      & 1.7   & 
HD84    & 2.1         & 6.0 & 2x60             & AH, SB\\

        &             &     & 5400-10000& 53   &  90    & 43      & 1.7   & 
HD84    & 2.1         & 6.0 & 2x60             & AH, SB\\

94.8      & 2000-02-01  & KII & 4510-7010 & 5    &  46    & 141     & 1.3   & 
G191    & 0.9         & 1.0 & 120              & WB, DS, AS, WV, MR, GM\\

95.8      & 2000-02-02  & L1  & 3948-7449 & 6   &  0      & 31      & 1.7   & 
HD19    & 2.5         & 2.9 & 4x900            & EG    \\

123.8     & 2000-03-01  & KI  & 3700-8500 & 10  &  0      & 46      & 1.2   & 
BD26    & 1.0         & 1.0 & 2x60             & MM, RE, JC\\

137.7     & 2000-03-15  & L3  & 3400-5400 & 7   &  33     & 34      & 1.6   & 
F34     & 3.0         & 2.0 & 800+320          & AF, AC\\

        &             &     & 5100-10500& 11  &  33     & 34      & 1.6   & 
BD26    & 3.0         & 2.0 & 800+320          & AF, AC\\

158.8     & 2000-04-05  & KIp & 4338-6872 & 7   &  66     & 66      & 2.3   & 
HD84    & 1.9         & 1.5 & 3x900            & AF, AB\\

162.8     & 2000-04-09  & KIp & 3720-8685 & 13  &  66     & 65      & 2.1   & 
F34     & 1.5         & 1.5 & 4x300            & RB, MB, RW\\

313.1     & 2000-09-06  & L3  & 3400-5400 & 7   &  150    & 154     & 1.5   & 
BD28    & 3.0         & 2.0 & 1800+300         & AF, RC, WL\\
        &             &     & 5100-10450& 11  &  150    & 154     & 1.5   & 
BD17    & 3.0         & 2.0 & 1800+300         & AF, RC, WL\\

333.1     & 2000-09-26  & L3  & 3400-5400 & 7   &  180    & 168     & 1.5   & 
BD28    & 2.0         & 2.0 & 2700             & AF, RC\\
        &             &     & 5100-7800 & 7   &  180    & 168     & 1.5   & 
BD17    & 2.0         & 2.0 & 2700             & AF, RC\\

516.7     & 2001-03-29  & KIp & 4340-6870 & 7   &  60     & 63      & 1.7   & 
HD84    & 1.0         & 1.0 & 700              & AF, AB\\

\enddata
\tablenotetext{a}{Days since discovery, 1999-10-29 (HJD 2,451,480.94), rounded
to the nearest tenth of a day.}

\tablenotetext{b}{L1 = Lick 1-m/Nickel Reflector + spectrograph (Stone \&
Shields 1990); L3(p) = Lick 3-m/Kast Double Spectrograph (Miller \& Stone 1993;
`p' denotes polarimeter attached; `h' denotes Hamilton echelle spectrograph
used instead of the Kast); K I(p) = Keck-I 10-m/Low Resolution Imaging
Spectrometer (LRIS [Oke et al. 1995]; `p' denotes polarimeter attached [Cohen
1996]); K II = Keck-II 10-m/LRIS; P = Palomar 5 m Hale/Double Spectrograph (Oke
\& Gunn 1983).}

\tablenotetext{c}{Observed wavelength range of spectrum.  In some cases, the
ends are very noisy, and are not shown in the figures.}

\tablenotetext{d}{Approximate spectral resolution derived from night-sky lines.}

\tablenotetext{e}{Position angle of the spectrograph slit.  }

\tablenotetext{f}{Optimal parallactic angle (Filippenko 1982) at the midpoint
of the exposures.}

\tablenotetext{g}{Average airmass of observations.}

\tablenotetext{h}{The standard stars are as follows: F34 = Feige~34, BD28 =
 BD+28$^\circ$4211---Stone (1977), Massey \& Gronwall (1990); G191 =
 G191B2B---Massey \& Gronwall (1990); HD19 = HD~19445, HD84 = HD~84937, BD26 =
 BD+26$^\circ$2606, BD17 = BD+17$^\circ$4711---Oke \& Gunn (1983).}

\tablenotetext{i}{Seeing is estimated from the FWHM of point sources on the CCD
chip.}  

\tablenotetext{j}{AB = Aaron Barth; AC = Alison Coil; AF = Alex Filippenko; AH
= Alan Harris; AQ = Andreas Quirrenbach; AR = Adam Riess; AS = Adam Stanford;
CG = Carol Grady; DA = David Ardila; DL = Douglas Leonard; DS = Daniel Stern;
EG = Elinor Gates; GM = George Miley; JC = James Colbert; MB = Michael
Brotherton; MM = Matt Malkan; MR = Michiel Reuland; PG = Peter Garnavich; RB =
Robert Becker; RC = Ryan Chornock; RE = Rick Edelson; RW = Richard White; SB =
Schelte Bus; SF = Sabine Frink; WB = Wil van Breugel; WL = Weidong Li; WV =
Willem de Vries.}

\end{deluxetable}
\clearpage


As discussed in \S~\ref{sec:introduction}, one area of uncertainty in the
application of EPM is estimating the photospheric velocity.  To test the
consistency of velocities measured from different lines, we shall derive
velocities from as many features as possible in the optical spectrum of
SN~1999em during the photospheric phase.  Our goal is to identify several
unblended, weak features and compare the velocity derived using them with that
obtained from the \fetwo\ lines, which have often been used in previous EPM
studies.

\begin{figure}
\ssp
\begin{center}
\rotatebox{0}{
\scalebox{0.9}{
\plotone{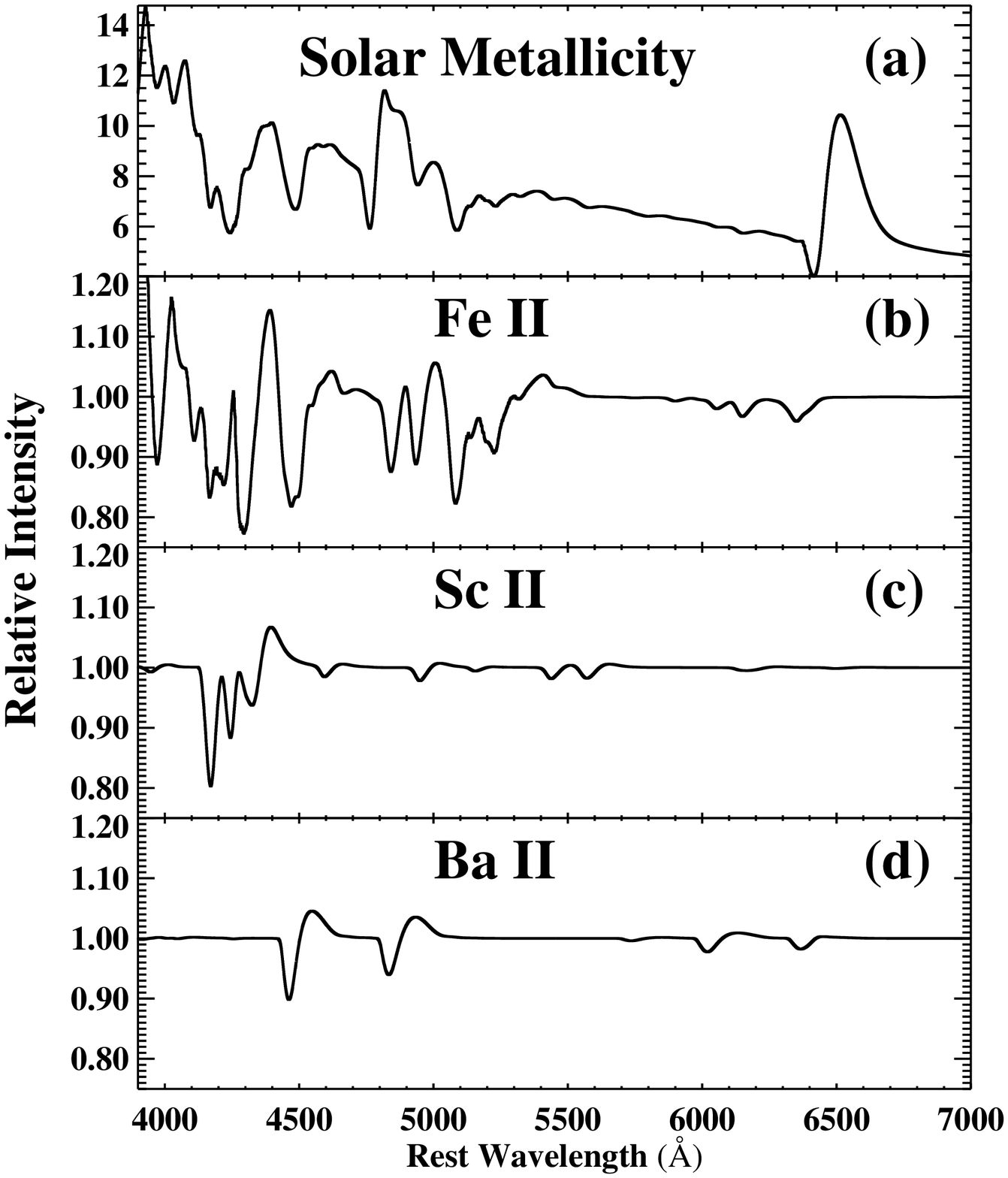}
}
}
\end{center}
\caption{ ({\it a}) Synthetic spectrum
of a $ 15 M_\odot$ RSG progenitor with solar metallicity 44 days after explosion
with $T_{\rm eff} = 5000$~K (model s15.44.2 of Weaver \& Woosley [1993]).
({\it b,c,d}) The relative strengths of line features due to \protect\ion{Fe}{2},
\protect\ion{Sc}{2}, and \protect\ion{Ba}{2}, respectively, in the same model.
}
\label{fig:5.6}
\end{figure}

\begin{figure}
\ssp
\vskip -0.5in
\hskip -0.3in
\rotatebox{90}{
\scalebox{0.8}{
\plotone{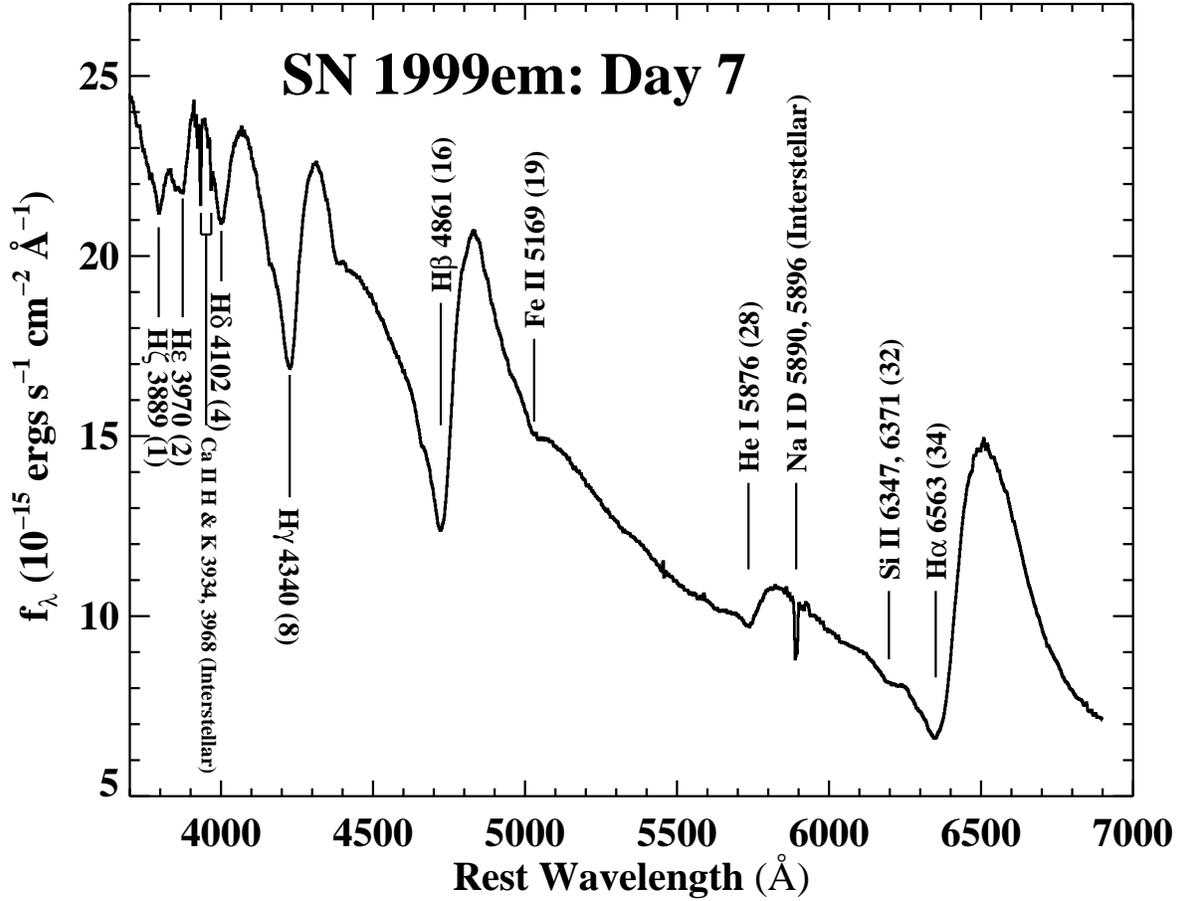}
}
}
\caption{Early-time flux spectrum of SN~1999em with prominent absorption
features identified; the feature numbers corresponding to the ions listed in
Table 4 are given in parenthesis.  At this early stage, the only strong lines
are due to hydrogen Balmer and \protect\ion{He}{1}. Weak
\protect\ion{Fe}{2} $\lambda 5169$\ and \protect\ion{Si}{2} $\lambda 6355$
features are also seen.  }
\label{fig:5.10}
\end{figure}

\begin{figure}
\ssp
\vskip -0.5in
\hskip -0.3in
\rotatebox{90}{
\scalebox{0.85}{
\plotone{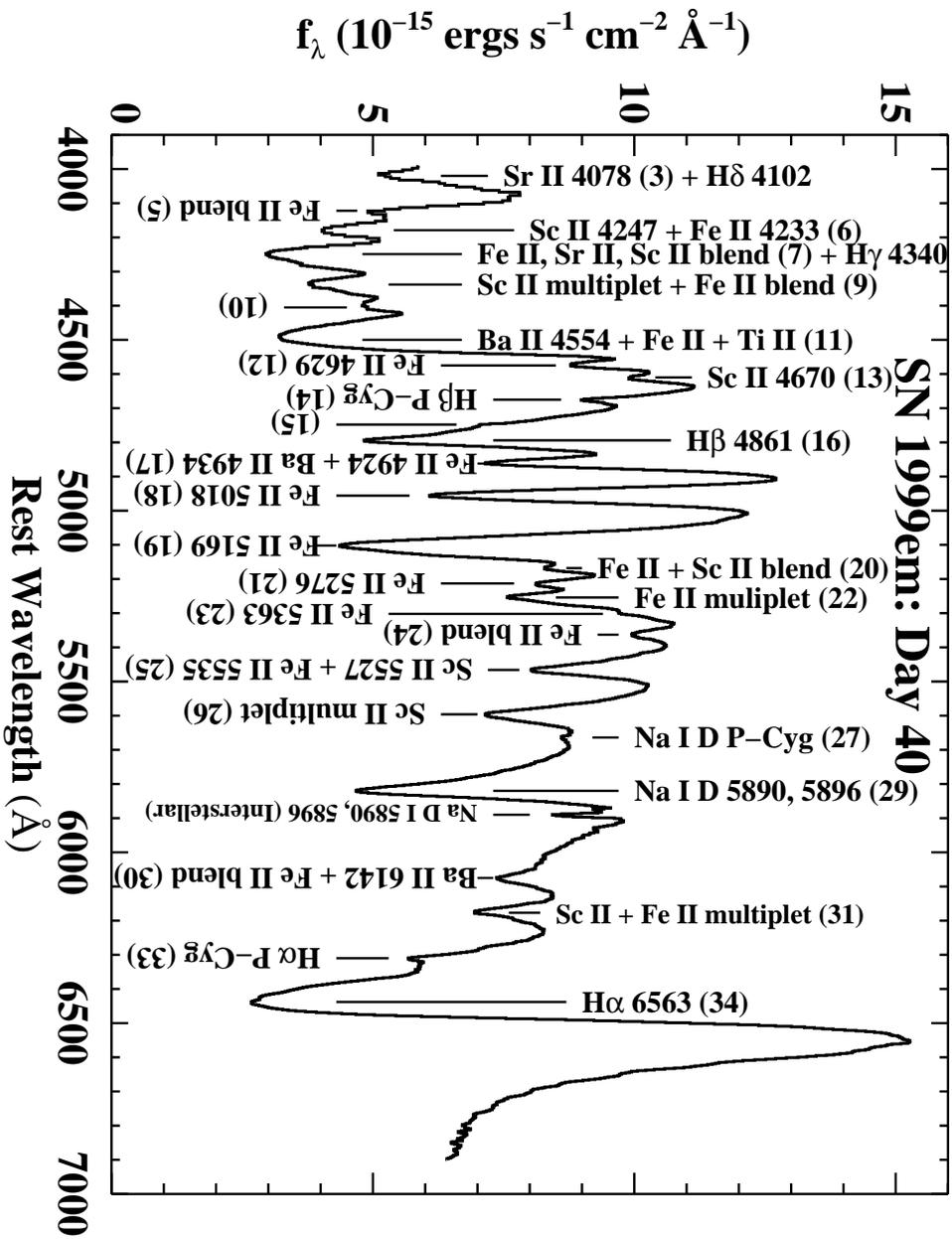}
}
}
\caption{Line
identifications for SN~1999em during the plateau phase, from Table 4.  }
\label{fig:5.11}
\end{figure}

\begin{figure}
\ssp
\vskip -0.5in
\hskip -0.5in
\rotatebox{90}{
\scalebox{0.9}{
\plotone{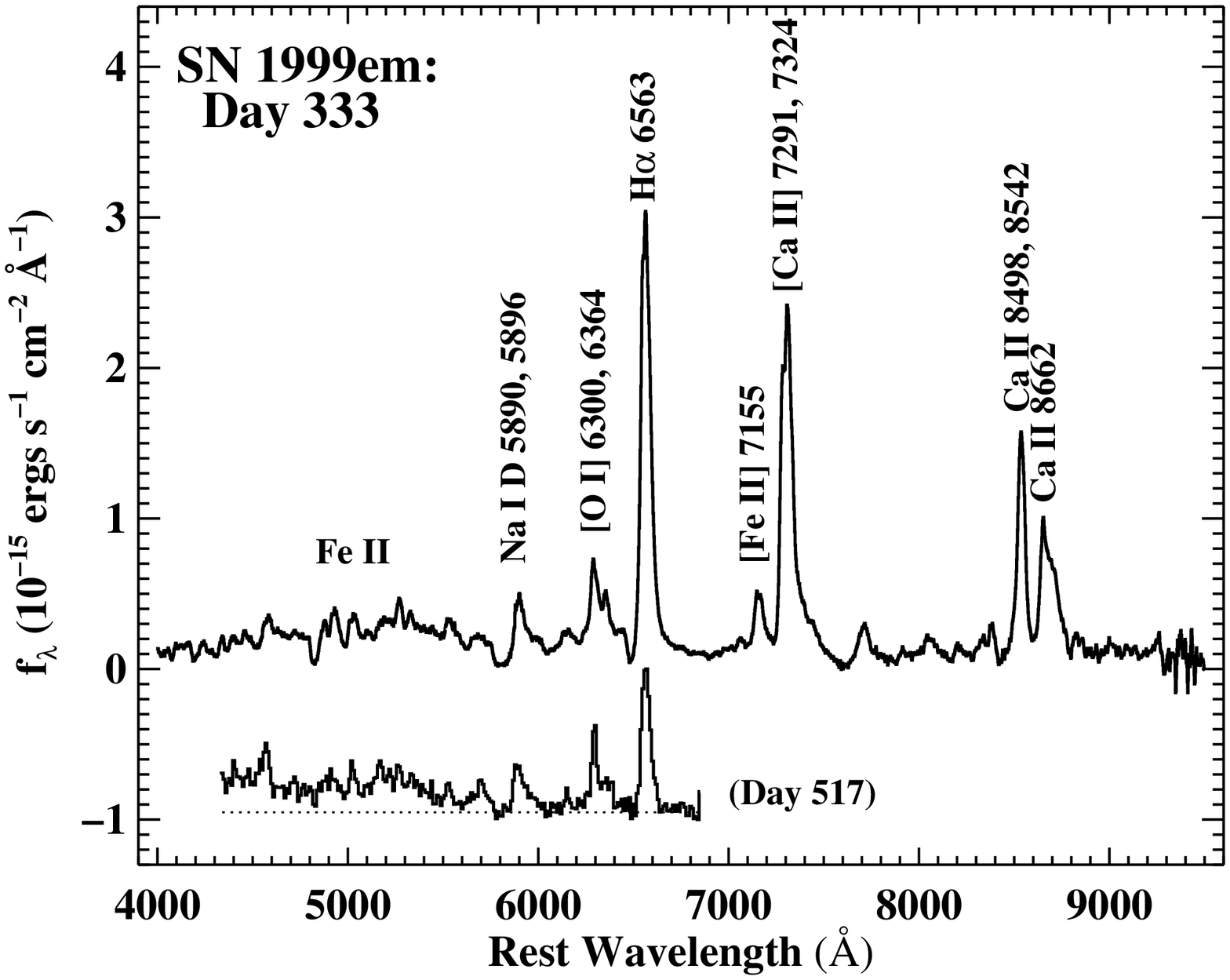}
}
}
\caption{
SN~1999em in the nebular phase, 333 days after discovery, with prominent
emission lines identified; redward of $\lambda 7780$ the displayed spectrum is
from day 313.  For comparison of features, the spectrum from day 517 is also
shown, having been scaled by a factor of 12.6 and then offset by -0.95; the
zero-flux level is indicated ({\it dotted line}). }
\label{fig:5.9}
\end{figure}

We first identified 36 distinct absorption features in the optical photospheric
spectrum of SN~1999em, focusing mainly on the region between 4000 and 7000 \AA\
since that range was best covered by our spectra.  We consulted previously
published line identifications (IDs) as a starting point for major features,
including Jeffery \& Branch (1990), Williams (1987), Elias et al. (1988),
McGregor (1988), Meikle et al. (1993), and Fassia et al. (1998).  We then
considered the theoretical models by Hatano et al. (1999), in which the
spectral signatures of the major ions predicted to occur in SN spectra of all
types are calculated (the results of the calculations are displayed in figures
retrievable from the authors'
web-site\footnote{\url{http://www.nhn.ou.edu/$\sim$baron/papers.html}.}).  To help
estimate the expected relative strengths of the different contributing species,
transition wavelengths and oscillator strengths were taken from the Kurucz
(1996) spectral line database.  We also studied the synthetic SN spectrum of
progenitor model s15.44.2 of Weaver \& Woosley (1993)\footnote{See Eastman
et al. (1994) for details of the physics involved in the model's calculation;
here we are primarily interested in its use to help identify line features.},
in which the non-LTE radiative transport code EDDINGTON (Eastman \& Pinto 1993)
is used to model the explosion of a 15 $M_\odot$ red supergiant (RSG)
progenitor of solar metallicity 44 days after shock breakout with an
effective temperature of 5000~K and a photosphere located at 5000~\kms.  The
optical portion of the synthetic spectrum is shown in Figure~\ref{fig:5.6}(a),
with individual species responsible for many of the line features shown in
Figure~\ref{fig:5.6}(b, c, d).  Immediately apparent is that \ion{Fe}{2}
features dominate the spectrum below 5500 \AA\ and consequently blend with many
of the features produced by other ions.  Such blended features are in general
unreliable for simple velocity measurements since the relative strengths of the
species making up the blend vary with photosphere location and temperature.
The resulting IDs of the line features are shown in
Figures~\ref{fig:5.10}~and~\ref{fig:5.11}, and all the likely contributors'
rest wavelengths, configurations, and energies are listed in Table 4.  The IDs
for major emission-line features in the nebular stage are shown in
Figure~\ref{fig:5.9}.


\begin{deluxetable}{lcrccr}
\tablenum{4}
\ssp
\tablewidth{480pt}
\tablecaption{SN II-P Photospheric-Phase Absorption Line Identifications}
\tablehead{\colhead{Ion} &
\colhead{Wavelength                  (\AA)\tablenotemark{a}} &
\colhead{log ({\it gf})} &
\colhead{Feature ID\tablenotemark{b}} &
\colhead{Designation} &
\colhead{{\it E} (eV)\tablenotemark{c}}}

\startdata

\protect\ion{H}{1}  & 3889.05 & $-1.19$   &
 1  &  \nodata\tablenotemark{d}  &  10.20 \nl

\protect\ion{H}{1}  & 3970.07 & $-0.99$   &
 2  &  \nodata\tablenotemark{d} &  10.20 \nl

\protect\ion{Sr}{2}  & 4077.71 &  0.17   &
 3  &  $5s\ ^2{\rm S} - 5p\ ^2{\rm P}^{\circ}$  &   0.00 \nl

\protect\ion{H}{1}  & 4101.73 & $-0.75$   &
 4,3 &  \nodata\tablenotemark{d} &  10.20 \nl

\protect\ion{Fe}{2}  & 4173.46 & $-2.18$   &
 5  &  $3d^6 (^3{\rm P}2)4s\ ^4{\rm P} - 3d^6 (^5{\rm D})4p\ ^4{\rm D}^{\circ} $ &  2.58 \nl

\protect\ion{Fe}{2}  & 4178.86 & $-2.48 $  &
 5  &  $3d^6 (^3{\rm P}2)4s\ ^4{\rm P} - 3d^6 (^5{\rm D})4p\ ^4{\rm F}^{\circ} $   &   2.58 \nl

\protect\ion{Fe}{2}  & 4180.98 & $-1.84 $  &
 5  &  $3d^6 (^1{\rm D}2)4s\ ^2{\rm D} - 3d^6 (^3{\rm P2})4p\ ^2{\rm D}^{\circ} $   &   4.74 \nl

\protect\ion{Fe}{2}  & 4233.17 &$ -2.00 $  &
 6  &  $3d^6 ( ^3{\rm P}2)4s \ ^4{\rm P} - 3d^6 ( ^5{\rm D})4p \ ^4{\rm D}^{\circ}$  &   2.58 \nl

\protect\ion{Sc}{2}  & 4246.82 & $ 0.32 $  &
 6  &  $3d4s \ ^1{\rm D} - 3d4p \ ^1{\rm D}^{\circ}$  &   0.32 \nl

\protect\ion{Fe}{2}  & 4296.57 &$ -3.01 $  &
 7  &  $3d^6 ( ^3{\rm P}2)4s \ ^4{\rm P} - 3d^6 ( ^5{\rm D})4p \ ^4{\rm F}^{\circ}$  &   2.70 \nl

\protect\ion{Fe}{2}  & 4303.18 &$ -2.49 $  &
 7  &  $3d^6 ( ^3{\rm P}2)4s \ ^4{\rm P} - 3d^6 ( ^5{\rm D})4p \ ^4{\rm D}^{\circ}$  &   2.70 \nl

\protect\ion{Sr}{2}  & 4305.44 &$ -0.14 $  &
 7  &  $5p\ ^2{\rm P} - 6s\ ^2{\rm S}$  &   3.04 \nl

\protect\ion{Sc}{2}  & 4314.08 &$ -0.10$   &
 7  &  $3d^2 \ ^3{\rm F} - 3d4p \ ^3{\rm D}^{\circ}$  &   0.62 \nl

\protect\ion{Sc}{2}  & 4320.73 &$ -0.26 $  &
 7  &  $3d^2 \ ^3{\rm F} - 3d4p \ ^3{\rm D}^{\circ}$  &   0.61 \nl

\protect\ion{Sc}{2}  & 4325.00 & $-0.44 $  &
 7  &  $3d^2 \ ^3{\rm F} - 3d4p \ ^3{\rm D}^{\circ}$  &   0.60 \nl

\protect\ion{H}{1}  & 4340.46 & $-0.45 $  &
 8,7&  \nodata\tablenotemark{d}   &  10.20 \nl

\protect\ion{Fe}{2}  & 4351.77 &$ -2.10 $  &
 9  &  $3d^6 ( ^3{\rm P}2)4s \ ^4{\rm P}- 3d^6 ( ^5{\rm D})4p\ ^4{\rm D}^{\circ}$  &   2.70 \nl

\protect\ion{Sc}{2}  & 4374.46 & $-0.44 $  &
 9  &  $3d^2 \ ^3{\rm F}- 3d4p \ ^3{\rm F}^{\circ}$  &   0.62 \nl

\protect\ion{Fe}{2}  & 4385.39 & $-2.57$   &
 9  &  $3d^6 ( ^3{\rm P}2)4s \ ^4{\rm P}- 3d^6 ( ^5{\rm D})4p \ ^4{\rm D}^{\circ}$  &   2.78 \nl

\protect\ion{Sc}{2}  & 4400.39 & $-0.51$   &
 9  &  $3d^2 \ ^3{\rm F}- 3d4p \ ^3{\rm F}^{\circ}$  &   0.61 \nl

\protect\ion{Sc}{2}  & 4415.56 & $-0.64 $  &
 9  &  $3d^2 \ ^3{\rm F}- 3d4p \ ^3{\rm F}^{\circ}$  &   0.60 \nl

\protect\ion{Fe}{2}  & 4508.29 &$ -2.21$   &
 11  &  $3d^6 ( ^3{\rm F}2)4s \ ^4{\rm F}- 3d^6 ( ^5{\rm D})4p \ ^4{\rm D}^{\circ}$  &   2.86 \nl

\protect\ion{Fe}{2}  & 4522.63 & $-2.03  $ &
 11  &  $3d^6 ( ^3{\rm F}2)4s \ ^4{\rm F}- 3d^6 ( ^5{\rm D})4p \ ^4{\rm D}^{\circ}$  &   2.84 \nl

\protect\ion{Fe}{2}  & 4549.47 & $-1.75 $  &
 11  &  $3d^6 ( ^3{\rm F}2)4s \ ^4{\rm F}- 3d^6 ( ^5{\rm D})4p \ ^4{\rm D}^{\circ}$  &   2.83 \nl

\protect\ion{Ti}{2}  & 4549.62 &$ -0.45 $  &
 11  &  $3d^3 \ ^2{\rm H}- 3d^2 ( ^3{\rm F})4p \ ^2{\rm G}^{\circ}$  &   1.58 \nl

\protect\ion{Ba}{2}  & 4554.03 & $ 0.17 $  &
 11  &  $6s\ ^2{\rm S} - 6p\ ^2{\rm P}^{\circ} $  &   0.00 \nl

\protect\ion{Fe}{2}  & 4555.89 & $-2.29 $  &
 11  &  $3d^6 ( ^3{\rm F}2)4s \ ^4{\rm F}- 3d^6 ( ^5{\rm D})4p \ ^4{\rm F}^{\circ}$  &   2.83 \nl

\protect\ion{Fe}{2}  & 4583.84 & $-2.02 $  &
 11  &  $3d^6 ( ^3{\rm F}2)4s \ ^4{\rm F}- 3d^6 ( ^5{\rm D})4p \ ^4{\rm D}^{\circ}$  &   2.81 \nl

\protect\ion{Fe}{2}  & 4629.34 &$ -2.37  $ &
 12  &  $3d^6 ( ^3{\rm F}2)4s \ ^4{\rm F}- 3d^6 ( ^5{\rm D})4p \ ^4{\rm F}^{\circ}$  &   2.81 \nl

\protect\ion{Sc}{2}  & 4670.41 &$ -0.37 $  &
 13  &  $3d^2\ ^1{\rm D} - 3d4p\ ^1{\rm F}^{\circ}$  &   1.36 \nl

\protect\ion{H}{1}  & 4861.32 &$ -0.02$   &
 16,14  &  \nodata\tablenotemark{d}  &  10.20 \nl

\protect\ion{Fe}{2}  & 4923.93 & $-1.32 $  &
 17  &  $3d^5 4s^2\ ^6{\rm S} - 3d^6 ( ^5{\rm D})4p\ ^6{\rm P}^{\circ}$  &   2.89 \nl

\protect\ion{Ba}{2}  & 4934.08 &$ -0.15 $  &
 17  &  $6s\ ^2{\rm S} - 6p\ ^2{\rm P}^{\circ}$  &   0.00 \nl

\protect\ion{Fe}{2}  & 5018.44 & $-1.22$   &
 18  &    $3d^5 4s^2\ ^6{\rm S}- 3d^6 ( ^5{\rm D})4p\ ^6{\rm P}^{\circ}$  & 2.89 \nl

\enddata
\end{deluxetable}

\begin{deluxetable}{lcrccr}
\tablenum{4}
\ssp
\tablewidth{480pt}
\tablecaption{SN II-P Photospheric-Phase Absorption Line Identifications -- {\it Continued} }
\tablehead{\colhead{Ion} &
\colhead{Wavelength                  (\AA)\tablenotemark{a}} &
\colhead{log ({\it gf})} &
\colhead{Feature ID\tablenotemark{b}} &
\colhead{Designation} &
\colhead{{\it E} (eV)\tablenotemark{c}}}

\startdata

\protect\ion{Sc}{2}  & 5031.02 & $-0.26 $  &
 18  &  $3d^2\ ^1{\rm D} - 3d4p\ ^1{\rm P}^{\circ}$   &   1.36 \nl

\protect\ion{Fe}{2}  & 5169.03 & $-0.87 $  &
 19  &  $3d^5 4s^2\ ^6{\rm S}- 3d^6 ( ^5{\rm D})4p\ ^6{\rm P}^{\circ}$  &   2.89 \nl

\protect\ion{Fe}{2}  & 5197.58 & $-2.10 $  &
 20  &  $3d^6 ( ^3{\rm G})4s\ ^4{\rm G} - 3d^6 ( ^5{\rm D})4p\ ^4{\rm F}^{\circ}$   &   3.23 \nl

\protect\ion{Fe}{2}  & 5234.62 & $-2.05 $  &
 20  &  $3d^6 ( ^3{\rm G})4s\ ^4{\rm G} - 3d^6 ( ^5{\rm D})4p\ ^4{\rm F}^{\circ}$  &   3.22 \nl

\protect\ion{Sc}{2}  & 5239.81 &$ -0.77$   &
 20  &  $4s^2\ ^1{\rm S} - 3d4p\ ^1{\rm P}^{\circ}$  &   1.46 \nl

\protect\ion{Fe}{2}  & 5276.00 & $-1.94$   &
 21  &  $3d^6 ( ^3{\rm G})4s\ ^4{\rm G} - 3d^6 ( ^5{\rm D})4p\ ^4{\rm F}^{\circ}$  &   3.20 \nl

\protect\ion{Fe}{2}  & 5316.62 & $-1.85 $  &
 22  &  $3d^6 ( ^3{\rm G})4s\ ^4{\rm G} - 3d^6 ( ^5{\rm D})4p\ ^4{\rm F}^{\circ}$  &   3.15 \nl

\protect\ion{Fe}{2}  & 5325.55 &$ -2.60 $  &
 22  &  $3d^6 ( ^3{\rm G})4s\ ^4{\rm G} - 3d^6 ( ^5{\rm D})4p\ ^4{\rm F}^{\circ}$  &   3.22 \nl

\protect\ion{Fe}{2}  & 5362.87 &$ -2.74 $  &
 23  &  $3d^6 ( ^3{\rm G})4s\ ^4{\rm G} - 3d^6 ( ^5{\rm D})4p\ ^4{\rm D}^{\circ}$  &   3.20 \nl

\protect\ion{Fe}{2}  & 5414.07 & $-3.79 $  &
 24  &  $3d^6 ( ^3{\rm G})4s\ ^4{\rm G} - 3d^6 ( ^5{\rm D})4p\ ^4{\rm D}^{\circ}$  &   3.22 \nl

\protect\ion{Fe}{2}  & 5425.26 & $-3.36$   &
 24  &  $3d^6 ( ^3{\rm G})4s\ ^4{\rm G} - 3d^6 ( ^5{\rm D})4p\ ^4{\rm F}^{\circ}$  &   3.20 \nl

\protect\ion{Fe}{2}  & 5432.97 & $-3.63$   &
 24  &  $3d^6 ( ^3{\rm H})4s\ ^2{\rm H} - 3d^6 ( ^5{\rm D})4p\ ^4{\rm F}^{\circ}$  &   3.27 \nl

\protect\ion{Sc}{2}  & 5526.79 & $ 0.13 $  &
 25  &  $3d^2\ ^1{\rm G} - 3d4p\ ^1{\rm F}^{\circ}$  &   1.77 \nl

\protect\ion{Fe}{2}  & 5534.85 & $-2.93 $  &
 25  &  $3d^6 ( ^3{\rm H})4s\ ^2{\rm H} - 3d^6 ( ^5{\rm D})4p\ ^4{\rm F}^{\circ}$  &   3.24 \nl

\protect\ion{Sc}{2}  & 5641.00 &$ -1.04 $  &
 26  &  $3d^2\ ^3{\rm P} - 3d4p\ ^3{\rm P}^{\circ}$  &   1.50 \nl

\protect\ion{Sc}{2}  & 5657.90 & $-0.50 $  &
 26  &  $3d^2\ ^3{\rm P} - 3d4p\ ^3{\rm P}^{\circ}$  &   1.51 \nl

\protect\ion{Sc}{2}  & 5658.36 &$ -1.17 $  &
 26  &  $3d^2\ ^3{\rm P} - 3d4p\ ^3{\rm P}^{\circ}$  &   1.50 \nl

\protect\ion{Sc}{2}  & 5667.15 &$ -1.24$   &
 26  &  $3d^2\ ^3{\rm P} - 3d4p\ ^3{\rm P}^{\circ}$  &   1.50 \nl

\protect\ion{Sc}{2}  & 5669.04 & $-1.12$   &
 26  &  $3d^2\ ^3{\rm P} - 3d4p\ ^3{\rm P}^{\circ}$  &   1.50 \nl

\protect\ion{Sc}{2}  & 5684.20 & $-1.05 $  &
 26  &  $3d^2\ ^3{\rm P} - 3d4p\ ^3{\rm P}^{\circ}$  &   1.51 \nl

\protect\ion{Ba}{2}  & 5853.67 &$ -1.00$   &
 29  &  $5d\ ^2{\rm D} - 6p\ ^2{\rm P}$  &   0.60 \nl

\protect\ion{He}{1}\tablenotemark{e}  & 5875.63 & $ 0.74$   &
 28,29  &  $1s2p\ ^3{\rm P}^{\circ} - 1s3d\ ^3{\rm D}$  &  20.97 \nl

\protect\ion{Na}{1}  & 5889.95 & $ 0.12$   &
 29,27  &  $3s\ ^2{\rm S} - 3p\ ^2{\rm P}^{\circ}$  &   0.00 \nl

\protect\ion{Na}{1}  & 5895.92 &$ -0.18$   &
 29,27  &  $3s\ ^2{\rm S} - 3p\ ^2{\rm P}^{\circ}$  &   0.00 \nl

\protect\ion{Ba}{2}  & 6141.71 & $-0.08$   &
 30  &  $5d\ ^2{\rm D} - 6p\ ^2{\rm P}^{\circ}$  &   0.70 \nl

\protect\ion{Fe}{2}  & 6147.74 &$ -2.72 $  &
 30  &  $3d^6 ( ^3{\rm D})4s\ ^4{\rm D} - 3d^6 ( ^5{\rm D})4p\ ^4{\rm P}^{\circ}$  &   3.89 \nl

\protect\ion{Fe}{2}  & 6149.26 &$ -2.72 $  &
 30  &  $3d^6 ( ^3{\rm D})4s\ ^4{\rm D} - 3d^6 ( ^5{\rm D})4p\ ^4{\rm P}^{\circ}$  &   3.89 \nl

\protect\ion{Fe}{2}  & 6238.39 & $-2.63$   &
 31  &  $3d^6 ( ^3{\rm D})4s\ ^4{\rm D} - 3d^6 ( ^5{\rm D})4p\ ^4{\rm P}^{\circ}$  &   3.89 \nl

\protect\ion{Sc}{2}  & 6245.64 & $-0.98$   &
 31  &  $3d^2\ ^3{\rm P} - 3d4p\ ^3{\rm D}^{\circ}$  &   1.51 \nl

\protect\ion{Fe}{2}  & 6247.56 & $-2.33$   &
 31  &  $3d^6 ( ^3{\rm D})4s\ ^4{\rm D} - 3d^6 ( ^5{\rm D})4p\ ^4{\rm P}^{\circ}$  &   3.89 \nl

\protect\ion{Sc}{2}  & 6279.75 & $-1.21$   &
 31  &  $3d^2\ ^3{\rm P} - 3d4p\ ^3{\rm D}^{\circ}$  &   1.50 \nl

\protect\ion{Si}{2}  & 6347.11 & $ 0.30$   &
 32  &  $3s^2 ( ^1{\rm S})4s\ ^2{\rm S} - 3s^2 ( ^1{\rm S})4p\ ^2{\rm P}^{\circ}$  &   8.12 \nl

\protect\ion{Si}{2}  & 6371.37 &$ -0.00 $  &
 32  &  $3s^2 ( ^1{\rm S})4s\ ^2{\rm S} - 3s^2 ( ^1{\rm S})4p\ ^2{\rm P}^{\circ}$  &   8.12 \nl

\protect\ion{H}{1}  & 6562.80 & $ 0.71$   &
 34,33  &  \nodata\tablenotemark{d}  &  10.20 \nl

\enddata
\end{deluxetable}

\renewcommand{\arraystretch}{0.6}

\begin{deluxetable}{lcrccr}
\tablenum{4}
\ssp
\tablewidth{480pt}
\tablecaption{SN II-P Photospheric-Phase Absorption Line Identifications -- {\it Continued} }
\tablehead{\colhead{Ion} &
\colhead{Wavelength                  (\AA)\tablenotemark{a}} &
\colhead{log ({\it gf})} &
\colhead{Feature ID\tablenotemark{b}} &
\colhead{Designation} &
\colhead{{\it E} (eV)\tablenotemark{c}}}

\startdata

\protect\ion{Ca}{2}  & 8498.02 &$ -1.31$   &
   35  &  $3d\ ^2{\rm D} - 4p\ ^2{\rm P}^{\circ}$  & 1.69 \nl

\protect\ion{Ca}{2}  & 8662.14 & $-0.62$   &
 36  &  $3d\ ^2{\rm D} - 4p\ ^2{\rm P}^{\circ}$  &   1.69 \nl

\enddata

\tablecomments{Lines producing or contributing to absorption features in the
optical spectrum of SN~1999em during the photospheric phase.  Rest wavelengths,
oscillators strengths, configurations, terms, and energies taken from the
Kurucz (1996) spectral line database.}

\tablenotetext{a}{Rest wavelength (air).}

<\tablenotetext{b}{Feature number(s) corresponding to the identifications shown in
Figures 9 and 10.}

\tablenotetext{c}{Energy of the lower energy state of the transition.}

\tablenotetext{d}{For \protect\ion{H}{1}, the contributing closely-spaced
configurations and terms are: $2p\ ^2{\rm P}^{\circ} - {\rm n}d\ ^2D$, $2s\
2{\rm S} - {\rm n}p\ ^2{\rm P}^{\circ}$, and $2p\ ^2{\rm P}^{\circ} - {\rm n}s\
^2{\rm S}$, where n is the principle quantum number of the upper state of the
transition (i.e., 3 for H$\alpha$, 4 for H$\beta$, etc.).}

\tablenotetext{e}{For brevity, we do not list the 6 closely-spaced components
of the \protect\ion{He}{1} $\lambda 5876$ multiplet, but rather list the mean
wavelength and total oscillator strength of the line.}

\end{deluxetable}
\clearpage


\begin{figure}
\ssp
\vskip -0.5in
\hskip +0.5in
\rotatebox{0}{
\scalebox{0.9}{
\plotone{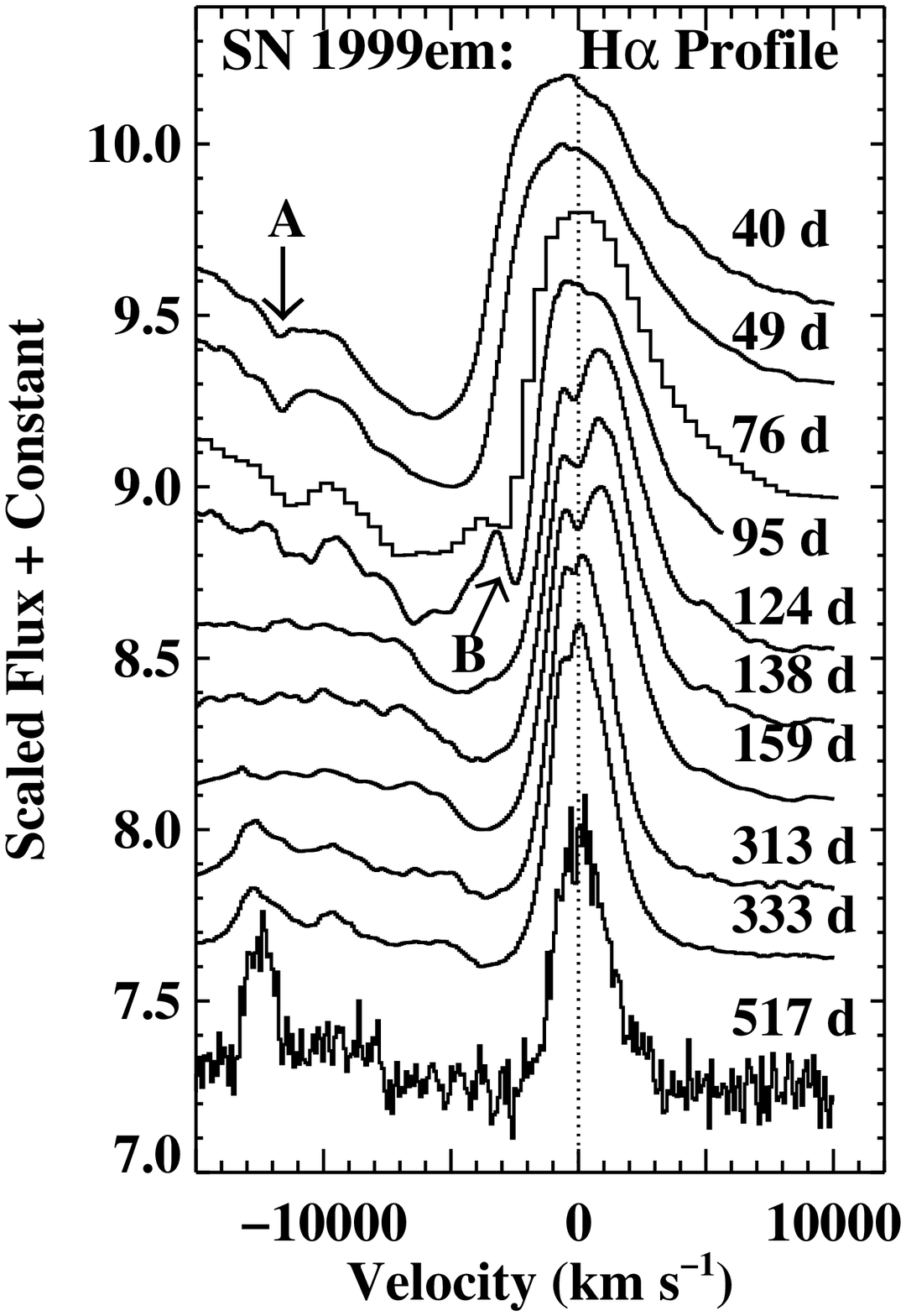}
}
}
\caption{ The development of the H$\alpha$ profile of SN~1999em, with days
since discovery (1999-11-29 UT) indicated.  Of particular interest are the
``notch'' near zero velocity that is first seen on day 124 (perhaps a hint on
day 95), a ``dip'' ({\it B}) on the blue side of the profile on days 76 and 95,
and a moderately strong absorption feature ({\it A}) at high velocity.  The
notch near zero velocity is seen in other strong emission lines (see L01), and
the high-velocity absorption feature ({\it A}) is seen in the \hbeta\ and
\protect\ion{Na}{1} $\lambda 5892$ profiles as well.  The plateau phase ended
about 95 days after discovery, the same period during which features {\it A}
and {\it B} disappear and the notch at zero velocity is first seen.  Note that
the spectrum from day 76 has very low resolution.}
\label{fig:5.18a}
\end{figure}

\subsubsection{Line Features Associated with P-Cygni Profiles}
\label{sec:chugai}

In all, there are five features (labeled 10, 14, 15, 27, and 33 in
Figure~\ref{fig:5.11}) for which no convincing line ID is found near the rest
wavelength inferred by assuming a velocity similar to that derived from the
more readily identified lines.  All five features appear at multiple epochs and
follow the general redward drift with time seen in the other lines.  We propose
that three of these features (14, 27, and 33) are actually associated with the
strong P-Cygni absorptions of \hbeta\ $\lambda4861$, \ion{Na}{1}~D $\lambda
5892$, and \halpha\ $\lambda 6563$, respectively, and represent the continued
evolution of the low-contrast features first identified in the spectrum from
day 7 (these features can be seen as small ``notches'' just blueward of the
\hbeta\ and $H\gamma$ absorption in Figure~\ref{fig:5.10}; due to contamination
from telluric features, the ``notch'' is not positively confirmed in \halpha\
on day 7 -- see L01).  From day 30 until the end of the plateau these features
mirror the other lines' decreasing velocity (about 2,000 \kms\ during this
period of time), but with a velocity about 8000 \kms\ higher.  It is {\it
possible} that the other two unidentified features (10 and 15) are also
high-velocity absorptions associated with strong lines: feature 10 with the
large metal blend of \ion{Ba}{2}, \ion{Ti}{2}, and \ion{Fe}{2} near $\lambda
4554$ (the candidate producing absorption most consistent with the other
high-velocity features is \ion{Fe}{2} $\lambda 4584$), and feature 15 with
\ion{Ba}{2} $\lambda 4934$.  These identifications are less secure, however,
given the relative weakness of the responsible lines.  We note that Fassia et
al. (1998) identify feature 33 with \ion{Sc}{2} $\lambda 6378$.  While this ID
does yield a velocity similar to other photospheric lines, the feature is
evident in neither the simulated spectra of Hatano et al. (1999) nor the
synthetic spectrum of Weaver \& Woosley (1993) that we analyzed.  The Kurucz
(1996) line list does confirm a relatively strong \ion{Sc}{2} feature at
$\lambda 6378.7$, but it lies 7.4 eV above the ground state.  For it to be
visible, many other features of similar strength and excitation should be
observed as well, and they are not.  We thus favor the identification of this
feature with \halpha.

Figure~\ref{fig:5.18a} shows the development of the \halpha\ profile from day
40 until our latest observation on day 517.  In addition to the high-velocity
absorption feature already discussed (labeled {\it A} in
Figure~\ref{fig:5.18a}), a curious ``absorption line,'' labeled {\it B},
develops between day 49 and day 76 and disappears by day 124.  If associated
with \halpha, the corresponding velocity is quite similar to the velocity
derived from weaker lines, such as \ion{Fe}{2} $\lambda 5018$.  A similar
feature was also observed in the Type II-P SN~1999gi (Leonard \& Filippenko
2001), as well as in SN~1988A (Turatto et al. 1993), and perhaps SN~1985L
(Filippenko \& Sargent 1986).  A likely explanation is that this feature is
produced by Chugai's (1991; see also Utrobin, Chugai, \& Andronova 1995)
proposed mechanism for the anomalous blueshifted ``absorption'' seen in
SN~1987A (i.e., the ``Bochum event''; see also Phillips \& Heathcote 1989).
Under this model, the feature is actually produced by a {\it lack} of
absorption by gas at velocities surrounding the ``dip,'' caused by ineffective
screening of continuum photons due to excitation stratification in the
envelope.

Finally, there is the development of a distinct ``notch'' in the \halpha,
\hbeta, and [\ion{Ca}{2}] $\lambda\lambda7291, 7324$ profiles near their rest
wavelengths, which develops immediately after the end of the plateau
(Figure~\ref{fig:5.18a}).  As discussed by L01, this feature might originate
within the expanding ejecta itself, perhaps due to an asymmetry in the
line-emitting region.  It is possible that it is produced by the ejecta
interacting with circumstellar material, although the lack of any other
evidence suggesting significant circumstellar interaction (e.g., Pooley et al.,
2001) makes us doubt this possibility.  The notch seems to diminish in strength
with time, and it may not be present on day 517, although the low
signal-to-noise ratio (S/N) of that spectrum renders it insensitive to
low-contrast features.  It is also interesting to note the many other features
that complicate the \halpha\ absorption profile as SN~1999em makes the
transition out of the plateau epoch, particularly evident in the day 95
spectrum; these features are also seen in a spectrum of SN~1999gi at a similar
epoch; in fact, the exact correspondence of {\it all} of the line featurs
between these two SNe II-P is quite remarkable (Leonard \& Filippenko 2001).

\subsubsection{Inferring Photospheric Velocity from Absorption Lines}
\label{sec:inferringvel}

From the lines identified in Figure~\ref{fig:5.11}, we find four weak,
unblended spectral features that should serve as good photospheric velocity
indicators to compare with the velocity determined by the stronger \fetwo\
lines: \ion{Fe}{2} $\lambda 4629$ (feature 12), \ion{Sc}{2} $\lambda 4670$
(feature 13), \ion{Fe}{2} $\lambda 5276$ (feature 21), and \ion{Fe}{2} $\lambda
5318$ (feature 22).  To translate the measured wavelength minima into
velocities we must remove the redshift of SN~1999em and assign rest wavelengths
to all of the line features.  The NASA/IPAC Extragalactic Database (NED)
recession velocity of NGC~1637 is 717~\kms.  However, since the host galaxy is
somewhat inclined ($i = 32^\circ$) and the location of the SN within the spiral
arm is not known, the NED velocity may differ significantly from the actual
velocity of the SN.  One way sometimes used to infer the recession velocity of
a SN is to measure the velocity of bright \ion{H}{2} regions on both sides of
the SN along the slit, and interpolate to the position of the SN.  When this is
done for SN~1999em on the frames that contain bright \ion{H}{2} regions,
however, values from 690 to 900 \kms\ result; in other words, we find
the inferred velocity to be a function of which \ion{H}{2} regions fall along
the slit, which is determined by the slit's position angle.  As shown in
Figure~\ref{fig5:10b}, the \ion{Na}{1} D interstellar absorption from NGC~1637
occurs over a range from about $700 {\rm\ to\ } 820$ \kms.  We are thus left
with some uncertainty in the exact recession velocity of SN~1999em.  This
uncertainty ultimately leads to a small systematic uncertainty in our derived
EPM distance to SN~1999em.  For our analysis, we assign a recession
velocity to SN~1999em of $800$ \kms\ and a $1\sigma$ uncertainty of $50$ \kms.

Assigning a single rest wavelength to each of the features identified in
Figures~\ref{fig:5.10} and ~\ref{fig:5.11} is complicated by the blending that
exists in many of the lines, and simply taking the rest wavelength of the
dominant line may lead to inaccuracies.  When a feature consists of a closely
spaced multiplet of a single ion, we estimate the rest wavelength as
\begin{equation}
\lambda_\circ = \frac{\sum_{i=1}^{n} \lambda_i \times (gf)_i }{\sum_{i=1}^{n} (gf)_i},
\label{eqn:multiplet}
\end{equation}

\noindent where $i$ is the line number and $(gf)$ is the oscillator strength.
Since it has been previously used as a photospheric velocity indicator (e.g.,
Schmidt et al. 1994a), we note that the 6 lines of the multiplet of \ion{Sc}{2}
comprising feature 26 yield $\lambda_\circ = 5665.6$ \AA, which is
significantly different from the rest wavelength of the strongest component,
$\lambda_\circ = 5657.9$ \AA; using $\lambda_\circ = 5657.9$ \AA, as was
evidently done in previous studies, would therefore result in derived
velocities that are $\sim 400$ \kms\ lower than is appropriate for this
feature.  For features that contain blends of lines from different ions (or
different energy levels within the same ion), the assigned rest wavelength was
guided by estimates of the relative strengths of the individual lines during
the photospheric phase.  In many cases, though (i.e., \ion{Fe}{2} $\lambda
4924$ and \ion{Sc}{2} $\lambda 5527$), the relative strengths of the individual
lines likely vary considerably, diminishing their reliability as simple
photospheric velocity indicators.

To assign an uncertainty to the inferred velocities, we attempt to quantify the
error resulting from uncertainty in both the wavelength scale and the
measurement process itself.  To gauge the uncertainty in our wavelength scale,
we cross-correlated the region around the strong \ion{Na}{1}~D $\lambda\lambda
5890, 5896$ host galaxy interstellar absorption line in every low-resolution
spectrum with this region in our day 95 spectrum, which has an excellent S/N.
The $1\sigma$ spread of wavelength offsets was found to be 0.6 \AA, which we
adopt as the uncertainty in our derived wavelength scale for every spectrum.

To measure the wavelength of maximum absorption and derive its measurement
uncertainty, we first estimated the local ``continuum'' shape by fitting a
straight line to points near the two local flux peaks on either side of the
absorption feature.  We then normalized the absorption feature by dividing it
by the interpolated continuum fit.  To estimate the wavelength of minimum flux
in the normalized line profile, we then fit a series of third-order
Savitsky-Golay smoothing polynomials (Press et al. 1992) to the data with
widths ranging from 1/2 to 3/2 the FWHM of the line profile and then averaged
the fits; this was found to produce more robust results than using a single
smoothing function alone.  The minimum of the smoothed spectrum was then found,
interpolated to the nearest 0.01 \AA.  To estimate the measurement error, we
added random noise at the level seen in the data to the smoothed spectrum and
used an automatic routine to find the absorption minimum in 1000 sets of noisy
artificial data.  The standard deviation of these measurements added in
quadrature to the 0.6 \AA\ uncertainty in our wavelength scale is the final
uncertainty reported for the line's absorption minimum.  We do not attempt to
quantify the systematic uncertainty in a feature's assigned rest wavelength,
but caution that for some blends this systematic error may dominate the total
(statistical and systematic) uncertainty.

We characterize each feature's strength by measuring its ``depth,'' defined as
\begin{equation}
d = (f_{\rm c} - f_{\rm min} ) / f_c, 
\label{eqn:depth}
\end{equation}

\noindent where $f_{\rm min}$ is the flux at the line's minimum and $f_c$ is
the value of the interpolated continuum flux at the location of the line's
minimum.  The measured depth in some cases depends significantly on the
strength of adjacent features since this affects the location of the points
used to fit the normalizing continuum.  In general, though, $d$ should be a
reasonably good indicator of line strength.  The rest wavelengths adopted for
all features and the measured velocities, velocity uncertainties, and line
depths are given in Table~5\notetoeditor{If possible, we would like only the
``abbreviated Table'' version of Table 5 to be displayed in the printed paper,
with the complete table to be given in electronic form only.}.

\begin{deluxetable}{rcccccccccc}
\ssp
\renewcommand{\arraystretch}{0.7}
\ptlandscape
\rotate
\tablenum{5}
\tablewidth{00pt}
\tablecaption{SN 1999em:  Absorption Line Measurements}
\tablehead{\colhead{}  &
\colhead{} &
\colhead{} &
\multicolumn{2}{c}{Day 2\tablenotemark{a}} &
\multicolumn{2}{c}{Day 3} &
\multicolumn{2}{c}{Day 4} &
\multicolumn{2}{c}{Day 5} \\ 
\cline{4-5} \cline{6-7} \cline{8-9} \cline{10-11}
\colhead{Line\tablenotemark{b}} &
\colhead{Ion\tablenotemark{c}}  &
\colhead{Wavelength\tablenotemark{d} (\AA)} &
\colhead{Vel.\tablenotemark{e}} &
\colhead{Depth\tablenotemark{f}} &
\colhead{Vel.} &
\colhead{Depth} &
\colhead{Vel.} &
\colhead{Depth} &
\colhead{Vel.} &
\colhead{Depth} } 

\startdata

 1 & H$\zeta$ & 3889.05 & \nodata & \nodata   & \nodata & \nodata   & \nodata & \nodata   & \nodata & \nodata   \nl
 2 & H$\epsilon$ & 3970.07 & \nodata & \nodata   & \nodata & \nodata   & \nodata & \nodata   & \nodata & \nodata   \nl
 3 & \ion{Sr}{2}: & 4077.71 & \nodata & \nodata   & \nodata & \nodata   & \nodata & \nodata   & \nodata & \nodata   \nl
 4 & H$\delta$ & 4101.77 & \nodata & \nodata   & \nodata & \nodata   & \nodata & \nodata   & \nodata & \nodata   \nl
 5 & \ion{Fe}{2}: & 4180.98 & \nodata & \nodata   & \nodata & \nodata   & \nodata & \nodata   & \nodata & \nodata   \nl
 6 & \ion{Sc}{2}: & 4241.00 & \nodata & \nodata   & \nodata & \nodata   & \nodata & \nodata   & \nodata & \nodata   \nl
 7 & \ion{Fe}{2}: & 4301.70 & \nodata & \nodata   & \nodata & \nodata   & \nodata & \nodata   & \nodata & \nodata   \nl
 8 & H$\gamma$ & 4340.49 & 9020 (240) & 0.08 & 8710 (560) & 0.10 & 8500 (340) & 0.12 & 8340 (180) & 0.18 \nl
 9 & \ion{Sc}{2}: & 4393.00 & \nodata & \nodata   & \nodata & \nodata   & \nodata & \nodata   & \nodata & \nodata   \nl
 10 & No ID & \nodata & \nodata & \nodata   & \nodata & \nodata   & \nodata & \nodata   & \nodata & \nodata   \nl
 11 & \ion{Ba}{2}: & 4554.03 & \nodata & \nodata   & \nodata & \nodata   & \nodata & \nodata   & \nodata & \nodata   \nl
 12 & \ion{Fe}{2} & 4629.34 & \nodata & \nodata   & \nodata & \nodata   & \nodata & \nodata   & \nodata & \nodata   \nl
 13 & \ion{Sc}{2} & 4670.41 & \nodata & \nodata   & \nodata & \nodata   & \nodata & \nodata   & \nodata & \nodata   \nl
 14 & H$\beta$ PCyg & 4861.36 & \nodata & \nodata   & \nodata & \nodata   & \nodata & \nodata   & \nodata & \nodata   \nl
 15 & No ID & \nodata & \nodata & \nodata   & \nodata & \nodata   & \nodata & \nodata   & \nodata & \nodata   \nl
 16 & H$\beta$ & 4861.36 & 10460 (360) & 0.22 & 10080 (110) & 0.27 & 10040 (120) & 0.29 & 9410 (90) & 0.33 \nl
 17 & \ion{Fe}{2}\tablenotemark{g} & 4923.93 & \nodata & \nodata   & \nodata & \nodata   & \nodata & \nodata   & \nodata & \nodata   \nl
 18 & \ion{Fe}{2} & 5018.44 & \nodata & \nodata   & \nodata & \nodata   & \nodata & \nodata   & \nodata & \nodata   \nl
 19 & \ion{Fe}{2} & 5169.03 & \nodata & \nodata   & \nodata & \nodata   & \nodata & \nodata   & \nodata & \nodata   \nl
 20 & \ion{Fe}{2}: & 5234.63 & \nodata & \nodata   & \nodata & \nodata   & \nodata & \nodata   & \nodata & \nodata   \nl
 21 & \ion{Fe}{2} & 5276.00 & \nodata & \nodata   & \nodata & \nodata   & \nodata & \nodata   & \nodata & \nodata   \nl
 22 & \ion{Fe}{2} & 5318.00 & \nodata & \nodata   & \nodata & \nodata   & \nodata & \nodata   & \nodata & \nodata   \nl
 23 & \ion{Fe}{2} & 5362.87 & \nodata & \nodata   & \nodata & \nodata   & \nodata & \nodata   & \nodata & \nodata   \nl
 24 & \ion{Fe}{2}: & 5425.00 & \nodata & \nodata   & \nodata & \nodata   & \nodata & \nodata   & \nodata & \nodata   \nl
 25 & \ion{Sc}{2}: & 5533.00 & \nodata & \nodata   & \nodata & \nodata   & \nodata & \nodata   & \nodata & \nodata   \nl
 26 & \ion{Sc}{2}: & 5665.60 & \nodata & \nodata   & \nodata & \nodata   & \nodata & \nodata   & \nodata & \nodata   \nl
 27 & \ion{Na}{1} PCyg & 5892.00 & \nodata & \nodata   & \nodata & \nodata   & \nodata & \nodata   & \nodata & \nodata   \nl
 28 & \ion{He}{1}\tablenotemark{h} & 5875.63 & 9330 (120) & 0.09 & 9240 (120) & 0.10 & 8580 (70) & 0.13 & 8300 (80) & 0.10 \nl
 29 & \ion{Na}{1}D:\tablenotemark{i} & 5892.00 & \nodata & \nodata   & \nodata & \nodata   & \nodata & \nodata   & \nodata & \nodata   \nl
 30 & \ion{Ba}{2}: & 6147.00 & \nodata & \nodata   & \nodata & \nodata   & \nodata & \nodata   & \nodata & \nodata   \nl
 31 & \ion{Sc}{2}: & 6254.00 & \nodata & \nodata   & \nodata & \nodata   & \nodata & \nodata   & \nodata & \nodata   \nl
 32 & \ion{Si}{2} & 6355.20 & \nodata & \nodata   & \nodata & \nodata   & \nodata & \nodata   & \nodata & \nodata   \nl
 33 & H$\alpha$ PCyg & 6562.85 & \nodata & \nodata   & \nodata & \nodata   & \nodata & \nodata   & \nodata & \nodata   \nl
 34 & H$\alpha$ & 6562.85 & 10810 (400) & 0.27 & 10540 (70) & 0.32 & 10210 (70) & 0.35 & 9560 (110)& 0.34   \nl
 35 & \ion{Ca}{2} & 8498.02 & \nodata & \nodata   & \nodata & \nodata   & \nodata & \nodata   & \nodata & \nodata   \nl
 36 & \ion{Ca}{2} & 8662.14 & \nodata & \nodata   & \nodata & \nodata   & \nodata & \nodata   & \nodata & \nodata   \nl

\enddata

\end{deluxetable}

\begin{deluxetable}{rcccccccccc}
\ssp
\rotate
\ptlandscape
\tablenum{5 -- {\it Continued}}
\tablecaption{}
\tablewidth{00pt}
\tablehead{\colhead{}  &
\multicolumn{2}{c}{Day 6} &
\multicolumn{2}{c}{Day 7} &
\multicolumn{2}{c}{Day 8} &
\multicolumn{2}{c}{Day 9} & 
\multicolumn{2}{c}{Day 10} \\ 
\cline{2-3} \cline{4-5} \cline{6-7} \cline{8-9} \cline{10-11}
\colhead{Line} &
\colhead{Vel.} &
\colhead{Depth} &
\colhead{Vel.} &
\colhead{Depth} &
\colhead{Vel.} &
\colhead{Depth} &
\colhead{Vel.} &
\colhead{Depth} &
\colhead{Vel.} &
\colhead{Depth} }

\startdata 
 1 & \nodata & \nodata   & 7270 (80) & 0.04  & \nodata & \nodata   & \nodata & \nodata   & \nodata & \nodata   \nl
 2 & \nodata & \nodata   & 7320 (180) & 0.05 & \nodata & \nodata   & \nodata & \nodata   & \nodata & \nodata   \nl
 3 & \nodata & \nodata   & \nodata & \nodata & \nodata & \nodata   & \nodata & \nodata   & \nodata & \nodata   \nl
 4 & \nodata & \nodata   & 7450 (40) & 0.11  & 6580 (200) & \nodata & 6210 (690) & \nodata & \nodata & \nodata   \nl
 5 & \nodata & \nodata   & \nodata & \nodata & \nodata & \nodata   & \nodata & \nodata   & \nodata & \nodata   \nl
 6 & \nodata & \nodata   & \nodata & \nodata & \nodata & \nodata   & \nodata & \nodata   & \nodata & \nodata   \nl
 7 & \nodata & \nodata   & \nodata & \nodata & \nodata & \nodata   & \nodata & \nodata   & \nodata & \nodata   \nl
 8 & 8350 (120) & 0.16   & 7990 (70) & 0.26  & 7910 (110) & 0.24 & 7580 (100) & 0.26 & 7290 (80) & 0.23 \nl
 9 & \nodata & \nodata   & \nodata & \nodata & \nodata & \nodata   & \nodata & \nodata   & \nodata & \nodata   \nl
 10 & \nodata & \nodata  & \nodata & \nodata & \nodata & \nodata   & \nodata & \nodata   & \nodata & \nodata   \nl
 11 & \nodata & \nodata  & \nodata & \nodata & \nodata & \nodata   & \nodata & \nodata   & 9380 (180) & 0.19 \nl
 12 & \nodata & \nodata  & \nodata & \nodata & \nodata & \nodata   & \nodata & \nodata   & \nodata & \nodata   \nl
 13 & \nodata & \nodata  & \nodata & \nodata & \nodata & \nodata   & \nodata & \nodata   & \nodata & \nodata   \nl
 14 & \nodata & \nodata  & 12780 (60) & 0.02 & \nodata & \nodata   & \nodata & \nodata   & \nodata & \nodata   \nl
 15 & \nodata & \nodata  & \nodata & \nodata & \nodata & \nodata   & \nodata & \nodata   & \nodata & \nodata   \nl
 16 & 9000 (100) & 0.29  & 8640 (40) & 0.37  & 8570 (60) & 0.38 & 8480 (60) & 0.38 & 7920 (60) & 0.39 \nl
 17 & \nodata & \nodata  & \nodata & \nodata & \nodata & \nodata   & \nodata & \nodata   & \nodata & \nodata   \nl
 18 & \nodata & \nodata  & \nodata & \nodata & \nodata & \nodata   & 7180 (570) & 0.03 & 7030 (200) & 0.06 \nl
 19 & \nodata & \nodata  & 8030 (130) & 0.02 & 8080 (280) & 0.07 & 7690 (180) & 0.09 & 7260 (100) & 0.15 \nl
 20 & \nodata & \nodata  & \nodata & \nodata & \nodata & \nodata   & \nodata & \nodata   & \nodata & \nodata   \nl
 21 & \nodata & \nodata  & \nodata & \nodata & \nodata & \nodata   & \nodata & \nodata   & \nodata & \nodata   \nl
 22 & \nodata & \nodata  & \nodata & \nodata & \nodata & \nodata   & \nodata & \nodata   & \nodata & \nodata   \nl
 23 & \nodata & \nodata  & \nodata & \nodata & \nodata & \nodata   & \nodata & \nodata   & \nodata & \nodata   \nl
 24 & \nodata & \nodata  & \nodata & \nodata & \nodata & \nodata   & \nodata & \nodata   & \nodata & \nodata   \nl
 25 & \nodata & \nodata  & \nodata & \nodata & \nodata & \nodata   & \nodata & \nodata   & \nodata & \nodata   \nl
 26 & \nodata & \nodata  & \nodata & \nodata & \nodata & \nodata   & \nodata & \nodata   & \nodata & \nodata   \nl
 27 & \nodata & \nodata  & \nodata & \nodata & \nodata & \nodata   & \nodata & \nodata   & \nodata & \nodata   \nl
 28 & 8110 (280) & 0.08  & 7260 (70) & 0.09  & 7070 (220) & 0.07 & 6810 (220) & 0.04 & 5810 (90) & 0.04 \nl
 29 & \nodata & \nodata  & \nodata & \nodata & \nodata & \nodata   & \nodata & \nodata   & \nodata & \nodata   \nl
 30 & \nodata & \nodata  & \nodata & \nodata & \nodata & \nodata   & \nodata & \nodata   & \nodata & \nodata   \nl
 31 & \nodata & \nodata  & \nodata & \nodata & \nodata & \nodata   & \nodata & \nodata   & \nodata & \nodata   \nl
 32 & \nodata & \nodata  & 7890 (120) & 0.03 & \nodata & \nodata   & \nodata & \nodata   & 6640 (120) & 0.07 \nl
 33 & \nodata & \nodata  & \nodata & \nodata & \nodata & \nodata   & \nodata & \nodata   & \nodata & \nodata   \nl
 34 & 9550 (160) & 0.34  & 9820 (50) & 0.48  & 9560 (60) & 0.43 & 9260 (60) & 0.44 & 9010 (40) & 0.51 \nl
 35 & \nodata & \nodata  & \nodata & \nodata & \nodata & \nodata   & \nodata & \nodata   & \nodata & \nodata   \nl
 36 & \nodata & \nodata  & \nodata & \nodata & \nodata & \nodata   & \nodata & \nodata   & \nodata & \nodata   \nl

\enddata

\end{deluxetable}

\clearpage

\begin{deluxetable}{rcccccccccc}
\ssp
\rotate
\ptlandscape
\tablenum{5 -- {\it Continued}}
\tablecaption{}
\tablewidth{00pt}
\tablehead{\colhead{}  &
\multicolumn{2}{c}{Day 11} &
\multicolumn{2}{c}{Day 21} &
\multicolumn{2}{c}{Day 24} &
\multicolumn{2}{c}{Day 26} & 
\multicolumn{2}{c}{Day 30} \\ 
\cline{2-3} \cline{4-5} \cline{6-7} \cline{8-9} \cline{10-11}
\colhead{Line} &
\colhead{Vel.} &
\colhead{Depth} &
\colhead{Vel.} &
\colhead{Depth} &
\colhead{Vel.} &
\colhead{Depth} &
\colhead{Vel.} &
\colhead{Depth} &
\colhead{Vel.} &
\colhead{Depth} }

\startdata 

 1 & \nodata & \nodata   & \nodata & \nodata   & \nodata & \nodata   & \nodata & \nodata   & \nodata & \nodata   \nl
 2 & \nodata & \nodata   & \nodata & \nodata   & \nodata & \nodata   & \nodata & \nodata   & \nodata & \nodata   \nl
 3 & \nodata & \nodata   & 3670 (220) & \nodata & 3170 (90) & \nodata & 2890 (90) & \nodata & 3250 (190) & \nodata \nl
 4 & \nodata & \nodata   & \nodata & \nodata   & \nodata & \nodata   & \nodata & \nodata   & \nodata & \nodata   \nl
 5 & \nodata & \nodata   & \nodata & \nodata   & \nodata & \nodata   & 5190 (80) & 0.24 & 4570 (340) & 0.17 \nl
 6 & \nodata & \nodata   & \nodata & \nodata   & 5030 (130) & 0.22 & 4920 (240) & 0.17 & 4640 (180) & 0.21 \nl
 7 & \nodata & \nodata   & 4040 (130) & 0.44 & 4040 (160) & 0.39 & 3690 (180) & 0.38 & 3550 (170) & 0.39 \nl
 8 & 7150 (150) & 0.29 & \nodata & \nodata   & \nodata & \nodata   & \nodata & \nodata   & \nodata & \nodata   \nl
 9 & \nodata & \nodata   & \nodata & \nodata   & \nodata & \nodata   & 4840 (250) & 0.11 & \nodata & \nodata   \nl
 10 & \nodata & \nodata   & \nodata & \nodata   & \nodata & \nodata   &  [4383.7 (6.9)]  & 0.05 & \nodata & \nodata   \nl
 11 & 8260 (60) & 0.18 & 5390 (110) & 0.45 & \nodata & \nodata   & 5190 (80) & 0.48 & \nodata & \nodata   \nl
 12 & \nodata & \nodata   & \nodata & \nodata   & \nodata & \nodata   & \nodata & \nodata   & 4240 (210) & 0.04 \nl
 13 & \nodata & \nodata   & \nodata & \nodata   & \nodata & \nodata   & \nodata & \nodata   & \nodata & \nodata   \nl
 14 & \nodata & \nodata   & \nodata & \nodata   & \nodata & \nodata   & \nodata & \nodata   & 12230 (220) & 0.09 \nl
 15 & \nodata & \nodata   & \nodata & \nodata   & \nodata & \nodata   & \nodata & \nodata   & \nodata & \nodata   \nl
 16 & 7870 (50) & 0.36 & 6630 (120) & 0.55 & 6160 (80) & 0.56 & 5870 (90) & 0.60 & 5240 (80) & 0.54 \nl
 17 & \nodata & \nodata   & 5470 (250) & 0.09 & 4850 (200) & 0.13 & 4670 (130) & 0.15 & 4580 (60) & 0.19 \nl
 18 & 7130 (90) & 0.11 & 5690 (80) & 0.26 & 5120 (70) & 0.31 & 4900 (80) & 0.34 & 4630 (50) & 0.36 \nl
 19 & 7250 (60) & 0.22 & 5540 (90) & 0.37 & 5200 (130) & 0.41 & 5050 (100) & 0.44 & 4660 (60) & 0.48 \nl
 20 & \nodata & \nodata   & \nodata & \nodata   & \nodata & \nodata   & \nodata & \nodata   & 4360 (110) & 0.04 \nl
 21 & \nodata & \nodata   & \nodata & \nodata   & \nodata & \nodata   & \nodata & \nodata   & 4370 (210) & 0.05 \nl
 22 & 7240 (360) & 0.01 & 5320 (120) & 0.09 & 4890 (290) & 0.11 & 4870 (220) & 0.14 & 4410 (150) & 0.11 \nl
 23 & \nodata & \nodata   & \nodata & \nodata   & \nodata & \nodata   & \nodata & \nodata   & \nodata & \nodata   \nl
 24 & \nodata & \nodata   & \nodata & \nodata   & 4930 (230) & 0.04 & \nodata & \nodata   & \nodata & \nodata   \nl
 25 & \nodata & \nodata   & 5510 (350) & 0.07 & 4840 (330) & 0.07 & 4650 (90) & 0.10 & 4450 (70) & 0.11 \nl
 26 & \nodata & \nodata   & \nodata & \nodata   & \nodata & \nodata   & \nodata & \nodata   & \nodata & \nodata   \nl
 27 & \nodata & \nodata   & \nodata & \nodata   & \nodata & \nodata   & \nodata & \nodata   & \nodata & \nodata   \nl
 28 & \nodata & \nodata   & \nodata & \nodata   & \nodata & \nodata   & \nodata & \nodata   & \nodata & \nodata   \nl
 29 & \nodata & \nodata   & 5620 (220) & 0.10 & 5170 (170) & 0.16 & 4790 (80) & 0.21 & 4470 (50) & 0.29 \nl
 30 & \nodata & \nodata   & \nodata & \nodata   & \nodata & \nodata   & \nodata & \nodata   & 4270 (190) & 0.04 \nl
 31 & \nodata & \nodata   & \nodata & \nodata   & 5450 (180) & 0.06 & 4900 (110) & 0.06 & 4630 (110) & 0.07 \nl
 32 & 6620 (110) & 0.07 & \nodata & \nodata   & \nodata & \nodata   & \nodata & \nodata   & \nodata & \nodata   \nl
 33 & \nodata & \nodata   & \nodata & \nodata   & \nodata & \nodata   & \nodata & \nodata   & \nodata & \nodata   \nl
 34 & 8880 (40) & 0.48 & 7680 (130) & 0.70 & 7160 (30) & 0.70 & 7100 (50) & 0.75 & 6650 (40) & 0.76 \nl
 35 & \nodata & \nodata   & \nodata & \nodata   & \nodata & \nodata   & \nodata & \nodata   & \nodata & \nodata   \nl
 36 & \nodata & \nodata   & \nodata & \nodata   & \nodata & \nodata   & \nodata & \nodata   & \nodata & \nodata   \nl

\enddata

\end{deluxetable}

\clearpage

\begin{deluxetable}{rcccccccccc}
\ssp
\rotate
\ptlandscape
\tablenum{5 -- {\it Continued}}
\tablecaption{}
\tablewidth{00pt}
\tablehead{\colhead{}  &
\multicolumn{2}{c}{Day 34} &
\multicolumn{2}{c}{Day 37} &
\multicolumn{2}{c}{Day 40} &
\multicolumn{2}{c}{Day 44} & 
\multicolumn{2}{c}{Day 47} \\ 
\cline{2-3} \cline{4-5} \cline{6-7} \cline{8-9} \cline{10-11}
\colhead{Line} &
\colhead{Vel.} &
\colhead{Depth} &
\colhead{Vel.} &
\colhead{Depth} &
\colhead{Vel.} &
\colhead{Depth} &
\colhead{Vel.} &
\colhead{Depth} &
\colhead{Vel.} &
\colhead{Depth} }

\startdata 

 1 & \nodata & \nodata   & \nodata & \nodata   & \nodata & \nodata   & \nodata & \nodata   & \nodata & \nodata   \nl
 2 & \nodata & \nodata   & \nodata & \nodata   & \nodata & \nodata   & \nodata & \nodata   & \nodata & \nodata   \nl
 3 & 3210 (370) & \nodata & \nodata & \nodata   & \nodata & \nodata   & 3510 (100) & 0.35 & 3350 (80) & 0.37 \nl
 4 & \nodata & \nodata   & \nodata & \nodata   & \nodata & \nodata   & \nodata & \nodata   & \nodata & \nodata   \nl
 5 & 4400 (130) & 0.22 & 4220 (100) & 0.17 & \nodata & \nodata   & 4180 (190) & 0.28 & 3870 (80) & 0.24 \nl
 6 & 4340 (100) & 0.24 & 4410 (100) & 0.20 & \nodata & \nodata   & 3990 (110) & 0.29 & 4000 (90) & 0.35 \nl
 7 & 3600 (230) & 0.44 & 3550 (140) & 0.39 & 3660 (60) & 0.35 & 3370 (80) & 0.42 & 3100 (220) & 0.40 \nl
 8 & \nodata & \nodata   & \nodata & \nodata   & \nodata & \nodata   & \nodata & \nodata   & \nodata & \nodata   \nl
 9 & 3810 (170) & 0.19 & 3690 (80) & 0.14 & 3730 (60) & 0.23 & 3650 (100) & 0.27 & 3650 (130) & 0.24 \nl
 10 &  [4394.6 (3.4)]  & 0.08 &  [4395.5 (3.4)]  & 0.06 &  [4405.2 (0.9)]  & 0.09 &  [4404.1 (1.9)]  & 0.09 & \nodata & \nodata   \nl
 11 & 4220 (50) & 0.51 & 3810 (90) & 0.54 & 3580 (70) & 0.59 & 3630 (70) & 0.56 & 3700 (120) & 0.56 \nl
 12 & 4020 (110) & 0.07 & 3540 (190) & 0.08 & 3480 (60) & 0.12 & 3540 (160) & 0.12 & 3410 (80) & 0.12 \nl
 13 & 3900 (100) & 0.02 & 3740 (280) & 0.04 & 3660 (60) & 0.07 & 3660 (80) & 0.08 & 3410 (70) & 0.10 \nl
 14 & 11960 (90) & 0.08 & 11710 (80) & 0.08 & 11700 (60) & 0.11 & \nodata & \nodata   & 11650 (100) & 0.12 \nl
 15 & \nodata & \nodata   & \nodata & \nodata   &  [4748.0 (0.9)]  & 0.05 & \nodata & \nodata   &  [4753.2 (1.6)]  & 0.07 \nl
 16 & 5050 (40) & 0.50 & 4270 (120) & 0.48 & 4160 (60) & 0.49 & 4020 (50) & 0.42 & 3870 (50) & 0.32 \nl
 17 & 4270 (50) & 0.25 & 3750 (110) & 0.25 & 3790 (50) & 0.33 & 3840 (50) & 0.33 & 3720 (50) & 0.35 \nl
 18 & 4310 (80) & 0.42 & 3800 (80) & 0.46 & 3720 (50) & 0.51 & 3770 (90) & 0.54 & 3680 (50) & 0.54 \nl
 19 & 4410 (80) & 0.49 & 3910 (130) & 0.55 & 3890 (70) & 0.56 & 3960 (120) & 0.55 & 3870 (50) & 0.54 \nl
 20 & 4170 (150) & 0.03 & 3720 (110) & 0.04 & 3680 (60) & 0.06 & 3770 (90) & 0.08 & 3420 (70) & 0.08 \nl
 21 & 4040 (150) & 0.04 & 3590 (150) & 0.05 & 3610 (70) & 0.09 & 3430 (60) & 0.09 & 3510 (90) & 0.11 \nl
 22 & 4090 (50) & 0.12 & 3560 (140) & 0.14 & 3620 (60) & 0.16 & 3680 (110) & 0.16 & 3480 (90) & 0.17 \nl
 23 & \nodata & \nodata   & \nodata & \nodata   & 3440 (60) & 0.02 & \nodata & \nodata   & \nodata & \nodata   \nl
 24 & 3980 (190) & 0.03 & 3340 (110) & 0.05 & 3430 (90) & 0.07 & 3450 (50) & 0.10 & 3240 (80) & 0.09 \nl
 25 & 4140 (50) & 0.16 & 3660 (140) & 0.19 & 3650 (60) & 0.23 & 3610 (50) & 0.27 & 3440 (50) & 0.27 \nl
 26 & 4290 (60) & 0.15 & 3690 (130) & 0.18 & 3710 (80) & 0.24 & 3610 (40) & 0.27 & 3410 (40) & 0.30 \nl
 27 & 12010 (110) & 0.03 & 11670 (310) & 0.02 & 11820 (60) & 0.02 & \nodata & \nodata   & \nodata & \nodata   \nl
 28 & \nodata & \nodata   & \nodata & \nodata   & \nodata & \nodata   & \nodata & \nodata   & \nodata & \nodata   \nl
 29 & 4170 (50) & 0.38 & 3750 (80) & 0.43 & 3680 (70) & 0.49 & 3580 (40) & 0.51 & 3540 (40) & 0.52 \nl
 30 & 3770 (90) & 0.09 & 3620 (190) & 0.08 & 3510 (70) & 0.11 & 3350 (60) & 0.14 & 3150 (40) & 0.17 \nl
 31 & 4200 (60) & 0.12 & 4010 (120) & 0.13 & 3740 (80) & 0.17 & 3750 (50) & 0.19 & 3640 (60) & 0.22 \nl
 32 & 4050 (70) & 0.05 & \nodata & \nodata   & 3760 (50) & 0.03 & \nodata & \nodata   & \nodata & \nodata   \nl
 33 & 11900 (50) & 0.10 & 11760 (80) & 0.07 & 11770 (40) & 0.13 & 11730 (50) & 0.11 & 11730 (50) & 0.14 \nl
 34 & 6230 (40) & 0.77 & 5860 (190) & 0.78 & 5570 (60) & 0.78 & 5120 (50) & 0.76 & 4940 (40) & 0.74 \nl
 35 & \nodata & \nodata   & 4870 (180) & 0.78 & \nodata & \nodata   & 4780 (60) & 0.59 & 4470 (120) & 0.57 \nl
 36 & \nodata & \nodata   & 4880 (150) & 0.44 & \nodata & \nodata   & 4930 (330) & 0.24 & 4340 (160) & 0.35 \nl

\enddata

\end{deluxetable}

\clearpage

\begin{deluxetable}{rcccccccccc}
\ssp
\rotate
\ptlandscape
\tablenum{5 -- {\it Continued}}
\tablecaption{}
\tablewidth{00pt}
\tablehead{\colhead{}  &
\multicolumn{2}{c}{Day 49} &
\multicolumn{2}{c}{Day 76} &
\multicolumn{2}{c}{Day 95} &
\multicolumn{2}{c}{Day 96} & 
\multicolumn{2}{c}{Day 124} \\ 
\cline{2-3} \cline{4-5} \cline{6-7} \cline{8-9} \cline{10-11}
\colhead{Line} &
\colhead{Vel.} &
\colhead{Depth} &
\colhead{Vel.} &
\colhead{Depth} &
\colhead{Vel.} &
\colhead{Depth} &
\colhead{Vel.} &
\colhead{Depth} &
\colhead{Vel.} &
\colhead{Depth} }

\startdata 
 1 & \nodata & \nodata   & \nodata & \nodata   & \nodata & \nodata   & \nodata & \nodata   & \nodata & \nodata   \nl
 2 & \nodata & \nodata   & \nodata & \nodata   & \nodata & \nodata   & \nodata & \nodata   & \nodata & \nodata   \nl
 3 & \nodata & \nodata   & 3660 (290) & 0.70 & \nodata & \nodata   & \nodata & \nodata   & 3270 (230) & 0.51 \nl
 4 & \nodata & \nodata   & \nodata & \nodata   & \nodata & \nodata   & \nodata & \nodata   & \nodata & \nodata   \nl
 5 & \nodata & \nodata   & \nodata & \nodata   & \nodata & \nodata   & \nodata & \nodata   & \nodata & \nodata   \nl
 6 & \nodata & \nodata   & \nodata & \nodata   & \nodata & \nodata   & \nodata & \nodata   & \nodata & \nodata   \nl
 7 & 3230 (90) & 0.28 & 2980 (290) & 0.47 & \nodata & \nodata   & 2320 (420) & 0.42 & 2250 (240) & 0.23 \nl
 8 & \nodata & \nodata   & \nodata & \nodata   & \nodata & \nodata   & \nodata & \nodata   & \nodata & \nodata   \nl
 9 & 3610 (80) & 0.26 & 3290 (330) & 0.27 & \nodata & \nodata   & \nodata & \nodata   & 2820 (250) & 0.33 \nl
 10 &  [4407.7 (1.1)]  & 0.12 & \nodata & \nodata   & \nodata & \nodata   & \nodata & \nodata   &  [4422.6 (1.7)]  & 0.22 \nl
 11 & 3410 (70) & 0.62 & 2790 (160) & 0.69 & \nodata & \nodata   & 2450 (500) & 0.66 & 2350 (80) & 0.66 \nl
 12 & 3220 (40) & 0.17 & 2290 (70) & 0.22 & 2010 (90) & 0.34 & 1900 (100) & 0.38 & \nodata & \nodata   \nl
 13 & 3280 (40) & 0.14 & 2310 (60) & 0.21 & 1850 (90) & 0.31 & 2060 (80) & 0.23 & \nodata & \nodata   \nl
 14 & 11370 (40) & 0.13 & 10550 (170) & 0.12 & 11240 (100) & 0.11 & 10110 (440) & 0.10 & 10700 (90) & 0.13 \nl
 15 &  [4753.5 (0.7)]  & 0.07 &  [4765.0 (1.6)]  & 0.25 &  [4755.5 (1.6)]  & 0.05 &  [4775.2 (1.9)]  & 0.19 &  [4783.2 (0.9)]  & 0.12 \nl
 16 & 3520 (40) & 0.32 & 2310 (70) & 0.28 & 2060 (50) & 0.39 & 2120 (160) & 0.34 & 2010 (40) & 0.42 \nl
 17 & 3410 (40) & 0.40 & 2550 (60) & 0.50 & 2240 (70) & 0.61 & 2330 (80) & 0.61 & 1830 (50) & 0.52 \nl
 18 & 3440 (40) & 0.59 & 2790 (50) & 0.70 & 2500 (60) & 0.76 & 2600 (70) & 0.73 & 2030 (40) & 0.58 \nl
 19 & 3660 (40) & 0.59 & 2900 (130) & 0.70 & 2660 (110) & 0.64 & 2660 (190) & 0.64 & 1960 (40) & 0.62 \nl
 20 & 3330 (40) & 0.09 & \nodata & \nodata   & 2100 (110) & 0.16 & 2110 (200) & 0.17 & \nodata & \nodata   \nl
 21 & 3230 (40) & 0.13 & \nodata & \nodata   & 2110 (80) & 0.25 & 2120 (160) & 0.21 & \nodata & \nodata   \nl
 22 & 3240 (40) & 0.19 & 2620 (290) & 0.42 & 1950 (40) & 0.35 & 2030 (200) & 0.32 & 1710 (30) & 0.36 \nl
 23 & 3110 (40) & 0.04 & \nodata & \nodata   & 1570 (50) & 0.14 & \nodata & \nodata   & \nodata & \nodata   \nl
 24 & 3190 (50) & 0.11 & 2240 (110) & 0.20 & 1850 (40) & 0.24 & 1970 (80) & 0.22 & 1340 (40) & 0.42 \nl
 25 & 3320 (30) & 0.32 & 2460 (70) & 0.44 & 2140 (50) & 0.51 & 2180 (60) & 0.47 & 1450 (40) & 0.17 \nl
 26 & 3360 (40) & 0.35 & 2580 (60) & 0.52 & 2360 (30) & 0.61 & 2250 (60) & 0.58 & 1790 (40) & 0.40 \nl
 27 & 11780 (50) & 0.02 & \nodata & \nodata   & \nodata & \nodata   & \nodata & \nodata   & \nodata & \nodata   \nl
 28 & \nodata & \nodata   & \nodata & \nodata   & \nodata & \nodata   & \nodata & \nodata   & \nodata & \nodata   \nl
 29 & 3370 (30) & 0.59 & 3010 (90) & 0.70 & 3240 (30) & 0.76 & 3140 (60) & 0.76 & 5620 (40) & 0.89 \nl
 30 & 3100 (30) & 0.19 & 2610 (70) & 0.41 & 2640 (40) & 0.52 & 2410 (80) & 0.53 & 2720 (30) & 0.60 \nl
 31 & 3470 (30) & 0.25 & 2800 (50) & 0.38 & 2350 (40) & 0.46 & 2190 (50) & 0.45 & 1770 (40) & 0.31 \nl
 32 & \nodata & \nodata   & \nodata & \nodata   & \nodata & \nodata   & \nodata & \nodata   & \nodata & \nodata   \nl
 33 & 11640 (30) & 0.15 & 11300 (90) & 0.17 & \nodata & \nodata   & \nodata & \nodata   & \nodata & \nodata   \nl
 34 & 4830 (80) & 0.73 & 5440 (120) & 0.63 & 5100 (610) & 0.65 & 5600 (150) & 0.64 & 4440 (50) & 0.83 \nl
 35 & \nodata & \nodata   & 4150 (260) & 0.85 & \nodata & \nodata   & \nodata & \nodata   & 2890 (130) & 0.82 \nl
 36 & \nodata & \nodata   & 3660 (160) & 0.54 & \nodata & \nodata   & \nodata & \nodata   & \nodata & \nodata   \nl

\enddata

\end{deluxetable}

\clearpage

\begin{deluxetable}{rcccccccccc}
\ssp
\rotate
\ptlandscape
\tablenum{5 -- {\it Continued}}
\tablecaption{}
\tablewidth{00pt}
\tablehead{\colhead{}  &
\multicolumn{2}{c}{Day 138} &
\multicolumn{2}{c}{Day 159} &
\multicolumn{2}{c}{Day 163} &
\multicolumn{2}{c}{Day 313} & 
\multicolumn{2}{c}{Day 333} \\ 
\cline{2-3} \cline{4-5} \cline{6-7} \cline{8-9} \cline{10-11}
\colhead{Line} &
\colhead{Vel.} &
\colhead{Depth} &
\colhead{Vel.} &
\colhead{Depth} &
\colhead{Vel.} &
\colhead{Depth} &
\colhead{Vel.} &
\colhead{Depth} &
\colhead{Vel.} &
\colhead{Depth} }

\startdata 
 1 & \nodata & \nodata   & \nodata & \nodata   & \nodata & \nodata   & \nodata & \nodata   & \nodata & \nodata   \nl
 2 & \nodata & \nodata   & \nodata & \nodata   & \nodata & \nodata   & \nodata & \nodata   & \nodata & \nodata   \nl
 3 & \nodata & \nodata   & \nodata & \nodata   & \nodata & \nodata   & \nodata & \nodata   & \nodata & \nodata   \nl
 4 & \nodata & \nodata   & \nodata & \nodata   & \nodata & \nodata   & \nodata & \nodata   & \nodata & \nodata   \nl
 5 & \nodata & \nodata   & \nodata & \nodata   & \nodata & \nodata   & \nodata & \nodata   & \nodata & \nodata   \nl
 6 & \nodata & \nodata   & \nodata & \nodata   & \nodata & \nodata   & \nodata & \nodata   & \nodata & \nodata   \nl
 7 & \nodata & \nodata   & \nodata & \nodata   & \nodata & \nodata   & \nodata & \nodata   & \nodata & \nodata   \nl
 8 & \nodata & \nodata   & \nodata & \nodata   & \nodata & \nodata   & \nodata & \nodata   & \nodata & \nodata   \nl
 9 & \nodata & \nodata   & \nodata & \nodata   & \nodata & \nodata   & \nodata & \nodata   & 1800 (190) & 0.39 \nl
 10 & \nodata & \nodata   & \nodata & \nodata   & \nodata & \nodata   & \nodata & \nodata   &  [4436.9 (3.7)]  & 0.28 \nl
 11 & 1800 (140) & 0.62 & 1950 (190) & 0.61 & 1620 (140) & 0.57 & 1290 (390) & 0.67 & 2280 (190) & 0.50 \nl
 12 & \nodata & \nodata   & \nodata & \nodata   & \nodata & \nodata   & \nodata & \nodata   & \nodata & \nodata   \nl
 13 & \nodata & \nodata   & \nodata & \nodata   & \nodata & \nodata   & \nodata & \nodata   & \nodata & \nodata   \nl
 14 & \nodata & \nodata   & 10760 (70) & 0.12 & 10780 (110) & 0.11 & \nodata & \nodata   & \nodata & \nodata   \nl
 15 & \nodata & \nodata   & \nodata & \nodata   & \nodata & \nodata   & \nodata & \nodata   & \nodata & \nodata   \nl
 16 & 2190 (60) & 0.75 & 2360 (70) & 0.85 & 2490 (60) & 0.84 & 2190 (240) & 1.05 & 2370 (80) & 0.86 \nl
 17 & 1790 (100) & 0.53 & 1510 (40) & 0.53 & 1530 (50) & 0.48 & 1500 (100) & 0.58 & 1520 (60) & 0.36 \nl
 18 & 1910 (130) & 0.54 & 1780 (40) & 0.59 & 1780 (50) & 0.58 & 1780 (250) & 0.66 & 1910 (80) & 0.60 \nl
 19 & 1430 (220) & 0.67 & 960 (40) & 0.67 & 1000 (120) & 0.65 & 1910 (180) & 0.30 & 1910 (90) & 0.24 \nl
 20 & \nodata & \nodata   & \nodata & \nodata   & \nodata & \nodata   & \nodata & \nodata   & \nodata & \nodata   \nl
 21 & \nodata & \nodata   & \nodata & \nodata   & \nodata & \nodata   & \nodata & \nodata   & \nodata & \nodata   \nl
 22 & 1500 (130) & 0.41 & \nodata & \nodata   & 1430 (80) & 0.38 & \nodata & \nodata   & \nodata & \nodata   \nl
 23 & \nodata & \nodata   & \nodata & \nodata   & \nodata & \nodata   & \nodata & \nodata   & \nodata & \nodata   \nl
 24 & 1510 (300) & 0.40 & \nodata & \nodata   & 1150 (70) & 0.41 & \nodata & \nodata   & \nodata & \nodata   \nl
 25 & 1260 (80) & 0.18 & \nodata & \nodata   & 1290 (60) & 0.17 & \nodata & \nodata   & \nodata & \nodata   \nl
 26 & 1630 (70) & 0.44 & \nodata & \nodata   & 1500 (90) & 0.39 & 2450 (170) & 0.42 & 2260 (90) & 0.48 \nl
 27 & \nodata & \nodata   & 10500 (100) & 0.06 & \nodata & \nodata   & \nodata & \nodata   & \nodata & \nodata   \nl
 28 & \nodata & \nodata   & \nodata & \nodata   & \nodata & \nodata   & \nodata & \nodata   & \nodata & \nodata   \nl
 29 & 5330 (50) & 0.91 & 5650 (160) & 0.92 & 5480 (130) & 0.91 & 4840 (430) & 0.95 & 4060 (460) & 0.91 \nl
 30 & 2660 (60) & 0.58 & 2730 (60) & 0.49 & 2770 (80) & 0.46 & 4540 (270) & 0.54 & 4930 (150) & 0.53 \nl
 31 & 1660 (170) & 0.30 & 1610 (30) & 0.39 & 1630 (50) & 0.38 & 1770 (290) & 0.61 & 1700 (80) & 0.60 \nl
 32 & \nodata & \nodata   & \nodata & \nodata   & \nodata & \nodata   & \nodata & \nodata   & \nodata & \nodata   \nl
 33 & \nodata & \nodata   & \nodata & \nodata   & \nodata & \nodata   & \nodata & \nodata   & \nodata & \nodata   \nl
 34 & 3740 (110) & 0.88 & 3580 (30) & 0.92 & 3630 (90) & 0.92 & 3460 (70) & 0.98 & 3520 (30) & 0.96 \nl

\enddata
\end{deluxetable}

\clearpage

\begin{deluxetable}{rcccccccccc}
\ssp
\rotate
\ptlandscape
\tablenum{5 -- {\it Continued}}
\tablecaption{}
\tablewidth{00pt}
\tablehead{\colhead{}  &
\multicolumn{2}{c}{Day 138} &
\multicolumn{2}{c}{Day 159} &
\multicolumn{2}{c}{Day 163} &
\multicolumn{2}{c}{Day 313} & 
\multicolumn{2}{c}{Day 333} \\ 
\cline{2-3} \cline{4-5} \cline{6-7} \cline{8-9} \cline{10-11}
\colhead{Line} &
\colhead{Vel.} &
\colhead{Depth} &
\colhead{Vel.} &
\colhead{Depth} &
\colhead{Vel.} &
\colhead{Depth} &
\colhead{Vel.} &
\colhead{Depth} &
\colhead{Vel.} &
\colhead{Depth} }

\startdata 

 35 & 2980 (40) & 0.93 & \nodata & \nodata   & 2910 (180) & 0.88 & 2440 (190) & 0.91 & \nodata & \nodata   \nl
 36 & 2970 (40) & 0.63 & \nodata & \nodata   & 2660 (50) & 0.65 & 2250 (110) & 0.84 & \nodata & \nodata   \nl

\enddata

\tablecomments{Line velocities for absorption features measured directly from
spectra of SN 1999em with estimated $1\sigma$ statistical uncertainty shown in
parentheses, shown for four example days only.  Uncertainty includes
measurement error and uncertainty in the wavelength scale, but does not include
uncertainty in the rest wavelength estimation, which may be significant for
blended features.  All spectra were deredshifted by 800 km s$^{-1}$ prior to
the velocity determination.  Note that the emission-dominated nebular spectrum
from day 517 is not included in the tabulation.}

\tablenotetext{a}{Day since discovery, 1999 October 29 (HJD 2,451,480.94), rounded
to the nearest day.}

\tablenotetext{b}{Line number corresponding to feature indicated in Figures
9 and 10, and identified in Table 4.}  

\tablenotetext{c}{Main ion responsible for feature.  Colon indicates
significant blending of multiple ions; see Table 4 for components of blend.
Velocities derived using blended lines may not yield accurate photospheric
velocities, since the relative strengths of the components may change with
time.  }

\tablenotetext{d}{Rest wavelength (air) adopted for line feature.  In the case
of blended lines, a weighted average of the blended components is used; see
text for details.}

\tablenotetext{e}{Velocity measured in units of km s$^{-1}$, with uncertainty
listed in parentheses.  For line features lacking identification, the measured
rest wavelength and uncertainty are given in brackets.}

\tablenotetext{f}{Depth is defined as: $d = 1 - f_{\rm min}$, where $f_{\rm
min}$ is the normalized flux level at the location of the line's minimum with
the continuum set to 1.  It should be noted that the measured depth in some
cases may depend on the strength of adjacent features since this affects the
location of the points used to fit the normalizing continuum.}

\tablenotetext{g}{To maintain consistency with the method traditionally used to
determine the photospheric velocity by averaging the velocities derived from
\ion{Fe}{2} $\lambda\lambda 4924, 5018, 5169$ line measurements, we adopt the
rest wavelength of \ion{Fe}{2} $\lambda 4924$ to determine this feature's
velocity.  The consistently lower velocity derived compared with the other
\ion{Fe}{2} lines of the multiplet, however, suggests significant blending with
\ion{Ba}{2} $\lambda 4934$.}

\tablenotetext{h}{Becomes weak and probably blended with \ion{Na}{1} D by day 10.}

\end{deluxetable}

\clearpage

\subsection{The Reddening of SN~1999em}
\label{sec:reddening}

SN~1999em suffers from minimal Galactic reddening, $\Ebv_{\rm MW} = 0.04$ mag
(Schlegel, Finkbeiner, \& Davis 1998).  From theoretical fits to the early-time
continuum shape, Baron et al. (2000) estimate the total (Galactic + host
galaxy) reddening of SN~1999em to be $0.05 \lesssim \Ebv_{\rm tot} \lesssim
0.10$ mag, with an upper limit of $\Ebv_{\rm tot} < 0.15$ mag; the most
successful fits seem to favor $\Ebv_{\rm tot} = 0.10$ mag.  We checked the
consistency of this result using two techniques.

First, we compared the $BVI$ color of SN~1999em with that of SN~1987A at the
moment the \ion{He}{1} $\lambda 5876$ line disappears from both objects'
spectra.  Since exciting a significant amount of \ion{He}{1} to the high energy
level ($E = 20.97$ eV; see Table 4) needed to produce the $\lambda 5876$ line
is very temperature sensitive (generally greater than about $T_{\rm eff} =
10,000$~K; Hatano et al. 1999; Branch 1987; see also Figure~\ref{fig:5.17}), we
expect this line to disappear when the photosphere is at a well-defined
temperature.  Therefore, the intrinsic color of the two objects should be
similar at this early phase, before significant line blanketing from metal
lines begins.  We assume the reddening of SN~1987A to be $\Ebv = 0.15 \pm 0.05$
mag (see discussion in Arnett et al. 1989) and compare its color at this phase
with that of SN~1999em.  For SN~1999em, the \ion{He}{1} $\lambda 5876$ line
becomes too weak to measure $10 \pm 1$ days after discovery, roughly
corresponding with the epoch when the \ion{Fe}{2} $\lambda 5169$ absorption
first appears (strengthening the argument that at this phase the optical
spectrum is nearly a blackbody, with little line blanketing).  At that epoch,
its \bvi\ magnitudes are $14.01$, $13.85$, and $13.49$, respectively,
interpolated from the values given in Table 2.  Phillips et al. (1988) estimate
that the \ion{He}{1} $\lambda 5876$ line disappeared in the spectrum of
SN~1987A on JD~2,446,854, just four days after discovery.  From the photometry
of Hamuy et al. (1988), the $BVI$ magnitudes for SN~1987A on that date are
$4.83, 4.45, {\rm and\ } 4.01$, respectively.  Using the reddening law of
Savage \& Mathis (1979) and minimizing the color difference between the two SNe
at this epoch, we derive a reddening of $\Ebv = 0.03 \pm 0.05$ mag for
SN~1999em, in good agreement with the estimate of Baron et al. (2000).


\begin{figure}
\ssp
\vskip -0.7in
\hskip 0.1in
\rotatebox{0}{
\scalebox{0.9}{
\plotone{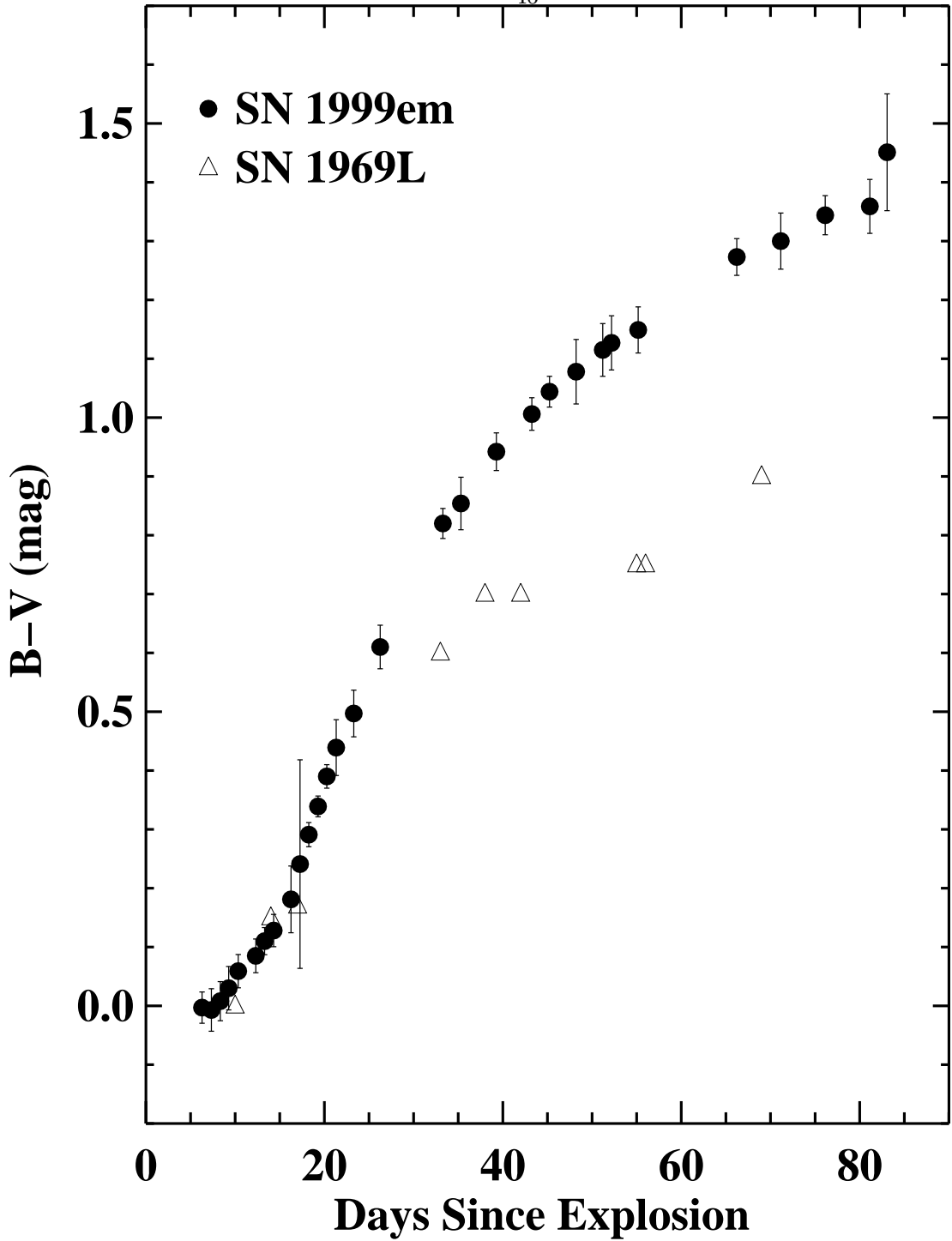}
}
}
\vskip 0.2in
\caption{ The temporal evolution of the $B - V$ color of SN~1999em compared
with that of the prototypical Type II-P event, SN~1969L, which is believed to
be nearly unreddened.  The time of explosion for both events is derived through
the EPM analysis (see \S~\ref{sec:sn1999emepmdistance} for SN~1999em and
Schmidt et al. [1992] for SN~1969L).  The good agreement between the early-time
color evolution of the two events suggests that SN~1999em might also suffer
little reddening.  }
\label{fig:5.16b}
\end{figure}


\begin{figure}
\ssp
\vskip -0.7in
\hskip -0.2in
\rotatebox{90}{
\scalebox{0.8}{
\plotone{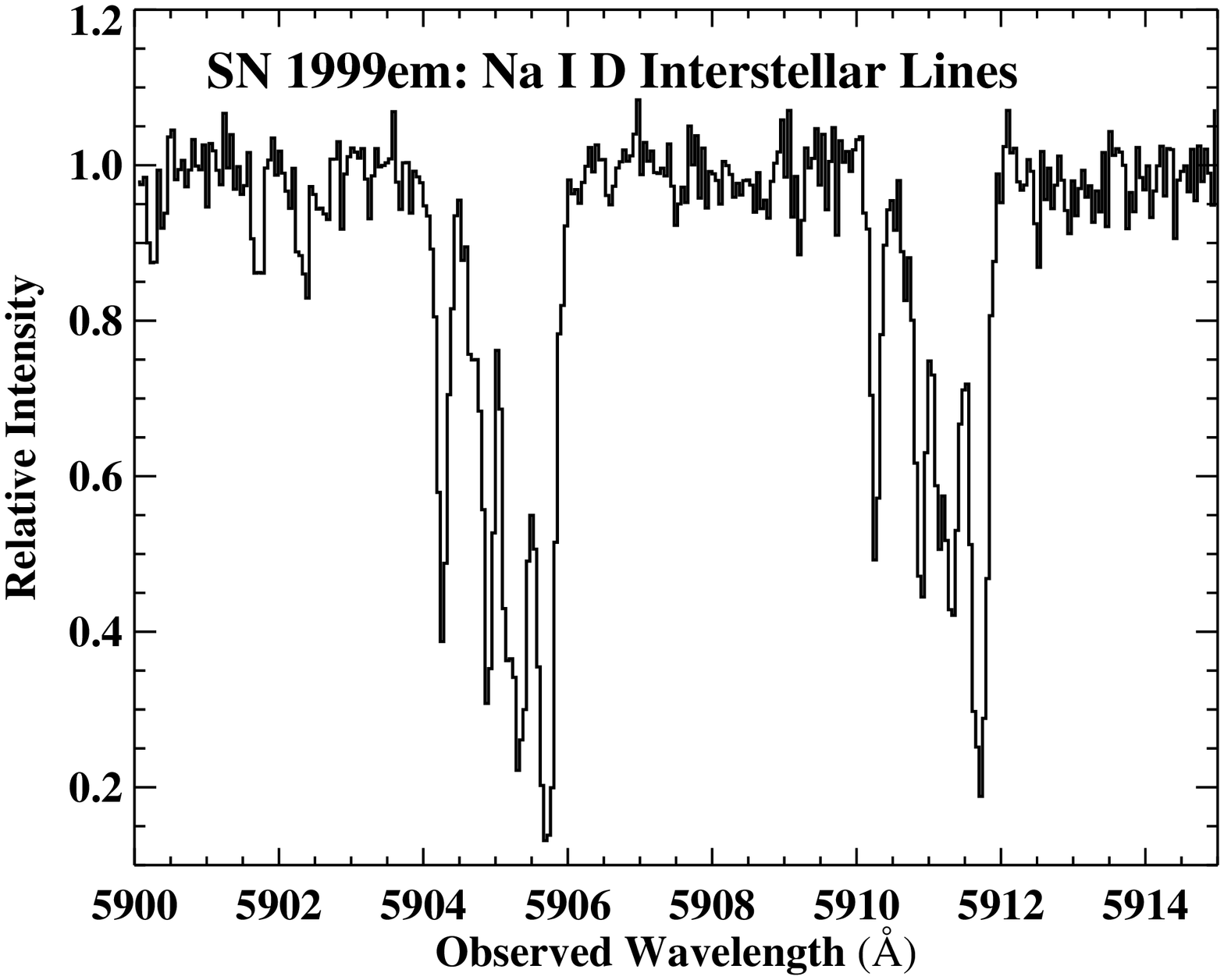}
}
}
\caption{Interstellar \ion{Na}{1} D ($\lambda\lambda 5890, 5896$) absorption
by gas in NGC~1637 along the line of sight to SN~1999em.}
\label{fig5:10b}
\end{figure}

A second method sometimes used to provide a reddening estimate for SNe~II-P
(e.g., Schmidt et al. 1993) is to assume that the early-time color evolution of
all SNe~II-P is similar, and compare the early color evolution of a SN~II-P
with that observed in SN~1969L, a Type~II-P believed to suffer nearly zero
host-galaxy reddening (Kirshner \& Kwan 1974; Schmidt et al. 1992), and only
$\Ebv_{\rm MW} = 0.06$ mag (Schlegel et al. 1998).  Using the time of explosion
for SN~1999em derived in \S~\ref{sec:sn1999emepmdistance} ($5.3\pm{1.4}$ days
before discovery), Figure~\ref{fig:5.16b} shows that at early times, SN~1999em
and SN 1969L have a very similar color.  This suggests similarly low reddenings
for the two events.  We do note, however, that the \bv\ color evolution of
SN~1999em at later times deviates substantially from that of SN~1969L, and may
indicate that SN~1999em's subsequent color evolution differed somewhat from the
norm.  Although limited by the small number of observations and rather poor
temporal coverage, Schmidt et al. (1992) demonstrate that the \bv\ color
evolution of the 4 SNe~II-P in their sample is quite similar to that of
SN~1969L during the recombination phase (none has photometric data taken prior
to 14 days after the explosion, however).  It is possible that the large
deviation between the later color evolution of SN~1999em and SN~1969L indicates
a real difference between SN~1999em and this prototypical SN~II-P.  We shall
further investigate the possibility of unusual photometric behavior for SN
1999em in \S~\ref{sec:typical}.

A final indication that there may be little reddening for SN~1999em comes
from the low value of the interstellar polarization implied by the
spectropolarimetry (L01).  One argument for higher extinction, however, comes
from the strength of the \ion{Na}{1} D ($\lambda\lambda 5890, 5896$)
interstellar absorption lines.  From the echelle spectrum taken 4 days after
discovery (Figure~\ref{fig5:10b}; a detailed analysis of the \ion{Na}{1} D
interstellar lines will be the subject of a future paper), we measure
$W_\lambda$ (D2; $\lambda 5890) = 0.86 $ \AA\ and $W_\lambda$ (D1; $\lambda
5896) = 0.65 $ \AA, for a total equivalent width of $W_\lambda$ (\ion{Na}{1} D)
= 1.51 \AA\ (cf. Jha et al. 1999).  If we adopt the rough correlation derived
by Barbon et al. (1990) between $W_\lambda$ (\ion{Na}{1} D) and reddening,
\begin{equation}
\Ebv = 0.25 W_\lambda {\rm (Na\ I\ D)},
\label{eqn4:barbon}
\end{equation}
we would predict $\Ebv \approx 0.38$ mag for SN~1999em, significantly greater
than the extinction implied by the other methods.  The relation given by
equation~(\ref{eqn4:barbon}) is known to have very large scatter (likely due to
variations in the dust-to-gas ratios among galaxies as well as the marginal
ability of \ion{Na}{1} D gas to serve as a tracer of the hydrogen gas column;
saturation effects can also play a large roll; see, e.g., Munari \& Zwitter
1997; Issa, MacLaren, \& Wolfendale 1990), however, and we are inclined to
favor the evidence supporting a lower total reddening; in \S~\ref{sec:sysunc}
we shall also see that the EPM distances derived for $\Ebv = 0.38$ mag using
the different filter combinations are very inconsistent with each other,
further arguing against such high extinction.  We therefore adopt the reddening
preferred by Baron et al. (2000) of $\Ebv = 0.10 $ mag and assign it an
uncertainty of $\pm\ 0.05$ mag.  The strength of the \ion{Na}{1} D lines,
however, does make us treat this value with some caution.

\subsection{The Photospheric Velocity and Color of SN~1999em During the Plateau}
\label{sec:photovel}


\begin{figure}
\ssp
\begin{center}
\rotatebox{90}{
\scalebox{0.7}{
\plotone{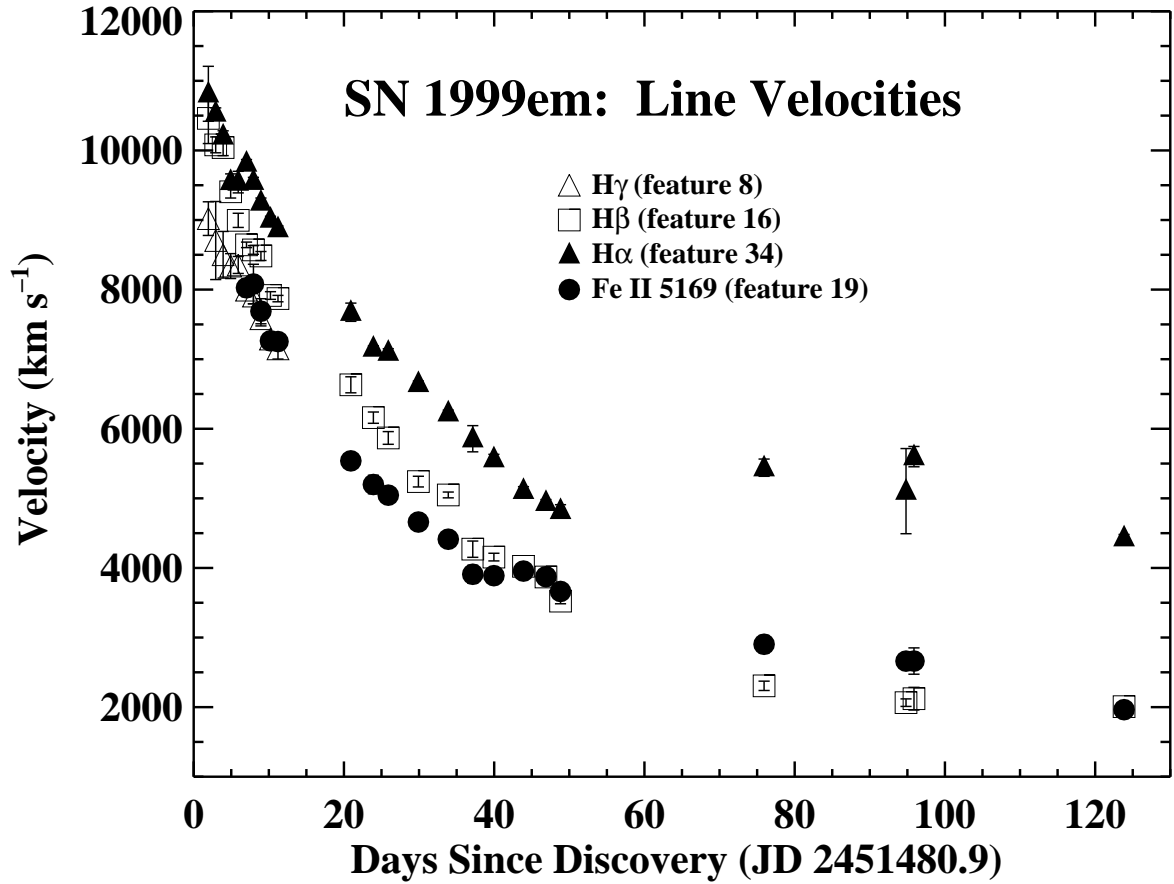}
}
}
\end{center}
\caption{
Line velocity determined from the P-Cygni absorption minima of 4 different
features during the first 130 days after discovery of SN~1999em.  }
\label{fig:5.13d}
\end{figure}

From Figure~\ref{fig:5.15} we see that the plateau phase of SN~1999em lasted
until about 95 days after discovery, and our application of EPM will therefore
be restricted to times somewhat earlier than the drop off the plateau.
Figure~\ref{fig:5.13d} shows the velocity inferred from the flux minima of the
first 3 hydrogen Balmer lines and \ion{Fe}{2} $\lambda 5169$ during the first
130 days after discovery.  As expected (\S~\ref{sec:introduction}), the
\halpha\ and \hbeta\ lines generally yield velocities significantly higher than
the weaker H$\gamma$ and \ion{Fe}{2} $\lambda 5169$ lines.  Curiously, after
about day 40, the \hbeta\ line yields velocities similar to, or even less than,
those derived from \ion{Fe}{2} $\lambda 5169$.  In the day 76, 95, and 96
spectra the \halpha\ absorption profile becomes corrupted by the excess
emission described in \S~\ref{sec:chugai} (Figure~\ref{fig:5.18a}), and the
inferred velocity increases dramatically.

Traditionally, the \ion{Fe}{2} $\lambda\lambda 4924, 5018, 5169$ line features
have been used to estimate photospheric velocity during the plateau phase of
SNe~II-P, with the occassional use of \ion{Sc}{2} $\lambda 5527$ and
\ion{Sc}{2} $\lambda 5658$ as well.  At very early times, before the iron lines
become available, it has been customary to either use the higher-order hydrogen
Balmer lines (i.e., H$\gamma$ and H$\delta$) or else to scale the \halpha\ and
\hbeta\ velocities using the ratio of their velocities to the iron features
when the iron features first appear.  Here, we shall follow two approaches.
First, we derive velocities following the usual methods for determining
photospheric velocity ($v_{\rm strong})$, relying mainly on the \ion{Fe}{2}
$\lambda\lambda 4924, 5018, 5169$ lines when they (or, at least, one of them)
become available (day 7 onwards), and the higher-order hydrogen Balmer lines at
earlier times.  Second, we estimate photospheric velocity by using the weakest
unblended features identified during each epoch ($v_{\rm weak}$).  We shall
then compare the EPM distance derived using each set of velocities.


\begin{figure}
\ssp
\begin{center}
\rotatebox{90}{
\scalebox{0.7}{
\plotone{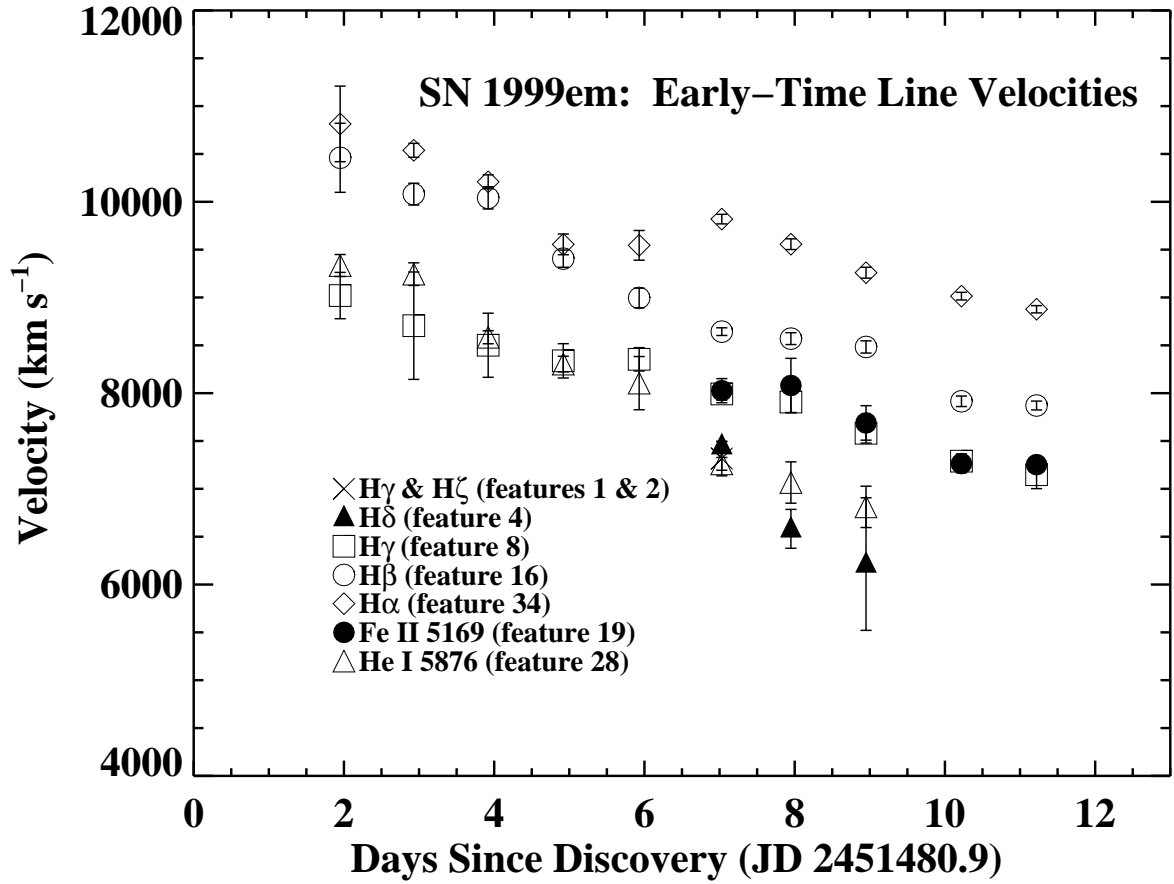}
}
}
\end{center}
\caption{The
velocity derived from absorption lines in the early-time spectrum of
SN~1999em.  Note that from day 7 onwards the velocity found for \protect\ion{Fe}{2}
and H$\gamma$ is somewhat higher than the weaker H$\delta$ and \protect\ion{He}{1}
$\lambda 5876$ lines suggest. }
\label{fig:5.13a}
\end{figure}

Figure~\ref{fig:5.13a} shows the velocities measured for features observed
during the first 11 days after discovery of SN~1999em.  Through day 5, \halpha\
and \hbeta\ yield similar velocities.  \ion{He}{1} $\lambda 5876$ and H$\gamma$
yield similar velocities through day 6, but significantly lower than those
measured by \halpha\ and \hbeta.  From day 7 onwards, the velocities inferred
from the weak \ion{He}{1} $\lambda 5876$ and H$\delta$ lines are somewhat lower
than those measured using \ion{Fe}{2} $\lambda 5169$ and H$\gamma$.  Our
estimates of $v_{\rm strong}$ and $v_{\rm weak}$, then, will be the same for
days 2 through 6, but will differ from day 7 onwards.  For days 2 through 6, we
take the weighted average of the \ion{He}{1} $\lambda 5876$ and H$\gamma$
velocities.  From day 7 onwards, $v_{\rm strong}$ is always determined by the
available \ion{Fe}{2} $\lambda\lambda 4924, 5018, 5169$ lines, and $v_{\rm
weak}$ is derived from the 4 weak, unblended features described in
\S~\ref{sec:inferringvel} (features 12, 13, 21, and 22 in
Figure~\ref{fig:5.11}).  

\begin{figure}
\ssp
\begin{center}
\rotatebox{90}{
\scalebox{0.7}{
\plotone{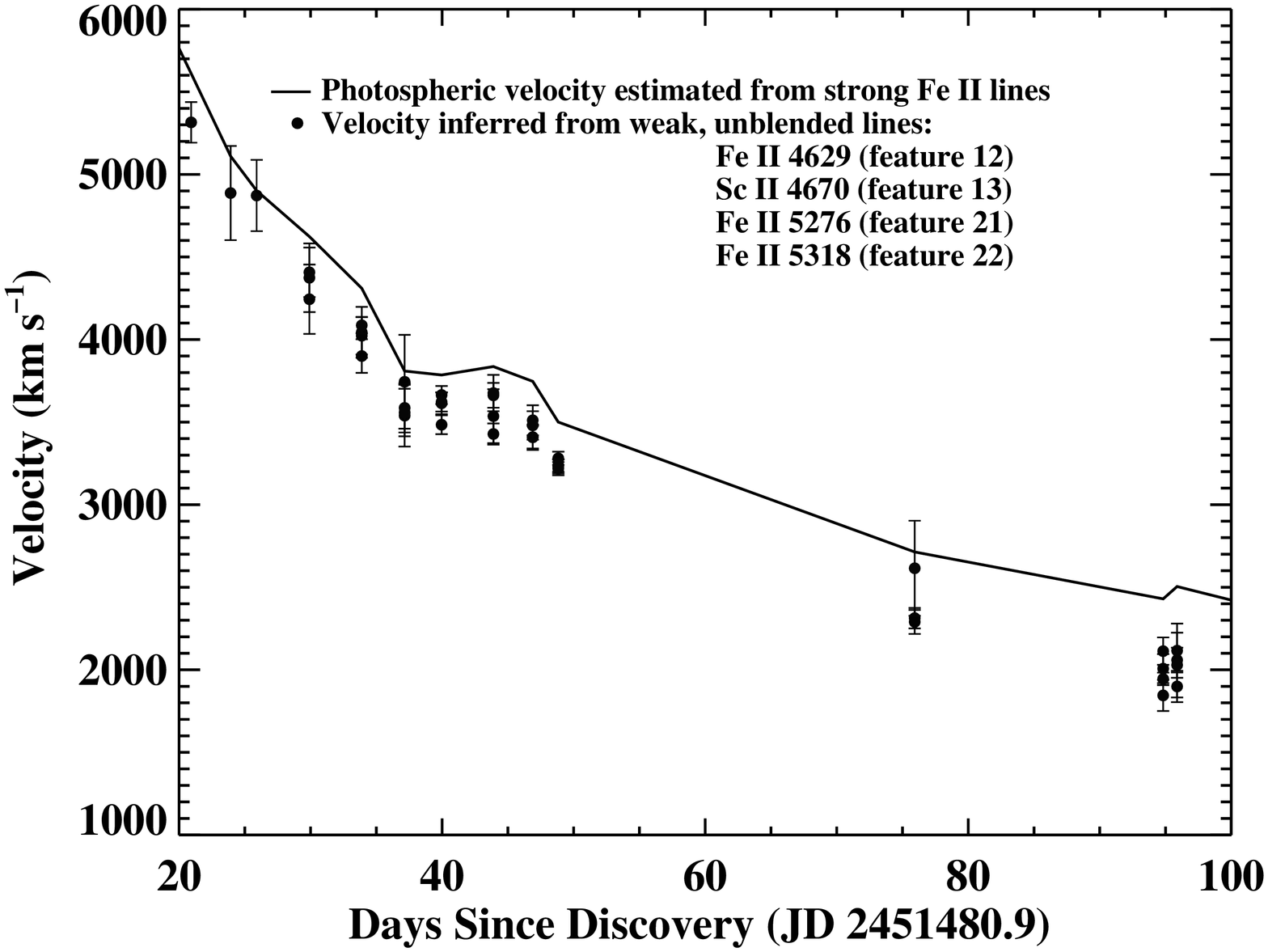}
}
}
\end{center}
\caption{Temporal evolution of the velocity inferred from weak, unblended
features compared with that derived from the stronger \protect\fetwo\
absorptions.  The weaker features give consistently lower velocities.}
\label{fig:5.12a}
\end{figure}

Figure~\ref{fig:5.12a} shows the velocities derived from these features
compared with those derived from the weighted average of the \fetwo\ lines from
day 20 to day 100.  Clearly, these weaker features yield systematically lower
velocities than those derived from the stronger \ion{Fe}{2} lines, and the
difference appears to grow with time.  To determine $v_{\rm weak}$, then, we
took the weighted mean of the available weak lines at each spectral epoch.
Table 6 lists the $v_{\rm strong}$ and $v_{\rm weak}$ values adopted for each
epoch; the reported uncertainty is the uncertainty of the weighted mean of the
lines added in quadrature to the 1$\sigma$ spread in the velocities of the
individual lines.


\begin{deluxetable}{lcccccc}
\ssp
\renewcommand{\arraystretch}{1.1}
\tablenum{6}
\tablewidth{00pt}
\tablecaption{Quantities Used in the EPM Analysis of SN 1999em }
\tablehead{\colhead{}  &
\colhead{$v_{\rm strong}$\tablenotemark{b}}  &
\colhead{$v_{\rm weak}$\tablenotemark{c}} &
\colhead{$B$($\sigma_B$)} &
\colhead{$V$($\sigma_V$)} &
\colhead{$I$($\sigma_I$)} \\
\colhead{Day\tablenotemark{a}}  &
\colhead{km s$^{-1}$}  &
\colhead{km s$^{-1}$} &
\colhead{mag} &
\colhead{mag} &
\colhead{mag} }

\startdata
 1.9 & 9276(201) & 9276(201) & 13.8018(0.0271) & 13.8085(0.0241) & 13.5845(0.0287)\\ 
 2.9 & 9221(192) & 9221(192) & 13.7951(0.0232) & 13.7887(0.0243) & 13.5620(0.0283)\\ 
 3.9 & 8581(70)  & 8581(70)  & 13.8214(0.0311) & 13.7927(0.0201) & 13.5572(0.0261)\\ 
 4.9 & 8310(77)  & 8310(77)  & 13.8434(0.0244) & 13.7877(0.0153) & 13.5411(0.0243)\\ 
 5.9 & 8314(171) & 8314(171) & 13.8803(0.0286) & 13.8095(0.0176) & 13.5349(0.0292)\\ 
 7.0 & 8026(178) & 7372(110) & 13.9225(0.0250) & 13.8371(0.0140) & 13.5294(0.0290)\\ 
 7.9 & 8080(402) & 6809(372) & 13.9486(0.0191) & 13.8398(0.0130) & 13.4977(0.0251)\\ 
 8.9 & 7642(271) & 6759(317) & 13.9837(0.0151) & 13.8568(0.0231) & 13.5072(0.0211)\\ 
10.2 & 7221(155) & \nodata   & 14.0089(0.0463) & 13.8481(0.0404) & 13.4876(0.0435)\\ 
11.2 & 7215(94)  & \nodata   & 14.0354(0.0559) & 13.8389(0.0428) & 13.4702(0.0404)\\ 
20.9 & 5611(115) & 5315(174) & 14.4667(0.0230) & 13.8579(0.0290) & 13.4054(0.0270)\\ 
23.9 & 5110(122) & 4887(403) & 14.5791(0.0262) & 13.8804(0.0323) & 13.3818(0.0309)\\ 
25.9 & 4902(169) & 4872(305) & 14.6532(0.0209) & 13.8953(0.0197) & 13.3663(0.0253)\\ 
29.9 & 4623(52)  & 4358(134) & 14.7882(0.0371) & 13.9358(0.0250) & 13.3490(0.0350)\\ 
33.9 & 4310(79)  & 4047(86)  & 14.8673(0.0250) & 13.9272(0.0200) & 13.3118(0.0210)\\ 
37.1 & 3809(89)  & 3581(107) & 14.9257(0.0220) & 13.9209(0.0222) & 13.2919(0.0275)\\ 
39.9 & 3785(86)  & 3596(86)  & 14.9741(0.0180) & 13.9298(0.0190) & 13.2845(0.0320)\\ 
43.9 & 3836(69)  & 3550(140) & 15.0025(0.0467) & 13.9123(0.0369) & 13.2226(0.0500)\\ 
46.9 & 3747(103) & 3443(64)  & 15.0541(0.0300) & 13.9270(0.0350) & 13.2200(0.0390)\\ 
48.8 & 3500(141) & 3242(34)  & 15.0800(0.0344) & 13.9387(0.0281) & 13.2148(0.0362)\\ 
75.9 & 2714(164) & 2312(78)  & 15.4031(0.0373) & 14.0396(0.0271) & 13.2176(0.0333)\\ 
94.8 & 2430(190) & 1965(87)  & 15.7184(0.0227) & 14.2587(0.0224) & 13.3670(0.0328)\\ 
95.8 & 2504(170) & 2012(106) & 15.7386(0.0210) & 14.2770(0.0220) & 13.3795(0.0320)\\

\enddata

\tablecomments{Although only data up through day 76 were included when deriving
the EPM distance, we list the values of the measured
parameters through the early part of the transition to the nebular phase for completeness. }

\tablenotetext{a}{Days since discovery, 1999-10-29 UT (HJD 2,451,480.94).}
\tablenotetext{b}{Velocity derived largely from \ion{Fe}{2} $\lambda\lambda
4924, 5018, 5169$ lines.}

\tablenotetext{c}{Velocity derived from weak, unblended lines; see text for
details.}
\end{deluxetable}


\begin{figure}
\ssp
\begin{center}
\rotatebox{90}{
\scalebox{0.7}{
\plotone{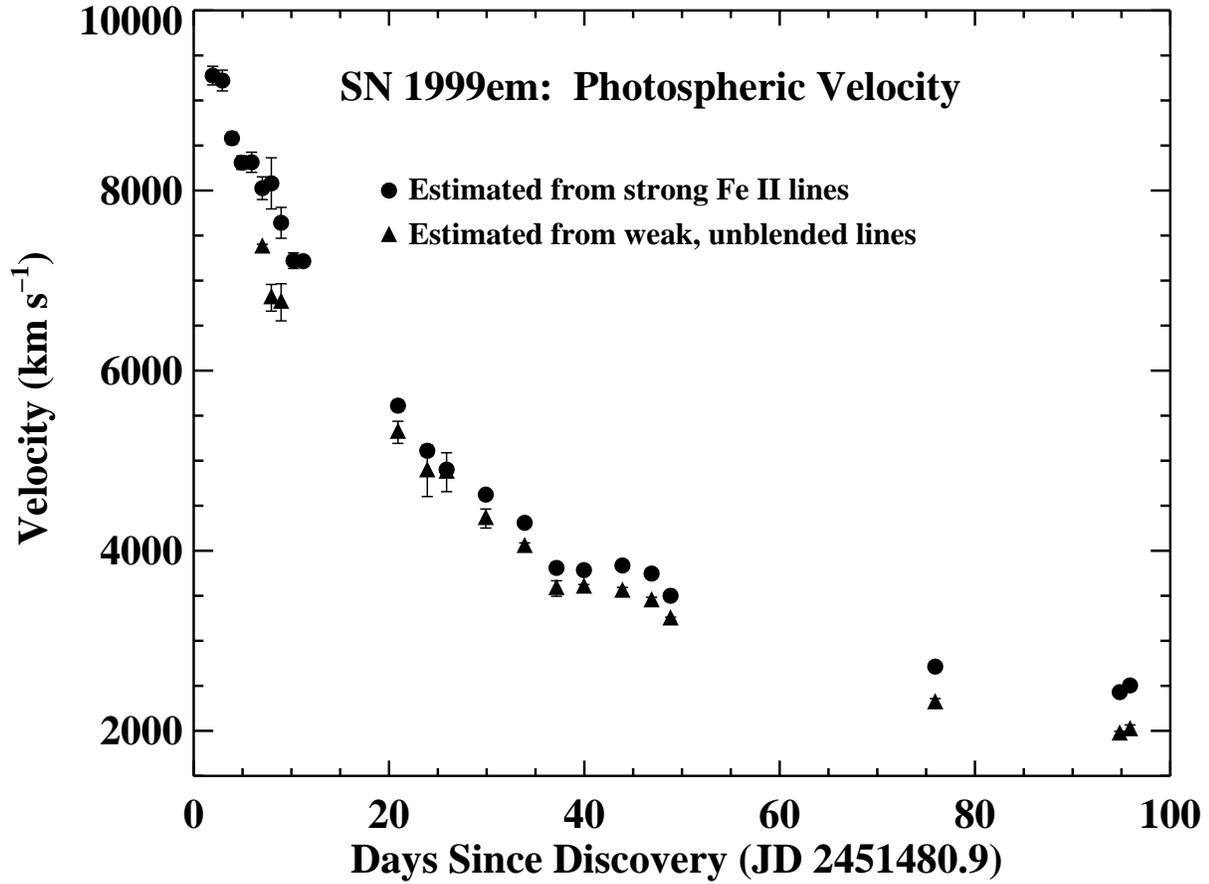}
}
}
\end{center}
\caption{Comparison of photospheric velocity derived for SN~1999em by using the
average of the strong \protect\fetwo\ lines ($v_{\rm strong}$) or the weaker
unblended features ($v_{\rm weak}$) shown in Figure~\ref{fig:5.12a}.}
\label{fig:5.21}
\end{figure}


\begin{figure}
\ssp
\begin{center}
\rotatebox{90}{
\scalebox{0.7}{
\plotone{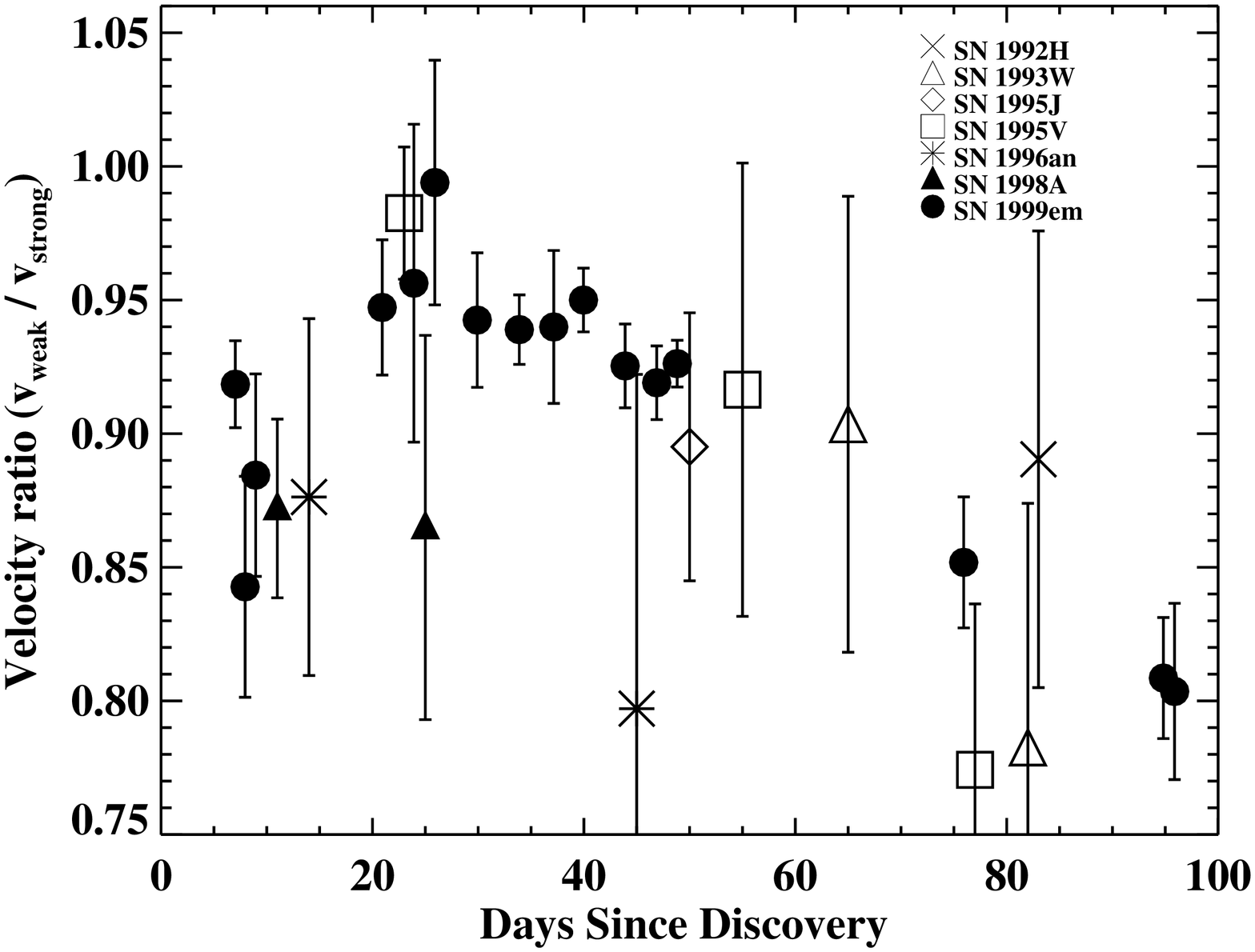}
}
}
\end{center}
\caption{The temporal evolution of the ratio of the velocity derived from weak
unblended lines to that derived from the strong \protect\ion{Fe}{2}
$\lambda\lambda 4924, 5018, 5169$ features for 7 SNe~II.  Note that the
relative phases of the SNe are somewhat uncertain since some may have been
discovered significantly after explosion.  }
\label{fig:5.21f}
\end{figure}

Figure~\ref{fig:5.21} compares $v_{\rm strong}$ and $v_{\rm weak}$ during the
plateau phase of SN~1999em.  Since lower photospheric velocity estimates will
result in a smaller derived distance (equation~\ref{eqn:epmdistance}), it is
useful to see whether such a velocity offset is ubiquitous among SNe~II.
Figure~\ref{fig:5.21f} shows the ratio of $v_{\rm weak}$ to $v_{\rm strong}$ as
a function of time in the spectra of 6 other SNe~II that had the high S/N
needed to make the measurements of these weak features.  The good agreement of
the data from other SNe~II with SN~1999em suggests that EPM distances derived
using the strong iron lines as photospheric velocity indicators will generally
yield distances $5 - 10\%$ larger than those using the weaker lines.

In order to derive an EPM distance to SN~1999em, we need its \bvi\ magnitude at
each of the spectral epochs.  The spectral observations of SN~1999em listed in
Table~3 are not, in general, coincident with the \bvi\ photometric observations
listed in Table~2.  We therefore used a linear interpolation from the nearest
photometric observations before and after the spectral observations.  The
reported photometric uncertainty represents the quadrature sum of the
uncertainty in the interpolation (i.e., uncertainty in the location of the
interpolated point due to the uncertainty in the slope of the line connecting
the two photometric data points) with the photometric uncertainty of the
nearest photometric point.  The $\!BVI$ magnitude and uncertainty adopted for
SN~1999em at the times of spectral observations are given in Table~6.

\section{The EPM Applied to SN~1999em}
\label{sec:sn1999emepmdistance}

Following the procedure outlined in \S~\ref{sec:introduction}, we derive the
theoretical angular size ($\theta$), flux dilution factor ($\zeta$), and color
temperature of SN~1999em for the three filter combinations $\!BV,\ \!BVI, {\rm\
and\ } VI$, and list the results in Table 7.  The uncertainty given for
$\theta, \zeta,$ and $T$ at each epoch is the 1$\sigma$ spread found for each
parameter derived from 1000 simulated sets of photometry characterized by the
photometric uncertainty listed in Table 6.

\begin{deluxetable}{lcccccccrr}
\renewcommand{\arraystretch}{1.0}
\ssp
\tablenum{7}
\ptlandscape
\rotate
\tablewidth{0pt}
\tablecaption{Quantities Derived from the EPM Analysis of SN 1999em}
\tablehead{\colhead{}  &
\multicolumn{3}{c}{Angular Size ($10^8$ km Mpc$^{-1}$)\tablenotemark{a} } &
\multicolumn{3}{c}{Dilution Factor}  &
\multicolumn{3}{c}{Temperature (K)} \\
\cline{2-4} \cline{5-7} \cline{8-10}
\colhead{Day\tablenotemark{b} }  &
\colhead{$\theta_{BV}$} &
\colhead{$\theta_{BVI}$} &
\colhead{$\theta_{VI}$} &
\colhead{$\zeta_{BV}$} &
\colhead{$\zeta_{BVI}$} &
\colhead{$\zeta_{VI}$} &
\colhead{$T_{BV}$} &
\colhead{$T_{BVI}$} &
\colhead{$T_{VI}$} }

\startdata

  1.9 &  6.13(1.01) &  7.57(0.50) &  8.46(0.70) & 0.411(0.020) & 0.429(0.005) & 0.444(0.005) & 17488(2121)  & 14050(646)  & 12150(798)  \\
  2.9 &  6.54(0.94) &  7.79(0.50) &  8.59(0.69) & 0.404(0.017) & 0.427(0.005) & 0.444(0.005) & 16809(1726)  & 13834(625)  & 12094(773)  \\
  3.9 &  7.14(1.09) &  8.12(0.51) &  8.76(0.66) & 0.394(0.016) & 0.424(0.005) & 0.443(0.004) & 15780(1739)  & 13384(637)  & 11915(701)  \\
  4.9 &  7.91(0.81) &  8.56(0.47) &  9.01(0.54) & 0.384(0.009) & 0.420(0.004) & 0.441(0.003) & 14709(1063)  & 12882(524)  & 11697(531)  \\
  5.9 &  8.25(0.97) &  8.99(0.50) &  9.50(0.68) & 0.380(0.010) & 0.417(0.003) & 0.439(0.003) & 14177(1192)  & 12361(480)  & 11185(614)  \\
  7.0 &  8.55(0.83) &  9.45(0.48) & 10.07(0.67) & 0.376(0.007) & 0.414(0.002) & 0.437(0.002) & 13703(945)  & 11829(416)  & 10641(527)  \\
  7.9 &  9.18(0.67) & 10.12(0.37) & 10.78(0.56) & 0.371(0.004) & 0.412(0.001) & 0.436(0.001) & 13012(667)  & 11250(282)  & 10135(394)  \\
  8.9 &  9.59(0.82) & 10.31(0.31) & 10.85(0.58) & 0.369(0.003) & 0.412(0.000) & 0.436(0.001) & 12529(716)  & 11019(229)  & 10032(431)  \\
 10.2 & 10.52(1.70) & 10.81(0.73) & 11.12(1.13) & 0.367(0.007) & 0.412(0.001) & 0.436(0.002) & 11723(1481)  & 10640(530)  &  9886(823)  \\
 11.2 & 11.45(1.79) & 11.28(0.79) & 11.34(1.09) & 0.367(0.010) & 0.412(0.002) & 0.436(0.002) & 10990(1465)  & 10294(605)  &  9779(781)  \\
 20.9 & 16.46(0.38) & 15.00(0.31) & 13.00(0.76) & 0.598(0.040) & 0.467(0.008) & 0.440(0.004) &  6523(249)  &  7612(151)  &  8820(439)  \\
 23.9 & 16.64(0.36) & 15.69(0.34) & 13.82(0.76) & 0.699(0.050) & 0.497(0.011) & 0.445(0.006) &  6005(215)  &  7110(162)  &  8375(403)  \\
 25.9 & 16.73(0.23) & 16.06(0.31) & 14.34(0.59) & 0.776(0.040) & 0.520(0.011) & 0.450(0.005) &  5706(141)  &  6822(127)  &  8108(278)  \\
 29.9 & 16.75(0.32) & 16.52(0.29) & 15.22(0.77) & 0.914(0.071) & 0.565(0.015) & 0.460(0.009) &  5286(184)  &  6378(119)  &  7649(327)  \\
 33.9 & 17.16(0.27) & 17.00(0.20) & 15.84(0.47) & 1.061(0.056) & 0.603(0.011) & 0.466(0.007) &  4948(114)  &  6085(77)  &  7442(201)  \\
 37.1 & 17.51(0.30) & 17.29(0.24) & 16.15(0.58) & 1.180(0.058) & 0.632(0.014) & 0.470(0.008) &  4725(100)  &  5899(84)  &  7348(235)  \\
 39.9 & 17.65(0.28) & 17.40(0.24) & 16.40(0.65) & 1.258(0.052) & 0.652(0.014) & 0.474(0.009) &  4598(81)  &  5778(79)  &  7238(248)  \\
 43.9 & 18.06(0.61) & 17.94(0.36) & 17.39(1.03) & 1.352(0.123) & 0.690(0.025) & 0.486(0.018) &  4460(172)  &  5582(120)  &  6957(378)  \\
 46.8 & 18.18(0.59) & 18.00(0.32) & 17.60(0.85) & 1.431(0.104) & 0.714(0.023) & 0.491(0.016) &  4354(132)  &  5472(100)  &  6854(319)  \\
 48.8 & 18.18(0.52) & 18.04(0.29) & 17.82(0.76) & 1.462(0.101) & 0.728(0.020) & 0.496(0.015) &  4315(125)  &  5410(86)  &  6756(271)  \\
 75.9 & 19.47(0.79) & 18.14(0.23) & 18.72(0.62) & 2.002(0.120) & 0.890(0.030) & 0.531(0.017) &  3789(93)  &  4857(82)  &  6242(207)  \\
 94.8 & 18.94(0.67) & 16.94(0.22) & 17.99(0.51) & 2.259(0.086) & 0.988(0.025) & 0.560(0.018) &  3605(56)  &  4609(57)  &  5925(169)  \\
 95.8 & 18.82(0.65) & 16.84(0.22) & 17.93(0.51) & 2.264(0.083) & 0.994(0.022) & 0.562(0.018) &  3602(54)  &  4596(49)  &  5900(167)  \\

\enddata
\tablecomments{Although only data up through day 76 were included when deriving
the EPM distance, we list the values of the measured
parameters through the early part of the transition to the nebular phase for completeness. }

\tablenotetext{a}{Angular size of the optical photosphere (i.e., the surface of
last scattering).}
\tablenotetext{b}{Days since discovery, 1999 October 29 UT (HJD 2,451,480.94).}

\end{deluxetable}

\begin{deluxetable}{lcc}
\renewcommand{\arraystretch}{1.5}
\ssp
\tablenum{8}
\tablewidth{00pt}
\tablecaption{EPM Distances to SN 1999em }
\tablehead{\colhead{Method\tablenotemark{a}} &
\colhead{Derived Distance}  &
\colhead{Time of Explosion\tablenotemark{b}} \\
\colhead{} &
\colhead{(Mpc)} &
\colhead{(d)} } 

\startdata

$BV_{\rm weak}$ & $7.68 \pm 0.24$ & $-3.70 \pm 0.72$  \\
$BVI_{\rm weak}$ & $8.27 \pm 0.18$ & $-5.69 \pm 0.50$  \\
$VI_{\rm weak}$ & $8.79 \pm 0.28$ & $-6.51 \pm 0.77$  \\
$BV_{\rm strong}$  & $8.18 \pm 0.41$ & $-3.83 \pm 1.15$  \\
$BVI_{\rm strong}$ & $9.20 \pm 0.27$ & $-6.08 \pm 0.74$  \\
$VI_{\rm strong}$ & $9.72 \pm 0.42$  & $-7.16 \pm 1.01$  \\

\enddata

\tablecomments{The EPM distance derived to SN 1999em using the flux dilution
factors of Eastman, Schmidt, \& Kirshner (1996) as updated by Hamuy et
al. (2001).  Uncertainties are the $1\sigma$ uncertainties resulting from 1000
simulated sets of data characterized by the uncertainties in $\theta$ and $v$
found in Tables 6 and 7.}

\tablenotetext{a}{The broadband filter combination ($BV$, $BVI$,  or $VI$)
used to derive the photospheric temperature, and the absorption lines used to
derive photospheric velocity, where ``strong'' indicates \ion{Fe}{2}
$\lambda\lambda 4924, 5018, 5169$ and ``weak'' indicates the weaker, unblended
features discussed in the text.}

\tablenotetext{b}{Days before discovery, 1999-10-29 UT (HJD 2,451,480.94).}

\end{deluxetable}

\begin{figure}
\ssp
\begin{center}
\rotatebox{90}{
\scalebox{0.7}{
\plotone{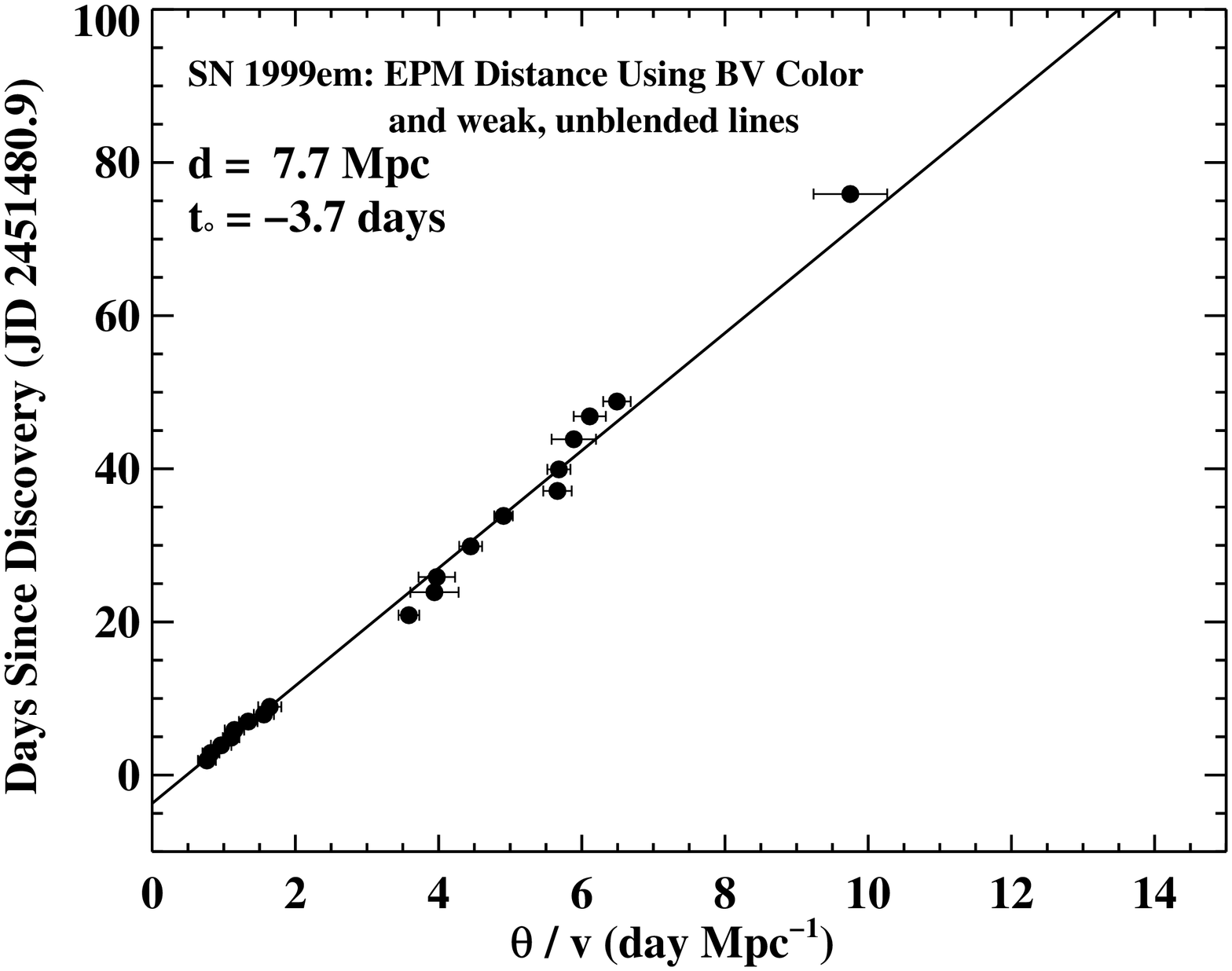}
} }
\end{center}
\caption{The parameters used to determine the distance to SN~1999em
(equation~[\ref{eqn:epmdistance}]), with the line of best fit determined by
using the criterion of least absolute deviations.  The slope of this line
yields the distance and the $y$-intercept the time of explosion of SN~1999em.
Here, photospheric temperature is estimated using $BV$ photometry, and
photospheric velocity is determined using $v_{\rm weak}$, the velocity derived
from weak, unblended features in the optical spectrum.}
\label{fig:5.22a}
\end{figure}

\begin{figure}
\ssp
\begin{center}
\rotatebox{90}{
\scalebox{0.7}{
\plotone{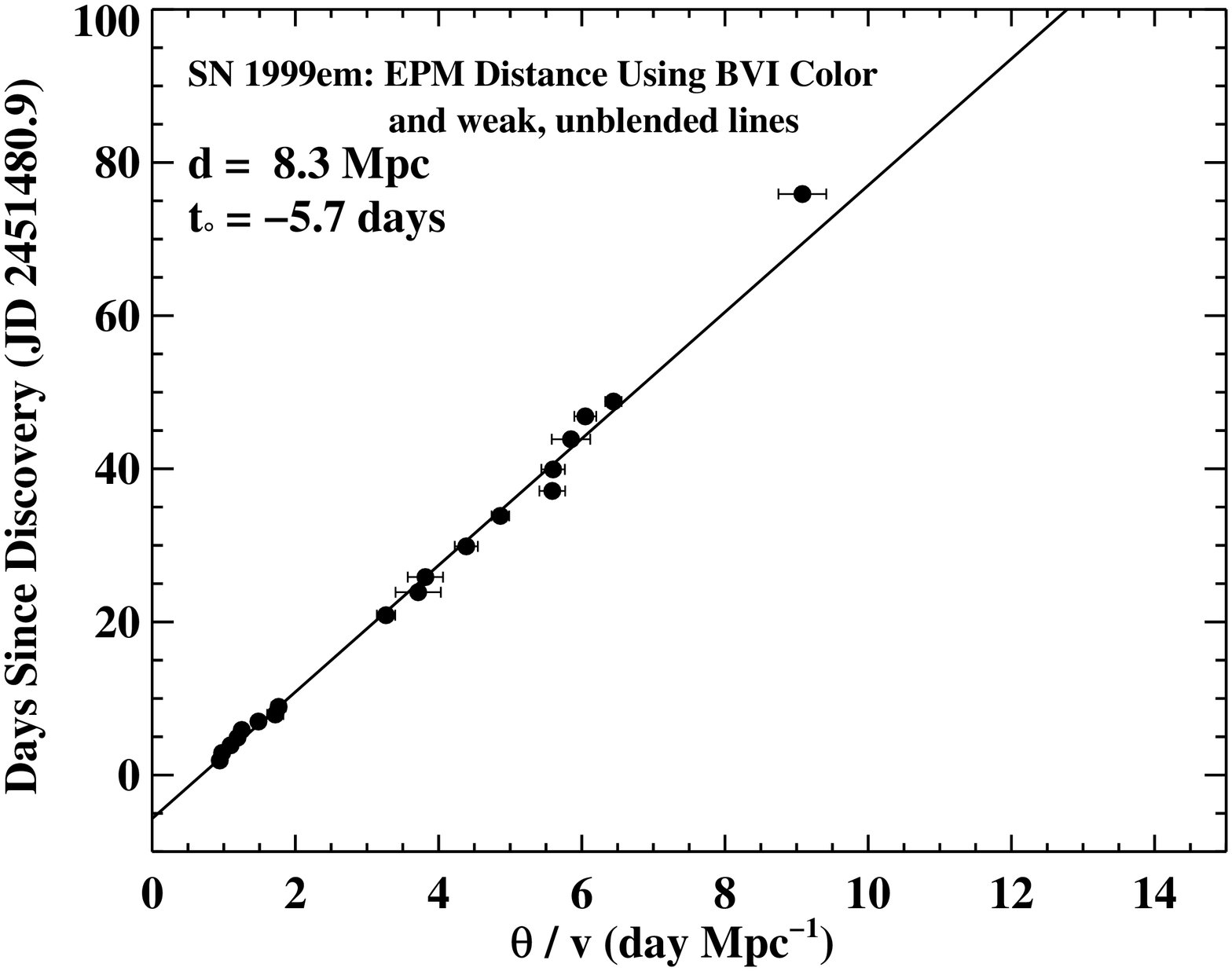}
}
}
\end{center}
\caption{As in
Figure~\ref{fig:5.22a}, except using $BVI$ photometry to determine photospheric
color temperature.}
\label{fig:5.22b}
\end{figure}

\begin{figure}
\ssp
\begin{center}
\rotatebox{90}{
\scalebox{0.7}{
\plotone{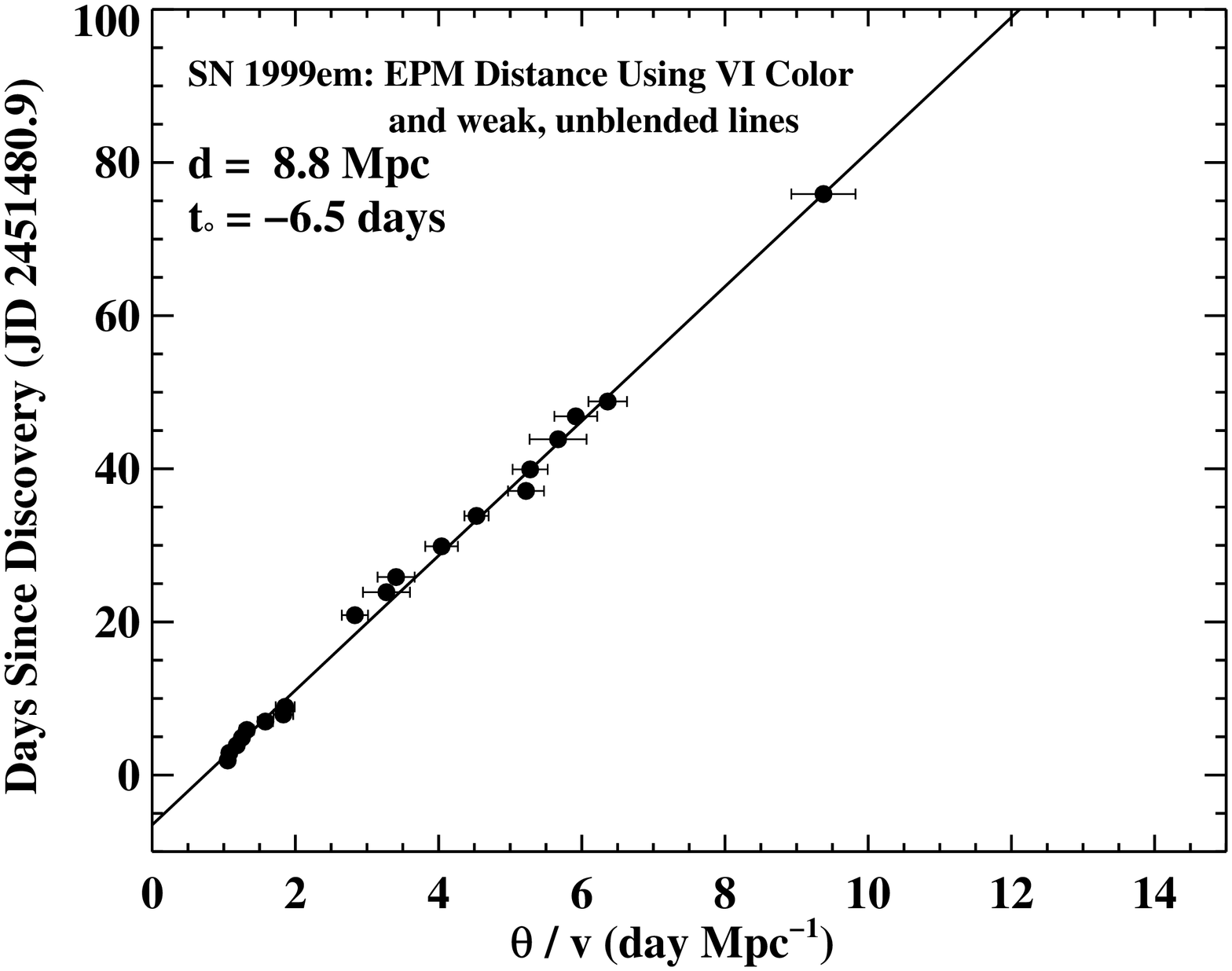}
}
}
\end{center}
\caption{As in
Figure~\ref{fig:5.22a}, except using $VI$ photometry to determine photospheric
color temperature.}
\label{fig:5.22c}
\end{figure}

\begin{figure}
\ssp
\begin{center}
\rotatebox{90}{
\scalebox{0.7}{
\plotone{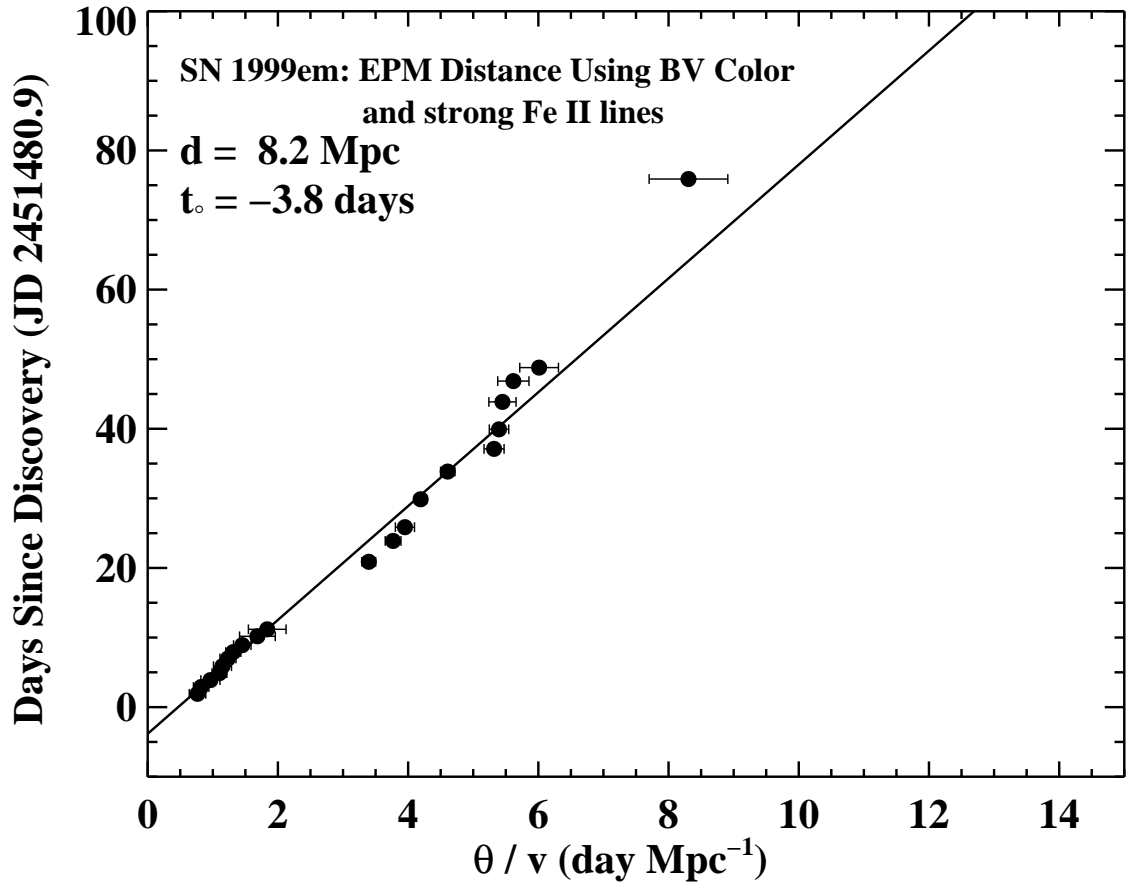}
}
}
\end{center}
\caption{As in
Figure~\ref{fig:5.22a}, except using $v_{\rm strong}$, the velocity determined by
using the \protect\fetwo\ line features, to estimate photospheric velocity.}
\label{fig:5.22d}
\end{figure}

\begin{figure}
\ssp
\begin{center}
\rotatebox{90}{
\scalebox{0.7}{
\plotone{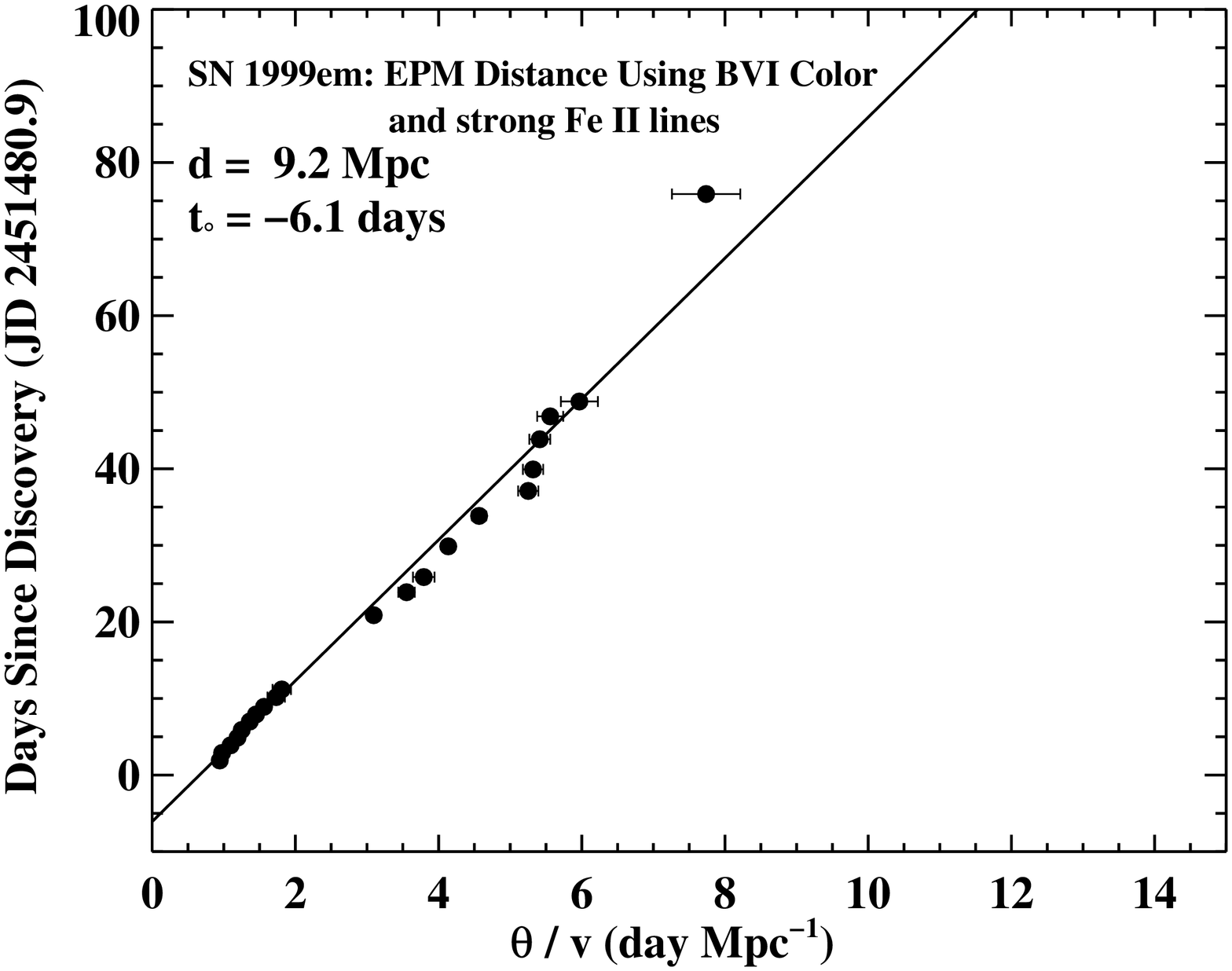}
}
}
\end{center}
\caption{As in
Figure~\ref{fig:5.22d}, except using $VI$ photometry to determine photospheric
color temperature.}
\label{fig:5.22e}
\end{figure}

\begin{figure}
\ssp
\begin{center}
\rotatebox{90}{
\scalebox{0.7}{
\plotone{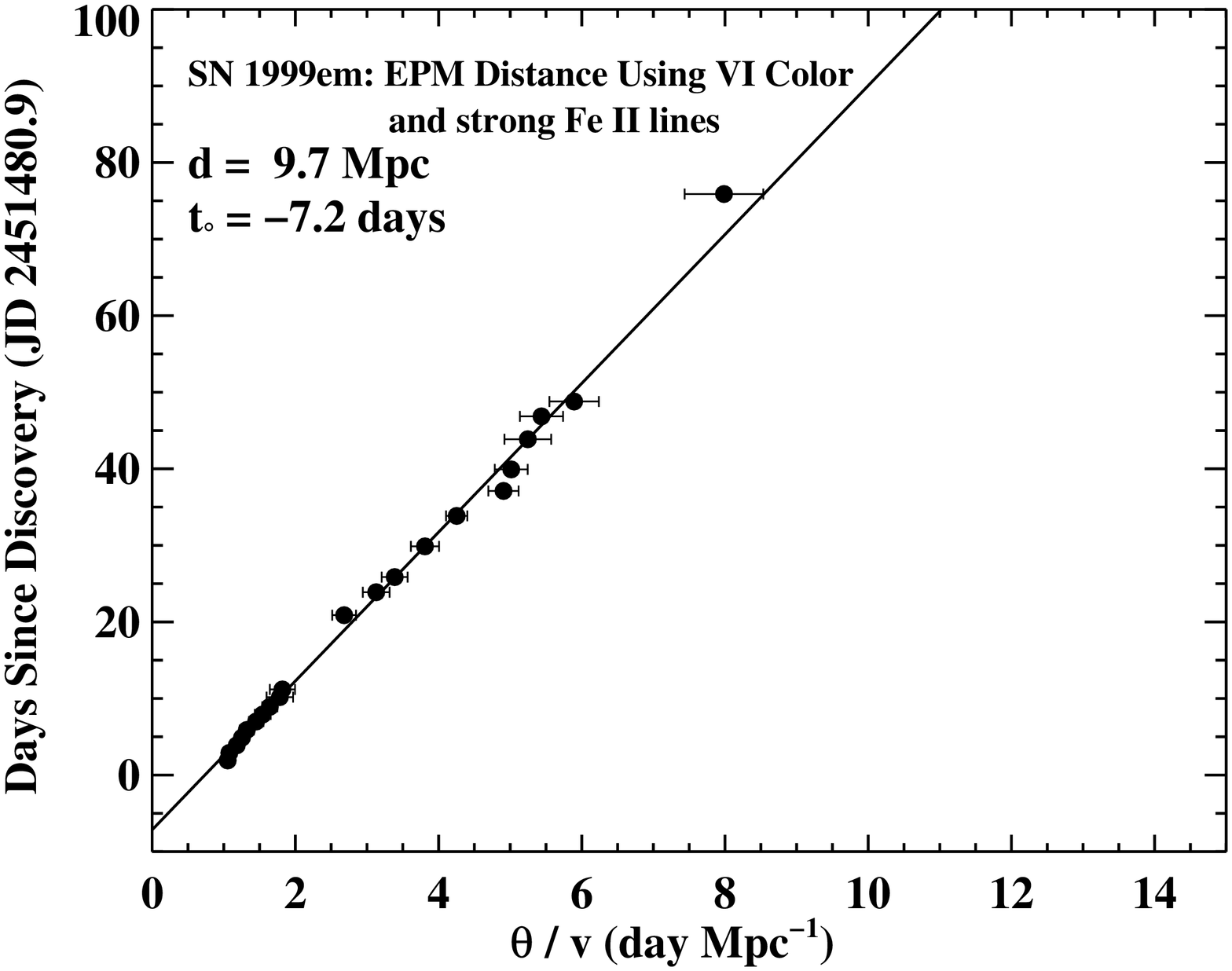}
}
}
\end{center}
\caption{As in
Figure~\ref{fig:5.22d}, except using $VI$ photometry to determine photospheric
color temperature.}
\label{fig:5.22f}
\end{figure}

To determine the distance to SN~1999em for each filter combination we first
need to decide how to weight the data, and then find the best-fitting line in
the plot of $t$ vs. $\theta/v$ (equation~[\ref{eqn:epmdistance}]).  Since the
uncertainty of $\theta/v$ increases in direct proportion to the value of
$\theta/v$, a weighting scheme based on the calculated uncertainty in each
point is necessarily biased towards the earlier data.  On the other hand, using
the method of least squares with equal weighting is susceptible to undue
influence by deviant points.  We therefore opt for the more robust fitting
method of using the criterion of the least absolute deviation for equally
weighted data points (i.e., 'medfit' from Numerical Recipes; Press et
al. 1992).  Since SN~1999em begins to drop off the plateau near day 95 after
discovery, we do not consider the data from day 95 onwards in the EPM analysis.
The distances and times of explosion resulting from the 3 filter combinations
and two photospheric velocity estimates are given in Table 8, along with
uncertainties derived from 1000 simulated sets of data.
Figures~\ref{fig:5.22a} through \ref{fig:5.22f} show the relevant plots for
each of the six distance estimates.


While the formal uncertainty of each derived distance listed in Table 8 is only
a few percent, discrepancies among the distances derived using the dilution
factors given for the various filter combinations run as high as 19\%.  It is
therefore clear that systematic, not statistical, error dominates the
uncertainty in the EPM distance to SN~1999em.  In general, $D_{VI} > D_{BVI} >
D_{BV}$.  We note that the 50 \kms\ uncertainty in the redshift of SN~1999em
(\S~\ref{sec:inferringvel}) contributes an uncertainty of only $\sim 2\%$ to
the derived distance.  As expected, distances derived using the stronger iron
features ($v_{\rm strong}$) are systematically about $10\%$ larger than those
derived using the weaker lines ($v_{\rm weak}$).  For the weak lines, the
(unweighted) average of the distances derived from the 3 filter combinations is
$D_{\rm weak} = 8.24 \pm 0.56$ Mpc, with $t_o = 5.30 \pm 1.45$ days before
discovery; the reported uncertainties are the standard deviations of the three
individual data points from the mean value.\footnote{We note that using the
method of least squares to determine the best-fitting slope and intercept of
the line described by equation~\ref{eqn:epmdistance} yields $D_{\rm weak} =
8.54 \pm 0.39$ Mpc, with $t_o = 6.20 \pm 0.92$ days before discovery.}
Similarly, for the strong lines, $D_{\rm strong} = 9.03 \pm 0.78$ Mpc with $t_o
= 5.69 \pm 1.70$ days before discovery.

Since we believe that photospheric velocity is slightly overestimated by using
the traditional \fetwo\ lines, we conclude that the most accurate distance to
SN~1999em is that found from the weak lines.  Rounding to the nearest tenths
place (in Mpc and days), then, yields $D = 8.2 \pm 0.6$ Mpc with an explosion
date of $5.3 \pm 1.4$ days prior to discovery as our best distance estimate.

\section{Discussion}
\label{sec:discussion}

\subsection{Sources of Systematic Uncertainty}
\label{sec:sysunc}

Our preferred EPM distance to SN~1999em is in good agreement with the recent
distance estimate to NGC~1637 of $D = 7.8^{+1.0}_{-0.9}$ Mpc found by Sohn \&
Davidge (1998) using the galaxy's brightest supergiant stars as well as the EPM
distance of $7.5 \pm 0.5$ Mpc recently reported by Hamuy et al. (2001) to SN
1999em.  The major difference in the application of the EPM between our study
and that of Hamuy et al. (2001) lies in the technique used to estimate
photospheric velocity: while we rely on the velocity indicated by the flux
minima of weak line features, Hamuy et al. cross-correlate the observed spectra
with the model spectra of E96.  A detailed discussion of the differences
between these two methods is beyond the scope of the present study; we do note
that the good agreement between the two resulting distances suggests that
systematic differences between the two techniques may be small.  It is also
encouraging that two completely independent sets of data yield such consistent
results. 

From the scatter among the EPM distances derived using the 3 different filter
combinations, it is clear that systematic effects dominate the uncertainty in 
the distance to this well-observed SN, and are at least of order $10\% - 20\%$.
One obvious potential source of systematic error is the uncertainty in the flux
dilution factor, $\zeta$.  The $5\% - 10\%$ uncertainty in $\zeta$ (see
\S~\ref{sec:introduction}) reported for the range of supernova atmospheres
modeled by E96 translates directly to distance uncertainties of $5\% - 10\%$.
The maximum deviation of any atmosphere studied by E96 from the mean $\zeta$ is
$24\%$.  If SN~1999em is such an extreme event and the degree of deviation
between the true dilution factor and the average value varies among the different
filter combinations studied, then the scatter among the derived distances may be
explained.

A second source of systematic error is the uncertainty in the extinction
assumed for SN~1999em, which affects each bandpass differently.  In general,
EPM-derived distances are fairly robust to uncertainty in extinction (E96;
Schmidt et al. 1992), due to the approximate cancellation of terms in equation
(\ref{eqn:theta2}): since dust both reddens and diminishes the light received,
both the measured flux, $f_\nu$, and the inferred blackbody flux, $B_\nu(T_c)$,
decrease with increasing extinction.  These two effects tend to cancel,
producing a derived angular size, $\theta$, that does not vary much with
changes in the reddening.  An additional point to consider, however, is that
$\zeta$ is also a strong function of $T_c$ for most filter combinations
(generally decreasing with increasing $T_c$).  The competition among these
terms makes it difficult to generalize about how changes in reddening will
alter the derived distance.  For SN~1999em, lowering \ebv\ to 0.05 mag produces
larger distances for all filter combinations ($D_{BV} = 8.5$ Mpc, $D_{BVI} =
8.9$ Mpc, and $D_{VI} = 9.0$ Mpc; all distances are derived using $v_{\rm
weak}$ for the photospheric velocity [\S~\ref{sec:photovel}]), while raising
the reddening to $\Ebv = 0.15$ mag lowers the derived distance for the $BV$ and
\bvi\ filter combinations ($D_{BV} = 7.0$ Mpc and $D_{BVI} = 7.8$ Mpc) and
keeps $D_{VI}$ at about the same distance ($D_{VI} = 8.8$ Mpc).  For $\Ebv =
0.05$ mag, taking the unweighted averages of the 3 filter combinations results
in a distance and time of explosion of $D = 8.8 \pm 0.2$ Mpc and $t_o = 7.0 \pm
1.3$ days before discovery, while taking $\Ebv = 0.15$ mag yields $D = 7.9 \pm
0.9$ Mpc and $t_o = 3.7 \pm 1.6$ days before discovery.  For the large
extinction ($\EBV = 0.38$) suggested by the strength of the interstellar
\ion{Na}{1} D lines (\S~\ref{sec:reddening}), we note that $D_{BV} = 4.7$ Mpc,
$D_{BVI} = 6.2$ Mpc, and $D_{VI} = 10.3$ Mpc.  The large discrepancies among
the three distances in the high-extinction case, along with very poor fits
obtained in the $t$ vs. $\theta/v$ graphs (i.e., the points show large,
systematic trends, and predict an explosion date somewhat {\it after}
discovery) and unrealistically high temperatures derived at early times
($T_{\rm BV} > 10^6$K), all argue strongly against such a high reddening.
While the lower extinction distances are much more self-consistent, we do note
that {\it no} value of the extinction is found to make $D_{BV} = D_{BVI} =
D_{VI}$.

\begin{figure}
\ssp
\begin{center}
\rotatebox{0}{
\scalebox{0.7}{
\plotone{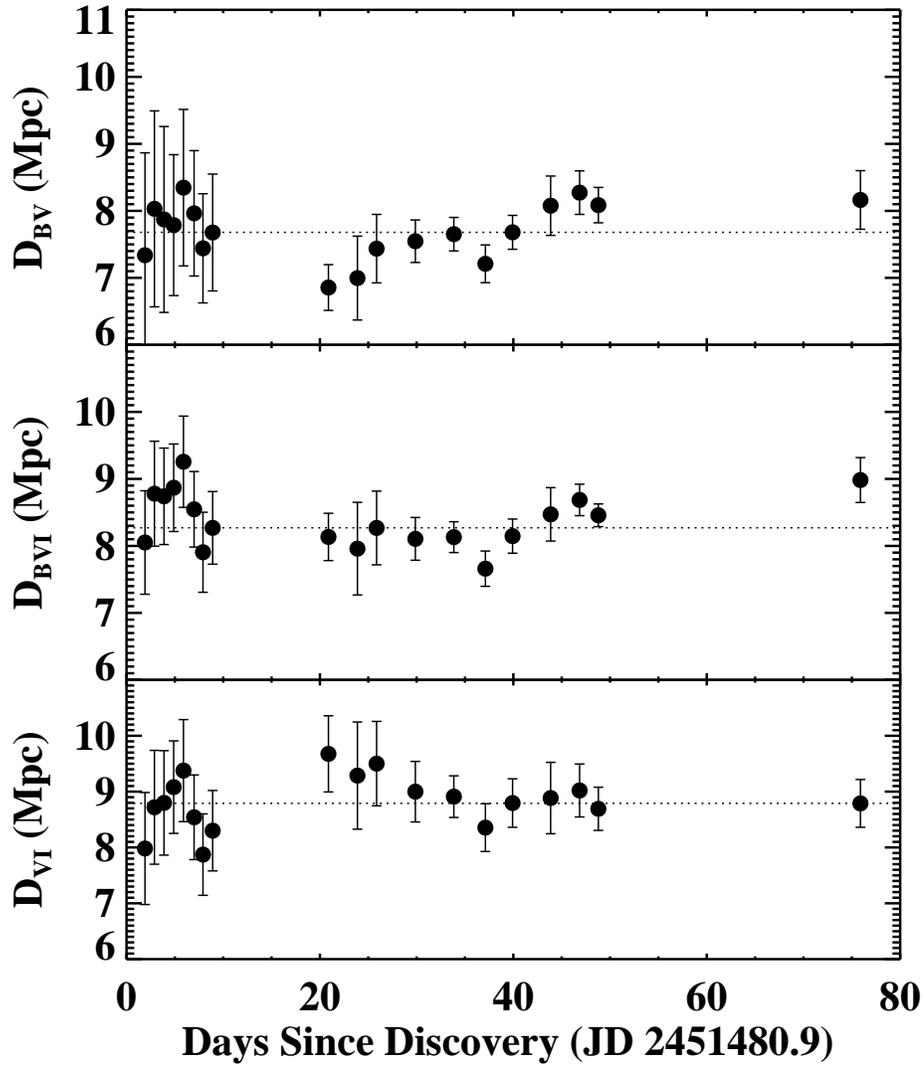}
}
}
\end{center}
\caption{The derived EPM distance to SN~1999em as a function of time for the
$BV$, \bvi, and $VI$ filter sets, with weak, unblended features used to estimate
photospheric velocity.  }
\label{fig:5.27}
\end{figure}

In addition to the uncertainty due to the dilution factor and extinction,
asphericity could certainly contribute as well.  However, it is not easy to see
how asphericity could affect the individual bandpasses differently, to produce
the filter-dependent discrepancies that are seen.  L01 found that the
polarization of SN~1999em increased with time, perhaps suggesting an increase
in asphericity deeper into the ejecta.  Barring a special viewing angle,
increasing asphericity should produce a changing EPM distance with time.
However, as shown in Figure~\ref{fig:5.27}, the derived distance to SN~1999em
stays reasonably stable for all 3 filter combinations.  Unless offsetting
effects exist, the lack of obvious and consistent temporal trends across the
three filter sets argues against the idea that the degree of asphericity
increases with increasing depth in the SN ejecta.  We therefore conclude that
the most likely sources of systematic error are the uncertainty in the flux
dilution factors and the extinction, and that these uncertainties clearly
dominate over the statistical uncertainty resulting from the application of the
EPM in any single filter combination alone.  It should be noted that the
uncertainty given for our preferred distance, $D = 8.2 \pm 0.6$ Mpc,
incorporates only the evident systematic uncertainty resulting from the
application of the EPM with the three different filter combinations, and
includes neither the uncertainty in extinction nor the disagreements among
different groups on the overall level of the dilution factor (e.g., Baron et
al. 1995).


\subsection{Comparison between EPM and Cepheid Distances}
\label{sec:cepheids}

Cepheid variables remain the most widely used and well studied extragalactic
primary distance indicators.  In particular, more than two dozen galaxies now
have Cepheid distances obtained with the {\it Hubble Space Telescope (HST)} by
the Distance Scale Key Project (e.g., Ferrarese et al. 2000; Freedman et
al. 2001) and other groups (e.g., Saha et al. 1999; Tanvir, Ferguson, \& Shanks
1999).  Observing strategies and reduction techniques are well defined (e.g.,
Tanvir 1999).  The slope and form of the Cepheid period-luminosity (PL)
relation is quite well established from observations of LMC variables (e.g.,
Udalski et al. 1999).  However, there is still some controversy over the
distance to the LMC itself, which sets the PL zero-point (e.g., Silbermann et
al. 1999).  Considering all sources of error, the Key Project has recently
announced ``final'' results for the Cepheid distances, with quoted random
uncertainties of about $5\%$ for each distance; systematic uncertainty,
incorporating uncertainty introduced by the LMC zero-point offset, metallicity
correction, photometric zero-point, and aperture correction is at the $10\%$
level (Freedman et al. 2001).

\begin{deluxetable}{lccccc}
\renewcommand{\arraystretch}{1.2}
\ssp
\tablenum{9}
\tablewidth{00pt}
\tablecaption{EPM and Cepheid Distances to Type II Supernovae}
\tablehead{\colhead{Supernova}  &
\multicolumn{2}{c}{Galaxy Name} &
\multicolumn{2}{c}{Cepheid Distance} &
\colhead{EPM Distance\tablenotemark{c}} \\
\cline{2-3} \cline{4-5}
\colhead{}  &
\colhead{SN host} &
\colhead{Cepheid galaxy} &
\colhead{D$_{\rm old}$ (Mpc)\tablenotemark{a}} &
\colhead{D$_{\rm new}$ (Mpc)\tablenotemark{b}} &
\colhead{(Mpc)} }

\startdata

SN 1968L  & NGC 5236	     & NGC 5253 group	  & $4.1 \pm 0.4$
	  & $3.15 \pm 0.20$    & $ 4.5^{+0.7}_{-0.8} $ \\

& & & & & \\

SN 1969L  & NGC 1058         & NGC 925 group      & $9.29 \pm 0.69$
	  & $9.16 \pm 0.17$  & $10.6^{+1.9}_{-1.1}$ \\

& & & & & \\

SN 1970G  & NGC 5457 (M101)  & NGC 5457 (M101)    & $7.5 \pm 0.8$
	  & $6.70 \pm 0.34$  & $7.4^{+1.0}_{-1.5}$ \\

& & & & & \\

SN 1973R  & NGC 3627         & NGC 3627           & \nodata
	  & $10.05 \pm 0.37$ & $15 \pm 7$ \\

& & & & & \\

SN 1979C  & NGC 4321 (M100)         & NGC 4321 (M100)          & $17.1 \pm 1.8$
	  & $15.21 \pm 0.49$ & $ 15 \pm 4$ \\

& & & & & \\

SN 1987A  & LMC              & LMC                & $ 0.051 \pm 0.003$
	  & $0.050 \pm 0.002$& $ 0.049 \pm 0.006$ \\

& & & & & \\

SN 1988A  & NGC 4579         & Virgo cluster      & $17.1 \pm 1.8$
	  & $15.28 \pm 0.35 $  & $ 20 \pm 3$ \\

& & & & & \\

SN 1989L  & NGC 7331         & NGC 7331           & \nodata
	  & $14.72 \pm 0.61$ & $ 17 \pm 4$  \\

\enddata

\tablenotetext{a}{Cepheid distance used by Eastman, Schmidt, \& Kirshner (1996).}
\tablenotetext{b}{Metallicity-corrected Cepheid distance and random uncertainty
reported by Freedman et al. (2001); systematic uncertainty in the Cepheid
distance scale is $\sim 10\%$.}
\tablenotetext{c}{Distances and uncertainties reported by Eastman, Schmidt, \&
Kirshner (1996).} 

\end{deluxetable}


Including the most recent additions, Cepheid distances now exist to 1 galaxy
that hosted a SN~II-P (SN~1973R), 5 galaxies that hosted photometrically or
spectroscopically peculiar SNe~II (SN 1970G, SN 1979C, SN 1987A, SN 1989L, and
SN 1993J), and 3 galaxies in the same {\it group} in which a SN~II-P occurred
(SN 1968L, SN 1969L, and SN 1988A).  Given the peculiar nature of some of these
SNe and the uncertainties involved in comparing distances derived to different
galaxies within the same group, it is quite impressive that E96 found $D_{\rm
Cepheids}/D_{\rm EPM} = 0.98 \pm 0.08$, based on the distances to 6 galaxies
(or galaxies in the same group) that had Cepheid distances at the time of their
study.  Table~9 lists the Cepheid distances that were used by E96, the most
recent Cepheid distances (taken from Freedman et al. 2001), and the EPM
distances to 8 galaxies; we do not include SN~1993J in M81 since detailed
models of its (likely unusual) early-time dilution factor evolution are needed
to produce an EPM distance truely independent of the Cepheid scale (see
Clocchiatti et al. 1995).  

Due to the recent modifications of the zero-point offset and refinement of the
Cepheid distance technique itself, it is seen that the Cepheid distances to
these galaxies have changed somewhat; in fact, nearly all of the Cepheid
distances have {\it decreased} from their previous values.  Using the revised
Cepheid distances for the 6 SNe that were compared by E96, and using the same
comparison technique (i.e., finding the simple weighted mean and $1\sigma$
uncertainty in the weighted mean making no special consideration for the added
uncertainty associated with comparing Cepheid distances to galaxies in the same
group as a galaxy that hosted a SN II), now yields $D_{\rm Cepheids}/D_{\rm
EPM} = 0.88 \pm 0.07$; including all 8 available galaxies yields $D_{\rm
Cepheids}/D_{\rm EPM} = 0.87 \pm 0.06$.  Of course, comparing group-member
galaxy distances is a risky business and certainly increases the actual
uncertainty of the comparison.  If we eliminate the three group member
comparisons (i.e., SN~1968L, SN~1969L, and SN~1988A), we then derive $D_{\rm
Cepheids}/D_{\rm EPM} = 0.96 \pm 0.09$.  Clearly, additional direct
comparisons, particularly to spectroscopically {\it normal} SNe II-P, are
needed to produce a more statistically meaningful comparison.  Despite these
being the ``final'' Cepheid distances, it should also be noted that other
workers find evidence for a significantly longer Cepheid distance scale (e.g.,
Parodi et al. 2000), and that even within the Key Project itself longer
distance scales have been given (e.g., Ferrarese et al. 2000).

Nonetheless, the ``best'' Cepheid distances currently available suggest that
EPM distances may be systematically too high.  Since systematic uncertainty
clearly remains at at least the $10\%$ level for Cepheids, and at the $10 -
20\%$ level for EPM, it is not clear how significant the observed discrepancy
actually is.  From our study of the effects that using strong lines to estimate
photospheric velocity has on the derived distance, we might conclude that some
of the difference between the EPM and Cepheid scales is due to inaccuracies in
the photospheric velocity estimates; indeed, the distances derived by using the
strong \fetwo\ lines would tend to produce EPM distances systmatically $5 -
10\%$ too large.  However, since it is not certain exactly which lines were
used in each of the previous EPM studies to estimate photospheric velocity (and
we note that using the \ion{Sc}{2} $\lambda 5658$ line with $\lambda_\circ =
5657.9$ \AA\ will produce velocities that are {\it lower} than those found from
the weak, unblended lines [\S~\ref{sec:inferringvel}]), it is difficult to
generalize.

\subsection{SNe II-P as Standard Candles}
\label{sec:stdcandles}

To obtain the EPM distance to a SN~II-P requires a well-spaced series of
high-quality spectral and photometric data, and even with an excellent data set
distance uncertainties of $10\% - 20\%$ currently still remain.  As we look
toward pushing EPM out to higher redshifts, it is clear that the most demanding
observational requirement is the need to obtain multiple spectral epochs in
order to determine the photospheric velocity.  This is markedly different from
the case for SNe~Ia, which require only a single spectral epoch near (or,
preferably, before) maximum light to identify it (e.g., Coil et al. 2000).  An
interesting suggestion has been made by H\"{o}flich et al. (2000) that
distances accurate to $\sim 30\%$ should be achievable by treating SNe~II-P as
{\it standard candles}.  Based on theoretical models with a wide range of
parameters (explosion energy, metallicity, mass loss of progenitor, etc.),
H\"{o}flich et al. (2000) find that although the {\it peak} luminosity varies
greatly among SNe~II-P, the mean absolute brightness during the plateau phase
is quite insensitive to the initial parameters, with $\overline{M}_V (\rm
plateau) \approx -17.6 \pm 0.6$ mag.  Treating SNe~II-P as standard candles is
particularly attractive since they have a unique light-curve shape, obviating
the need for even a single spectrum.  Only a few deep images, roughly every 50
days (in the SN rest frame), are needed to discover, identify, and measure the
mean plateau brightness; there is no need to follow the SN after the plateau
ends and the SN becomes faint.

\begin{deluxetable}{lcccccc}
\renewcommand{\arraystretch}{1.2}
\ssp
\tablenum{10}
\tablewidth{00pt}
\tablecaption{Average $V$-Band Magnitude of SNe II-P During the Plateau Phase }
\tablehead{\colhead{Supernova}  &
\colhead{$A_V$\tablenotemark{a}} &
\colhead{$\overline{m}_V$\tablenotemark{b}} &
\colhead{Distance} &
\colhead{$\overline{M}_V$\tablenotemark{c}} &
\multicolumn{2}{c}{References} \\
\cline{6-7}
\colhead{}  &
\colhead{(mag)} &
\colhead{(mag)} &
\colhead{(Mpc)} &
\colhead{(mag)} &
\colhead{Photometry} &
\colhead{Distance} }

\startdata

SN 1968L &  0.03  &  12.0  &  $4.5^{+0.7}_{-0.8}$  & $-16.3^{+0.3}_{-0.3}$
	 &  1  & 2   \\

 & & & & & & \\

SN 1969L &  0.18  &  13.2  & $10.6^{+1.9}_{-1.1}$  & $-16.9^{+0.2}_{-0.4}$
	 &  1  & 2   \\

 & & & & & & \\

SN 1973R &  2.80  &  11.8  & $15^{+7}_{-7}$    & $-19.1^{+1.0}_{-1.0}$
	 &  1  & 2   \\

 & & & & & & \\

SN 1988A &  0.15  &  14.6  & $20^{+3}_{-3}$     & $-16.9^{+0.3}_{-0.3}$
	 &  1  & 2   \\

 & & & & & & \\

SN 1990E &  1.5   &  14.2  & $18^{+3}_{-2}$     & $-17.1^{+0.2}_{-0.4}$
	 &  3            & 2   \\

 & & & & & & \\

SN 1990ae&  0.5   &  19.3  & $115^{+35}_{-25}$  & $-16.0^{+0.5}_{-0.7}$
	 &  1  & 2   \\

 & & & & & & \\

SN 1992H &  0.28  &  14.6  &  $20.2^{+4}_{-4}$     & $-16.9^{+0.4}_{-0.4}$
	 &  4             & 4\\

 & & & & & & \\

SN 1992am&  0.3   &  19.5  & $180^{+35}_{-25}$ & $-16.8^{+0.3}_{-0.4}$
	 & 5          & 2   \\

 & & & & & & \\

SN 1999em&  0.31  &  13.7  & $8.2^{+0.6}_{-0.6}$   & $-15.9^{+0.2}_{-0.2}$
	 & 6                             & 6             \\

\enddata

\tablecomments{All uncertainties are $1\sigma$.}

\tablenotetext{a}{Estimated total extinction in the $V$ band.  If the reference
	 from which the distance estimate was taken also included a derivation
	 of the {\it total} extinction of the SN based on properties of the SN,
	 then that value was used.  In cases where the extinction was estimated
	 for the Galaxy and host individually, and then summed, we have updated
	 the Galactic extinction to be the values provided by the dust maps of
	 Schlegel et al. (1998), replacing those given by Burstein \& Heiles
	 (1982).}

\tablenotetext{b}{Average apparent $V$-band magnitude during the plateau phase,
	 derived as the average of all $V$-band data between days 20 and 100
	 after explosion, corrected for extinction. }

\tablenotetext{c}{Average absolute $V$-band magnitude during the plateau
	 phase, with uncertainty derived solely from the uncertainty in the EPM
	 distance estimate. }

\tablenotetext{d}{Photometry data taken from Schmidt et al. (1992), and references
	 therein.}

\tablerefs{(1) Schmidt et al. 1992, and references therein; (2) Schmidt et
al. 1994a; (3) Schmidt et al. 1993; (4) Clocchiatti et al. 1996; (5) Schmidt et
al. 1994b; (6) This work.}

\end{deluxetable}

To date, EPM distances have been derived to 10 SNe~II-P, of which 8 have
published $V$-band photometry sampling the plateau epoch.  Adopting the EPM
distance, the published photometry, and the extinction used in the EPM analysis
(in some cases modified slightly due to the more accurate reddening values
provided by the recent dust maps of Schlegel et al. [1998]), we derive the
average $V$-band magnitude during the plateau for these 8 SNe~II-P, and list
the results in Table~10 along with SN~1999em.  To estimate the average $V$-band
magnitude during the plateau, we took the average of the $V$-band data between
days 20 and 100 after the date of explosion (as determined by the EPM
analysis).  With the exception of SN~1973R, which is anomalously
bright,\footnote{Note that the recent Cepheid distance to this galaxy is
significantly less than that estimated by EPM (Table~9).  If the Cepheid
distance is correct, then SN~1973R actually has $\overline{M}_V (\rm plateau) =
-18.2 $ mag, making it somewhat more consistent with the other SNe II-P.} all
of the values cluster within 1.2 mag of each other. In fact, the weighted mean
absolute $V$-band magnitude during the plateau epoch for SNe~II-P, not
including SN~1999em, is $\overline{M}_V~{\rm (plateau)} = -16.8^{+0.4}_{-0.5}$
mag, where the reported uncertainty is the weighted standard deviation of the
data from the mean.  Including SN~1999em lowers the average to
$\overline{M}_V~{\rm (plateau)} = -16.4^{+0.6}_{-0.6}$ mag.  The data thus
favor a somewhat fainter mean plateau $V$-band brightness than the models of
H\"{o}flich et al. (2000) predict, but with a similar spread around the mean,
much of which may be due to the uncertainty in the EPM technique itself.  From
just these 9 SNe~II-P, we would conclude that distances to SNe~II-P accurate to
$\pm 28\%$ ($1\sigma$) could be achieved with the standard-candle assumption,
which compares well with the current level of systematic uncertainty in EPM,
and is much less observationally taxing.  Clearly, additional distances to
SNe~II-P are needed to better quantify the uncertainty resulting from the
standard-candle assumption.  Although the use of SNe~II-P as standard candles
will not achieve the same individual accuracy as SNe~Ia, the removal of the
need to obtain a spectrum, coupled with their certain existence at high
redshift (before SNe~Ia even occur, perhaps), may yet make them the best SN
class for cosmology at redshifts beyond $2.0$.

\subsection{Is SN~1999em a ``Typical'' SN~II-P?}
\label{sec:typical}

In \S~\ref{sec:reddening} we raised the possibility that the \bv\ color
evolution of SN~1999em might be somewhat different from the normal evolution of
SNe~II-P during the plateau phase.  In \S~\ref{sec:stdcandles} we found that
SN~1999em was somewhat fainter during the plateau than other SNe II-P
with EPM distances; at our derived distance, SN~1999em also reached a peak
extinction-corrected $B$-band magnitude of only $M_B = -16.2 \pm 0.2$ mag,
which places it on the fainter end of the range observed for SNe~II-P (Patat et
al. 1994; Miller \& Branch 1990).  This naturally leads to the question: is
SN~1999em unusual?  If so, could this have affected the EPM analysis?  More to
the point, are the observed properties of SN~1999em similar to the theoretical
models used to derive the dilution factors ($\zeta$) by E96?

Since SN~1999em evolved more rapidly toward red colors than the prototypical
Type II-P SN~1969L (see Figure~\ref{fig:5.16b}), we first consider the temporal
evolution of its color temperature.  The lowest color temperatures for
any of the models studied by E96 are those for model s15.60.1, which has
$T_{BV} = 4132 {\rm~K\ } {\rm and\ } T_{BVI} = 5418 {\rm~K}$, and 10h.60.1,
which has $T_{VI} = 6638 {\rm~K};$\footnote{We note that a small change in the
model's color-temperatures will result from the use of the Hamuy et al. (2001)
filter functions in place of those of E96; for the low-temperature regime of
interest here, however, these differences should be minimal.} only 1 other
model studied (p6.40.1) has $T_{BV} < 5000 {\rm~K\ } {\rm and\ } T_{BVI} < 6000
{\rm~K}$.  However, a quick look at Figure~\ref{fig:5.24a} and Table~7 reveals
that SN~1999em drops below $T_{BV} = 5000 {\rm~K\ } {\rm and\ } T_{BVI} = 6000
{\rm~K}$ near day 30, and below $T_{BV} = 4132 {\rm\ K\ } {\rm and\ } T_{BVI} =
5418 {\rm\ K}$ around day 50.  This means that after about day 30, the color
temperature of SN~1999em is only represented by 2 of the 63 models studied by
E96, and after day 50 it has no representation in the model set used to derive
$\zeta(T_c)$, but rather relies on extrapolation.  At these low temperatures,
$\zeta_{BV}\ {\rm and\ } \zeta_{BVI}$ are changing very rapidly with
temperature.  A contribution to the distance uncertainty could thus
come from the lack of models near the observed color temperature of SN~1999em
during the late portion of the recombination phase.  We note that increasing
the reddening of SN~1999em to the upper limit given by Baron et al. (2000; see
also \S~\ref{sec:reddening}) of \ebv\ = 0.15 mag raises $T_{BVI}$ by $400$~K
and $T_{BV}$ by only about $200$~K on day 30.

From its relative faintness and rapid color evolution during the plateau there
is therefore some evidence that SN~1999em was a somewhat unusual II-P event.  It is
possible that the average $\zeta$ values derived by E96 do not adequately
represent the evolution seen in SN~1999em, and may point toward the need for
models specifically crafted for it to increase the accuracy of its distance.

One way to help place the theoretically derived flux dilution factor on a more
sound empirical footing is to directly compare the EPM distance to NGC~1637
with that derived using other primary extragalactic distance indicators.  At a
distance of 8.2 Mpc, NGC~1637 makes an ideal target for a Cepheid study using
{\it HST}: it is rather face-on ($i \approx 32^\circ$), has low foreground
extinction, and has a metallicity of $\log {\rm (O/H)}+12 \approx 9.1$ in the
region near the SN~(van Zee et al. 1998), typical of other {\it HST} galaxies
for which Cepheid distances have been derived (e.g., Ferrarese et al. 2000; see
also Tanvir 1999).  An approved Cycle~10 {\it HST} program to derive a
Cepheid-based distance to NGC~1637 is currently underway, and the results will
be directly compared with the EPM distances derived to SN~1999em.

\section{Conclusions}
\label{sec:conclusions}

We present 30 optical spectra and 49 photometric epochs of SN~1999em sampling
the first 517 days of its development.  SN~1999em displays the major
photometric and spectral characteristics of a SN~II-P, exhibiting a distinct
photometric plateau lasting about 100 days from explosion, and well-developed
P-Cygni spectral line features.  We suspect that SN~1999em suffers from minimal
reddening, and concur with the range of possible values found by Baron et
al. (2000), and adopt $\Ebv = 0.10 \pm 0.05$ mag.

We find a number of interesting features in the spectral evolution of the
P-Cygni line profiles, particularly \halpha.  We identify the dominant ions
responsible for most of the absorption features seen in the optical portion of
the spectrum of SN~1999em during the plateau phase.  The photospheric velocity
derived from four weak, unblended absorption features is found to be $5\% -
10\%$ lower than that derived using the stronger \fetwo\ lines, which have
traditionally been used to estimate the photospheric velocity in EPM studies.
The velocity offset between the weakest lines and the stronger \ion{Fe}{2}
features is shown to exist in other SNe~II as well.  If this is a generic
feature of SNe~II and weaker lines are indeed a better tracer of photospheric
velocity, then EPM estimates previously made with the stronger \fetwo\ lines
are skewed by $5\% - 10\%$ toward larger distances.  We note that the
comparison of EPM distances with the most recent Cepheid distances yields
$D_{\rm Cepheids}/D_{\rm EPM} = 0.87 \pm 0.06$, a discrepancy that is in the
same sense (i.e., $D_{\rm EPM} > D_{\rm Cepheids}$) as the aforementioned
velocity offset; difficulties in this comparison exist, however, due to the
peculiar nature of several of the SNe II, the use of Cepheid distances to
galaxies that are only in the same group in which the SN occurred (i.e., not
the host galaxy itself), and existing discrepancies in the Cepheid distance
scale.

We derive an EPM distance to SN~1999em using the flux dilution factors
($\zeta$) of E96 as modified by Hamuy et al. (2001).  We use three different
filter combinations ($BV, BVI, {\rm and\ } VI$) to determine the photospheric
color temperature.  The total spread in the distances derived using the
different filter combinations is about $15\%$ (when using the weak, unblended
absorption features to estimate photospheric velocity), in the sense $D_{VI} >
D_{BVI} > D_{BV}$.  These discrepancies cannot be explained by statistical
uncertainty in the measured quantities, and must be systematic in nature, most
likely due to uncertainty in the dilution factor and perhaps aggravated by the
lack of models used to derive the dilution factor with properties similar to
SN~1999em: SN~1999em became redder more quickly than the prototypical Type II-P
SN~1969L, and reached color temperatures lower than any model studied by E96
about half way through the plateau phase.  Taking the observed scatter among
the different filter combinations into account, our best estimate for the
distance to SN~1999em is $D = 8.2^{+0.6}_{-0.6}$~Mpc, with an explosion date of
$t_\circ = {\rm HJD}~2,451,475.6^{+1.4}_{-1.4}$, or $5.3\pm{1.4}$ days before
discovery.  This leads to an extinction-corrected peak $B$-band magnitude of
$M_B = -16.2 \pm 0.2$~mag, and an average $V$-band plateau brightness of
$\overline{M}_V (\rm plateau) = -15.9 \pm 0.2$ mag, which is somewhat fainter
than average SNe~II-P.

Finally, we investigate the theoretical prediction by H\"{o}flich et al. (2000)
that SNe~II-P should exhibit a rather small dispersion in their average
$V$-band magnitude during the plateau phase.  Their unique light-curve shape
(making spectral identification unnecessary) and certain existence at high
redshift make them attractive cosmological tools.  From 8 SNe~II-P with
previously published distances and SN~1999em, we find $\overline{M}_V~{\rm
(plateau)} = -16.4^{+0.6}_{-0.6}$.  If this small sample is representative of
all SNe~II-P, then distances to SNe~II-P accurate to $\sim 30\%$ ($1\sigma$)
may be possible without having to acquire the extensive data set required for
an EPM analysis, thereby increasing the viability of this class of objects as
cosmological beacons at $z > 2$.

\acknowledgments

We thank Wil J. M. van Breugel, Daniel Stern, David R. Ardila, Robert
H. Becker, Michael S. Brotherton, James W. Colbert, Willem H. de Vries, Richard
Edelson, George K. Miley, Michiel Reuland, Adam G. Riess, S. Adam Stanford, and
Richard L. White for assistance with the observations, and Shashi M. Kanbur,
Jeffery A. Newman, and Nial A. Tanvir for helpful discussions.  We thank the
referee, Brian Schmidt, for very helpful comments and suggestions that resulted
in an improved manuscript.  Some of the data presented herein were obtained at
the W. M. Keck Observatory, which is operated as a scientific partnership among
the California Institute of Technology, the University of California, and the
National Aeronautics and Space Administration.  The Observatory was made
possible by the generous financial support of the W. M. Keck Foundation. We are
grateful to the Keck staff for their support of the telescopes.  This research
has made use of the NASA/IPAC Extragalactic Database (NED), which is operated
by the Jet Propulsion Laboratory, California Institute of Technology, under
contract with NASA.  We have made use of the LEDA database
(\url{http://leda.univ-lyon1.fr}).  Our work was funded by NASA grants GO-7821,
GO-8243, and GO-8648 from the Space Telescope Science Institute, which is
operated by AURA, Inc., under NASA contract NAS 5-26555.  Additional funding
was provided to A. V. F. by NASA/Chandra grant GO-0-1009C, NSF grants
AST-9417213 and AST-9987438, and by the Guggenheim Foundation.  KAIT was made
possible by generous donations from Sun Microsystems, Inc., the Hewlett-Packard
Company, AutoScope Corporation, the National Science Foundation, the University
of California, Lick Observatory, and the Sylvia and Jim Katzman Foundation.

\end{document}